\documentclass[aps,prd,showpacs,superscriptaddress,preprintnumbers]{revtex4}
\usepackage{graphicx,float,wrapfig,subfigure}
\usepackage{amsfonts,amsmath,amssymb,amstext}
\usepackage{latexsym}
 \usepackage{bm}
\usepackage{color}
\usepackage[normalem]{ulem}

\newcommand{\tr}{\,\mbox{tr}}
\newcommand{\sign}{\,\mbox{sign}}
\definecolor{red}{rgb}{0.7,0,0}
\definecolor{blue}{rgb}{0,0,0.7}

\begin{document}

\title{Photon polarization tensor in a magnetized plasma: Absorptive part}
\date{\today}

\author{Xinyang Wang}
\email{wangxy@ujs.edu.cn}
\affiliation{Department of Physics, Jiangsu University, Zhenjiang 212013, People's Republic of China}

\author{Igor Shovkovy}
\email{igor.shovkovy@asu.edu}
\affiliation{College of Integrative Sciences and Arts, Arizona State University, Mesa, Arizona 85212, USA}
\affiliation{Department of Physics, Arizona State University, Tempe, Arizona 85287, USA}

\begin{abstract}
We calculate the absorptive part of the photon polarization tensor in a hot magnetized relativistic plasma. In the derivation, we utilize a Landau-level representation for the fermion Green's function in a mixed coordinate-momentum space and obtain a closed-form expression for the one-loop polarization tensor. At the leading order in the coupling, its absorptive part is determined by particle and antiparticle splitting processes ($e^{-} \leftrightarrow e^{-}+\gamma$ and $e^{+} \leftrightarrow e^{+}+\gamma $, respectively), as well as by particle-antiparticle annihilation processes ($e^{-} + e^{+}\leftrightarrow \gamma$). The interpretation in terms of quantum transitions between Landau levels is also given. By making use of the photon polarization tensor, we study the differential photon emission rate in the quantum limit of magnetized relativistic plasma.  At low energies, the photon emission has a prolate profile with the symmetry axis along the line of the magnetic field. At high energies, on the other hand, the photon emission has an oblate profile.  The underlying reasons for such emission profiles are given in both regimes. The general result for the photon polarization tensor is also used to calculate the longitudinal and transverse components of magneto-optical conductivity. 
\end{abstract}
\maketitle

\section{Introduction}
\label{sec:introduction}

Relativistic plasmas appear in a wide range of high-energy systems. The archetypal examples include the hot plasma of the early Universe and the quark-gluon plasma created in heavy-ion collisions. One also finds various forms of relativistic plasmas in astrophysics, ranging from cold dense matter inside compact stars to electromagnetic plasmas in astrophysics jets powered by black holes. Even in condensed matter physics, pseudorelativistic electron plasmas can be realized in Dirac and Weyl semimetals. In many cases, the corresponding plasmas could be also strongly magnetized. 

The properties of relativistic plasmas in weak magnetic fields have been studied extensively over the years and are well understood. On the other hand, the regime of strongly magnetized relativistic plasma is less explored. One of the specific long-standing and largely unresolved problems is photon polarization. Formally, the photon polarization tensor can be calculated by making use of the well-known Schwinger's result for the fermion Green's function in a background magnetic field \cite{Schwinger:1951nm}. While this appears to be a straightforward task conceptually, the calculation of the relevant Feynman diagrams is plugged with serious technical difficulties, stemming from an elaborate structure of the Green's function. From a physics viewpoint, there are several reasons for complications. One of them is the absence of the Lorentz symmetry, which is broken explicitly by the background magnetic field. At a nonzero temperature, the symmetry is further reduced by the thermal bath effects. Additionally, the quantum states of charged fermions in a magnetic field are much more complicated than the simple plane waves. They are classified by a discrete Landau level index that replaces the two transverse components of the particle momentum. The matters are made worse also because the Landau level states are highly degenerate. 

Before addressing the photon polarization effects in a hot magnetized plasma, it should be noted that even the case of a strongly magnetized vacuum is very challenging. The first calculations of the vacuum polarization in a homogeneous magnetic field were done in the 1970s \cite{Tsai:1974ap,Shabad:1975ik}, but their physical meaning was not always clear because of the complicated structure. Beyond the weak field case \cite{Tsai:1974fa,Urrutia:1977xb}, only the limit of an ultrastrong magnetic field, where the lowest Landau approximation becomes reliable, was understood well \cite{Loskutov:1976aa,Calucci:1993fi}. More recently, some progress in solving the problem of QED vacuum in a strong magnetic field beyond the lowest Landau approximation was reported in Refs.~\cite{Hattori:2012je,Hattori:2012ny,Karbstein:2013ufa,Ishikawa:2013fxa,Hattori:2020htm}, where the combination of analytical and numerical methods was employed. 

In the vacuum, the regime of a quantizing magnetic field is achieved when the field strength is comparable to the fermion mass or larger. In the case of quantum electrodynamics, this means that the magnetic field should be comparable to or larger than $B_c \simeq m_e^2/e\approx 4.414\times 10^{13}~\mbox{G}$. In hot (dense) matter, besides, the Landau energy scale should be comparable to or exceed the value of temperature (chemical potential). While such relativistic systems are not common, sufficiently strong magnetic fields are realized naturally under certain conditions. One example is the magnetars that host a strong enough field to turn the vacuum into a birefringent medium. The corresponding theoretical prediction stems from the well-known Heisenberg-Euler effective action \cite{Heisenberg:1935qt}. Tentatively, it has been supported by the observation of thermal emission with a large degree of polarization coming from the isolated neutron star RX J1856.5--3754 \cite{Mignani:2016fwz}.  

Superstrong magnetic fields can also appear in the early Universe and possibly affect its evolution \cite{Vachaspati:1991nm,Brandenburg:1996fc,Grasso:2000wj,Banerjee:2004df}. Moreover, some observable signatures of the primordial fields can be detected in present-day astronomy \cite{Neronov:1900zz,Tavecchio:2010mk,Dolag:2010ni,Taylor:2011bn,Tashiro:2013bxa,Tashiro:2013ita}. Similar conditions, albeit on much smaller spatial scales, could be also created in ultrarelativistic heavy-ion collisions. Indeed, such collisions produce not only a hot quark-gluon plasma but also strong magnetic fields. Theoretical predictions suggest that the strength of the magnetic field in noncentral heavy-ion collisions could be comparable to the QCD energy scales, i.e., $B \gtrsim m_\pi^2/e\approx 3.1\times 10^{18}~\mbox{G}$  \cite{Skokov:2009qp,Deng:2012pc}. The presence of such a strong field is likely to have numerous observable effects that lead to distinct features in various multi-particle correlators of charged particles \cite{Kharzeev:2015znc,Miransky:2015ava}.

To make further progress in studies of strongly magnetized plasmas, it is beneficial to have a more detailed knowledge of the photon polarization effects. Despite promising attempts reported by several research groups, however, they remain largely unknown in the regime of nonzero temperature and chemical potential. At nonzero temperature, for example, the polarization tensor was obtained in the lowest Landau level approximation in Ref.~\cite{Bandyopadhyay:2016fyd} and the weak-field limit in Ref.~\cite{Das:2019nzv,Ghosh:2019kmf}. Some calculations were also done within the framework of the Ritus method and the real-time formalism in Refs.~\cite{Sadooghi:2016jyf} and~\cite{Ghosh:2018xhh,Ghosh:2020xwp}. 

Here we report a partial solution to the general problem in the regime of a strongly magnetized thermal plasma. In particular, by making use of the Landau level representation for the fermion Green's function, we derive a relatively simple closed-form analytical expression for the absorptive part of the polarization tensor \cite{foot1} at nonzero energy and momentum. It should be emphasized that we consider a magnetized relativistic plasma in the quantizing limit when the effects of Landau level quantization play an important role. This is in contrast to the semiclassical regime that assumes a weak magnetic field when the quantization effects play no role. The absorptive part of the polarization function gets contributions from the following particle-particle, antiparticle-antiparticle, and particle-antiparticle processes:  (i) $e^{-}\rightarrow e^{-}+\gamma$, (ii) $e^{+}\rightarrow e^{+}+\gamma$, and (iii) $e^{-}+e^{+}\rightarrow \gamma$. By extending this work in the future, we hope to obtain also the real part of the polarization tensor within a similar framework.  

In this study, we will determine the photon polarization function in a magnetized plasma by direct calculation. We will find that it contains 4 symmetric and 2 antisymmetric tensor structures. As required by the gauge symmetry, all tensor structures are transverse with respect to the external photon momentum. Our results may have direct applications to several observable effects, including the photon emission and absorption, optical conductivity, and Landau damping in strongly magnetized relativistic plasmas. We expect that the polarization effects in this regime could be useful also for illuminating the anomalous properties of chiral plasmas \cite{Miransky:2015ava}.

To demonstrate possible applications of the main result, we use the absorptive part of the polarization tensor to calculate the differential photon emission rate in a hot QED plasma in a strong magnetic field. As we will argue below, one of the contributions is analogous to the synchrotron emission in the quantum limit \cite{Sokolov:1986nk}, while the other to the one-photon particle-antiparticle annihilation. This particular result was applied in Ref.~\cite{Wang:2020dsr} to a study of the hot quark-gluon plasma, produced in relativistic heavy-ion collisions. In particular, it was argued that the angular dependence of the photon emission in a strong magnetic field can resolve the so-called $v_2$-puzzle in the observed photon spectra. (For earlier studies, see also Refs.~\cite{Tuchin:2012mf,Basar:2012bp,Yee:2013qma,Tuchin:2014pka,Zakharov:2016mmc}.) To further verify the validity of our result, here we also calculate the optical conductivity of the magnetized hot QED plasma and compare it with similar results obtained for Dirac semimetals. 

This paper is organized as follows. In Sec.~\ref{sec:Polarization}, we start from the definition of the one-loop photon polarization tensor in terms of the fermion Green's functions in a magnetic field. In the same section, we also obtain a general expression for the absorptive part of the polarization tensor. The photon emission in a magnetized hot QED plasma is studied in detail in Sec.~\ref{sec:emission-QED}. The derivation of the magneto-optical conductivity and the corresponding numerical results are presented in Sec.~\ref{sec:Numerical-results}. The summary of the main results and conclusions are given in Sec.~\ref{sec:summary}. Some technical details and calculations are provided in several appendices at the end of the paper.

\section{Polarization function}
\label{sec:Polarization}

\begin{figure}[t]
\centering
\includegraphics[width=0.33\textwidth]{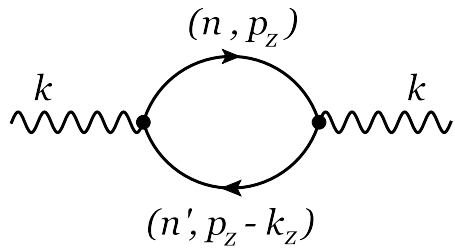}
\caption{The leading-order one-loop Feynman diagram for the photon polarization tensor in a magnetic field.}
\label{illustration}
\end{figure}

In a hot magnetized plasma, each charged particle species (or flavor) affects the photon polarization function. At leading order in the coupling, the individual contributions to the polarization are represented by the same one-loop Feynman diagram in Fig.~\ref{illustration}. This implies that the effects of different particle species are additive in this approximation. Therefore, without loss of generality, it is sufficient to analyze a partial contribution of a single fermion species of mass $m$ and charge $q$. In the case of QED plasma, for example, the values of the corresponding model parameters are $m=0.511~\mbox{MeV}$ and $q=-e$, where $e>0$ is the absolute value of the electron charge. Such a QED plasma, which is made of equal number densities of electrons $e^{-}$ and positrons $e^{+}$, can be viewed as a default choice in the analysis below. However, by changing the values of $m$ and $q$, the same results will also apply to other relativistic plasmas made of charged fermions (e.g., the quark-gluon plasma, where the polarization effects are determined primarily by the lightest up, down, and perhaps strange quarks). 

Let us briefly comment on the validity of the one-loop approximation used here. In the context of QED, which is the main emphasis in the current study, it is an excellent approximation because the fine structure constant is very small. Of course, the same approximation is not necessarily as good in QCD. Nevertheless, as argued in Ref.~\cite{Wang:2020dsr}, it may work with some caveats in the regime of a very strong magnetic field. Possible ways of improving the approximation by going beyond the leading order were also outlined in Ref.~\cite{Wang:2020dsr}, but the prospects of real progress in calculating subleading corrections are not clear at this point.

\begin{figure}[b]
\centering
\includegraphics[width=0.33\textwidth]{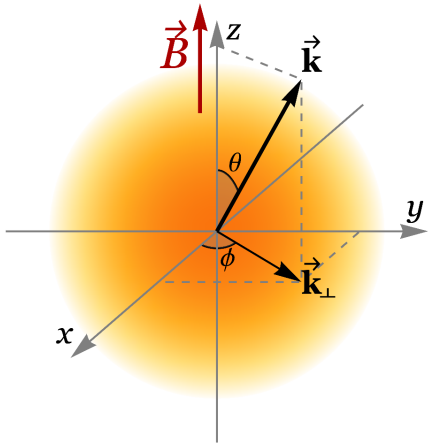}
\caption{A schematic illustration of the coordinate system used. The photon momentum and its projection on the plane perpendicular to the magnetic field are denoted by $\mathbf{k}$ and $\mathbf{k}_\perp$, respectively.}
\label{setup}
\end{figure}

Without loss of generality, we will assume that the magnetic field $\mathbf{B}$ points in the $+z$ direction, see Fig.~\ref{setup}. For the vector potential, we will use the simplest Landau gauge: $\mathbf{A}=(-B y,0,0)$, where $B$ is the magnetic field strength. As is easy to verify, the corresponding field strength tensor reads $F^{\mu\nu} = - \varepsilon^{0\mu\nu3} B$. In such a constant background field, the fermion Green's function takes the following form~\cite{Miransky:2015ava}:
\begin{equation}
G(t-t^\prime;\mathbf{r},\mathbf{r}^\prime) = e^{i\Phi (\mathbf{r}_\perp,\mathbf{r}_\perp^\prime)}\bar{G}(t-t^\prime;\mathbf{r}-\mathbf{r}^\prime), 
\label{quark-prop}
\end{equation} 
where $\mathbf{r}=(x,y,z)$ is the position vector and $\mathbf{r}_\perp=(x,y)$ is its projection on the plane perpendicular to the magnetic field. The explicit expression for the Schwinger phase is given by $\Phi(\mathbf{r}_\perp,\mathbf{r}_\perp^\prime)=-q B(x-x^{\prime})(y+y^{\prime})/2$. As one can see, the Schwinger phase is the only part of the Green's function that breaks the translation invariance. Note that this phase does not affect the calculation of the photon polarization tensor at the leading one-loop order. The same will not remain true at higher loops, however. 

It is convenient to write the translation invariant part of Green's function $\bar{G}(t;\mathbf{r})$ in Eq.~(\ref{quark-prop}) by using a mixed coordinate-momentum space representation~\cite{Miransky:2015ava}, 
\begin{equation}
\bar{G}(t;\mathbf{r}) = \int \frac{d\omega dp_z }{(2\pi)^2} e^{-i\omega t +i p_z z}
\bar{G}(\omega; p_z ;\mathbf{r}_\perp) ,
\end{equation}
where the Fourier transform is given as a sum over the Landau levels, i.e.,
\begin{equation}
\bar{G}(\omega, p_z ;\mathbf{r}_\perp) = i\frac{e^{-\mathbf{r}_\perp^2/(4\ell^2)}}{2\pi \ell^2}
\sum_{n=0}^{\infty}
\frac{\tilde{D}_{n}(\omega,p_z ;\mathbf{r}_\perp)}{\omega^2-p_z^2-m^2-2n|qB|}. 
\label{GDn-alt}
\end{equation}
The numerator in the $n$th Landau level contribution has the following explicit form:
\begin{equation}
\tilde{D}_{n}(\omega,p_z ;\mathbf{r}_\perp) = \left(\omega\gamma^0 -p_z\gamma^3 + m \right)
\left[\mathcal{P}_{+}L_n\left(\frac{\mathbf{r}_{\perp}^2}{2\ell^{2}}\right)
+\mathcal{P}_{-}L_{n-1}\left(\frac{\mathbf{r}_{\perp}^2}{2\ell^{2}}\right)\right]
-\frac{i}{\ell^2}(\mathbf{r}_{\perp}\cdot\bm{\gamma}_{\perp}) 
 L_{n-1}^1\left(\frac{\mathbf{r}_{\perp}^2}{2\ell^{2}}\right),
\end{equation}
where $L_{n}^\alpha(z)$ are the generalized Laguerre polynomials \cite{Gradshtein},
$\mathcal{P}_{\pm}\equiv \frac12 \left(1\pm i s_\perp \gamma^1\gamma^2\right)$ 
are the spin projectors, and $\ell =1/\sqrt{|q B|}$ is the magnetic length. By definition, 
$s_\perp=\sign (q B)$ and $L_{-1}^\alpha(z) \equiv 0$. 
 
By making use of the fermion Green's function in the Landau-level representation, it is straightforward to write down the momentum-space expression for the polarization function in a magnetized plasma. Within the Matsubara finite-temperature formalism, the corresponding one-loop result reads
\begin{equation}
\Pi^{\mu\nu}(i\Omega_m;\mathbf{k}) =  4\pi N_f \alpha  T \sum_{k=-\infty}^{\infty} \int \frac{dp_z}{2\pi} 
\int d^2 \mathbf{r}_\perp e^{-i \mathbf{r}_\perp\cdot \mathbf{k}_\perp} 
\tr \left[ \gamma^\mu \bar{G} (i\omega_k, p_z ;\mathbf{r}_\perp)  
\gamma^\nu \bar{G} (i\omega_k-i\Omega_m, p_z-k_z; -\mathbf{r}_\perp)\right],
\label{Pi_Omega_k-alt}
\end{equation}
where $\alpha  =q^2/(4\pi)$ is the coupling constant, $N_f$ is the number of active fermion flavors in the plasma, and the trace in the integrand runs over the Dirac indices. As per standard convention, the fermionic and bosonic Matsubara frequencies are $\omega_k= (2k+1)\pi T$ and $\Omega_m=2n \pi T$, respectively. By substituting Green's function (\ref{GDn-alt}) into the definition for $\Pi^{\mu\nu}$ and performing the Matsubara summation with the help of Eq.~(\ref{Matsubara-sum}) in Appendix~\ref{MatsubaraSums}, we derive the following expression for the polarization tensor: 
\begin{equation}
\Pi^{\mu\nu}(i\Omega_m;\mathbf{k}) =
-  \frac{\alpha N_{f}}{\pi  \ell^4} \sum_{n,n^\prime=0}^{\infty}\int \frac{dp_z}{2\pi} 
\sum_{\lambda=\pm 1} 
\frac{ (E_{n,p_z}-\lambda E_{n^{\prime},p_z-k_z})\left[ n_F(E_{n,p_z})-n_F(\lambda E_{n^{\prime},p_z-k_z})  \right]}{2 \lambda  E_{n,p_z}E_{n^{\prime},p_z-k_z}\left[(E_{n,p_z}-\lambda E_{n^{\prime},p_z-k_z})^2+\Omega_m^2\right]}
\sum_{i=1}^{4}I_{i}^{\mu\nu},
\label{Pi_deriv-1}
\end{equation}
where $E_{n}= \sqrt{m^2+p_z^2 + 2 n |q B|}$ is the fermion energy in the $n$th Landau level. Note that, in the last expression, we also calculated the Dirac traces and integrated over the transverse spatial coordinates ($\mathbf{r}_\perp$) by using the results in Appendices~\ref{prop-use} and \ref{AP-tr}. The explicit expressions for functions $I_{i}^{\mu\nu}$ can be found in Appendix~\ref{AP-tr}. 

By imposing a suitable regularization for the Landau-level sum in Eq.~(\ref{Pi_deriv-1}), in principle, one could use the brute-force numerical methods to evaluate the (retarded) photon polarization function in a hot magnetized plasma. Moreover, the partial contributions in the sum have a clear physical interpretation in terms of quantum transitions between Landau-level states. In practice, however, the corresponding calculation is not easy. Additional complications come from the need to perform the analytic continuation to real values of photon energy, $i\Omega_m\to \Omega+ i \epsilon$. A partial resolution of the problem is to extract the real and imaginary parts of the retarded polarization function and study them separately. As we will discuss below, the structure of the imaginary part  is much simpler than that of the real one. Thus, in this paper, we will concentrate primarily on the imaginary part and leave the real part of the polarization tensor for future studies. 

After replacing $i \Omega_m \to \Omega+ i \epsilon$, it is straightforward to extract the imaginary part of the retarded polarization tensor. The result reads
\begin{equation}
\mbox{Im} \left[\Pi_R^{\mu\nu}(\Omega;\mathbf{k}) \right] 
=  \frac{\alpha N_{f}}{2\ell^4} \sum_{n,n^\prime=0}^{\infty}\int \frac{dp_z}{2\pi} 
\sum_{\lambda,\eta=\pm 1} 
\frac{n_F(E_{n,p_z})-n_F(\lambda E_{n^{\prime},p_z-k_z}) }{2 \eta\lambda  E_{n,p_z}E_{n^{\prime},p_z-k_z}}
\sum_{i=1}^{4}I_{i}^{\mu\nu}
\delta\left(E_{n,p_z}-\lambda E_{n^{\prime},p_z-k_z}+\eta \Omega\right).
\label{Im-Pol-fun-general}
\end{equation}
(Strictly speaking, this is the absorptive rather then imaginary part of the polarization tensor \cite{foot1}.)
The $\delta$ function has a nonvanishing support when the following energy conservation equation
\begin{equation}
E_{n,p_z}-\lambda E_{n^{\prime},p_z-k_z}+ \eta\Omega=0
\label{energy-conservation}
\end{equation} 
is satisfied. Therefore, by finding the values of $p_z$ that solve the corresponding equation, the integration over $p_z$ can be performed analytically. 

Before proceeding further with the analysis, let us note that the imaginary part of the polarization 
tensor in Eq.~(\ref{Im-Pol-fun-general}) is an odd function of the photon frequency $\Omega$. Taking this 
into account, we will assume without loss of generality that $\Omega >0$ in the rest of the paper. 
From a physics viewpoint, the case of $\Omega >0$ could be associated with the photon emission 
processes while $\Omega <0$ with the photon absorption processes. In equilibrium, 
of course, the rates should be the same as required by the principle of detailed balance.

\begin{figure}[t]
\centering
  \subfigure[]{\includegraphics[width=0.25\textwidth]{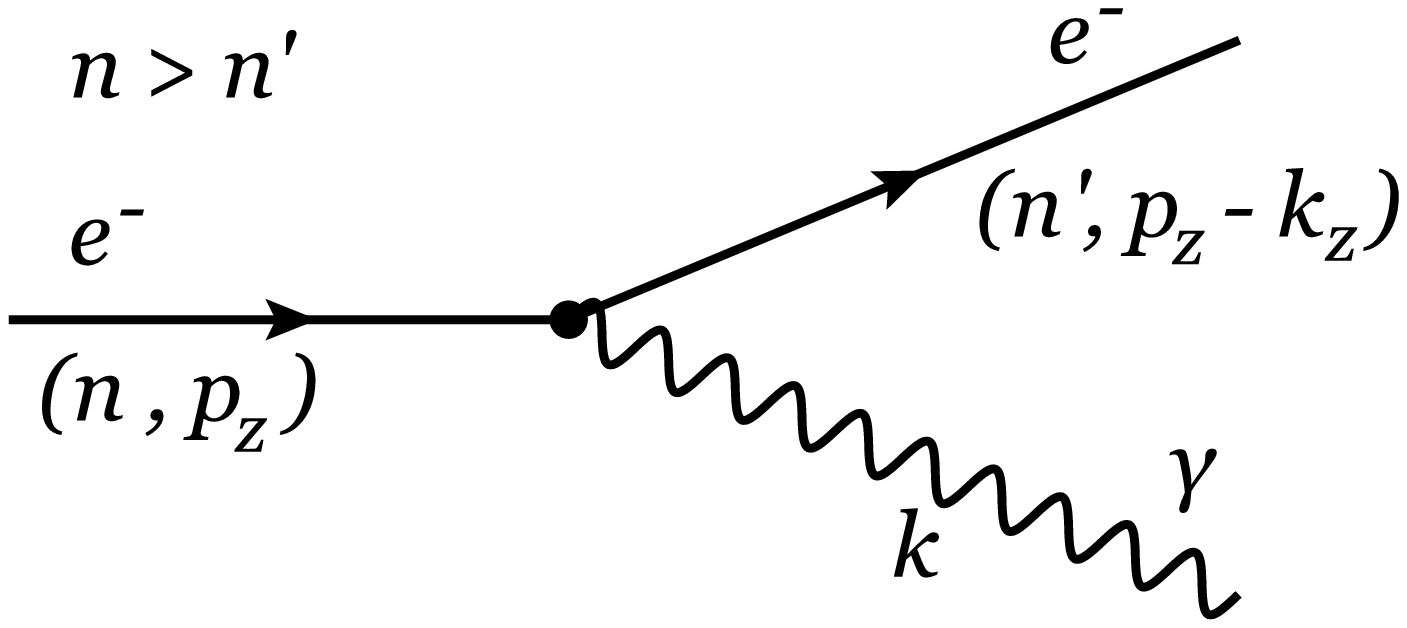}}
  \hspace{0.1\textwidth}
  \subfigure[]{\includegraphics[width=0.25\textwidth]{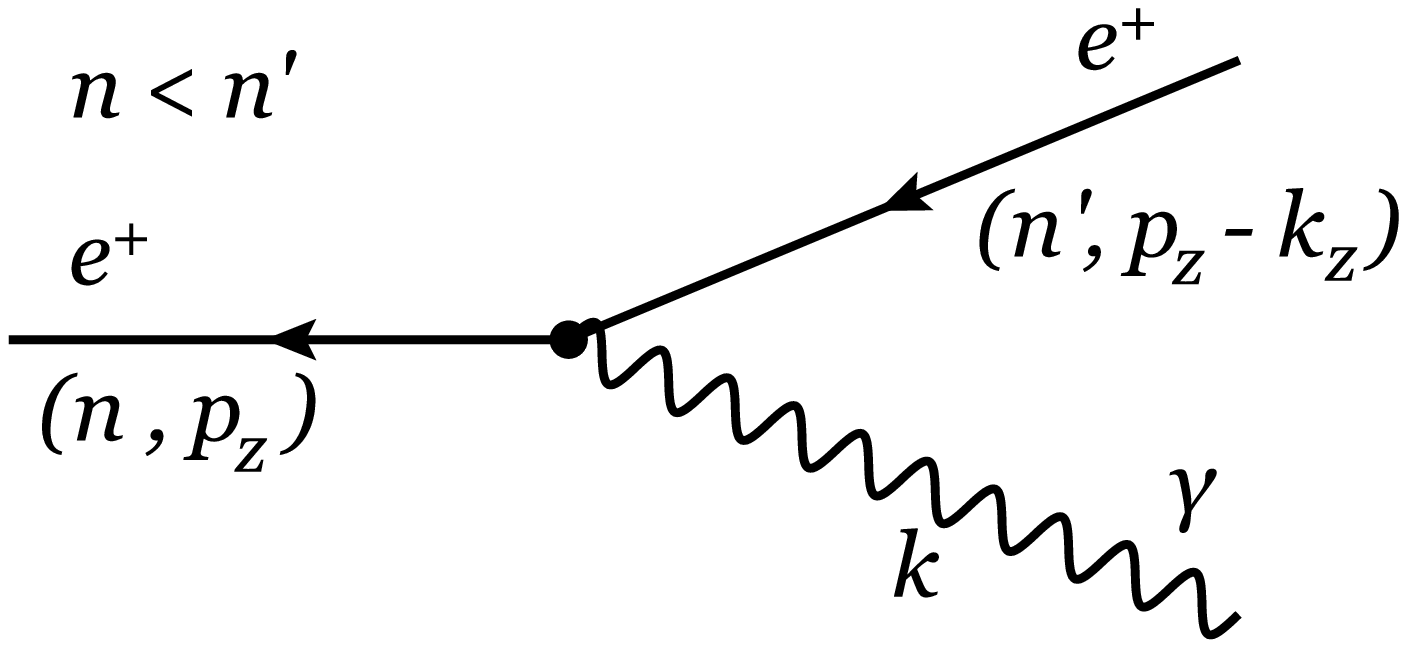}}
  \hspace{0.1\textwidth}
  \subfigure[]{\includegraphics[width=0.25\textwidth]{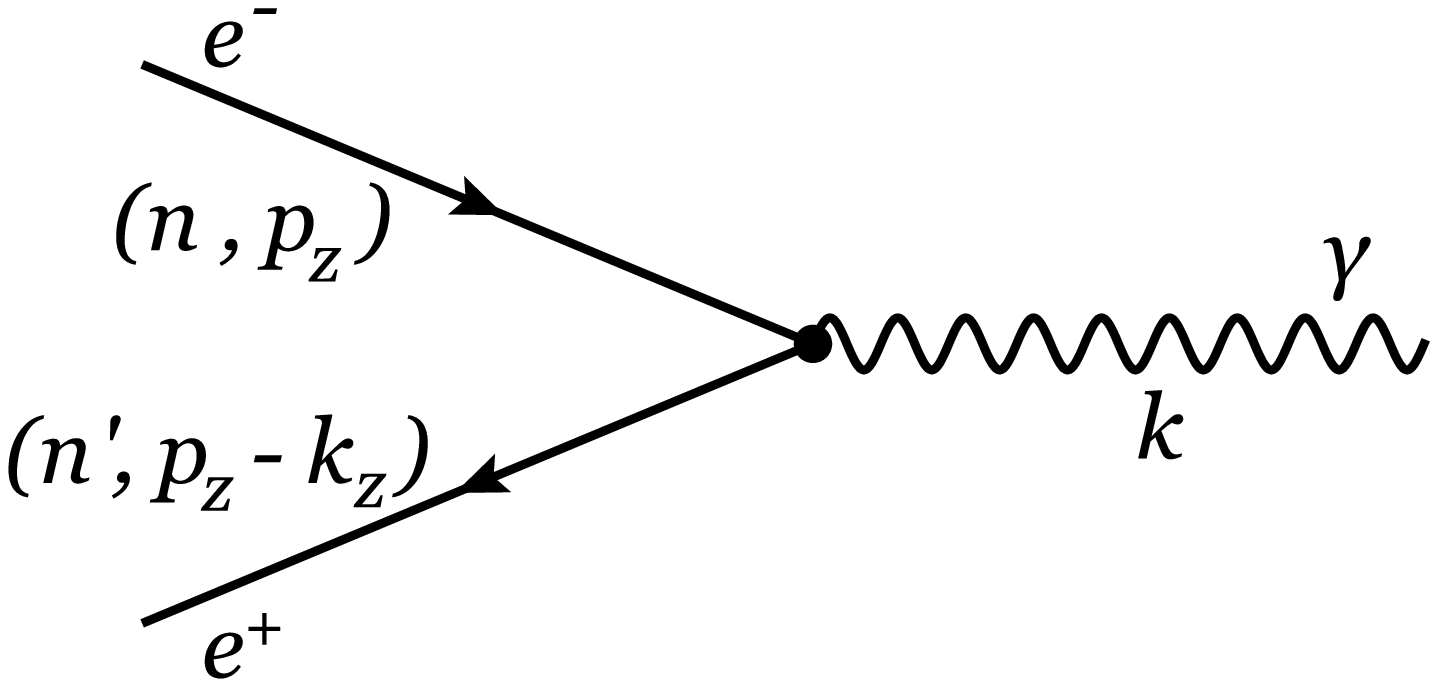}}\\
   \subfigure[]{\includegraphics[width=0.25\textwidth]{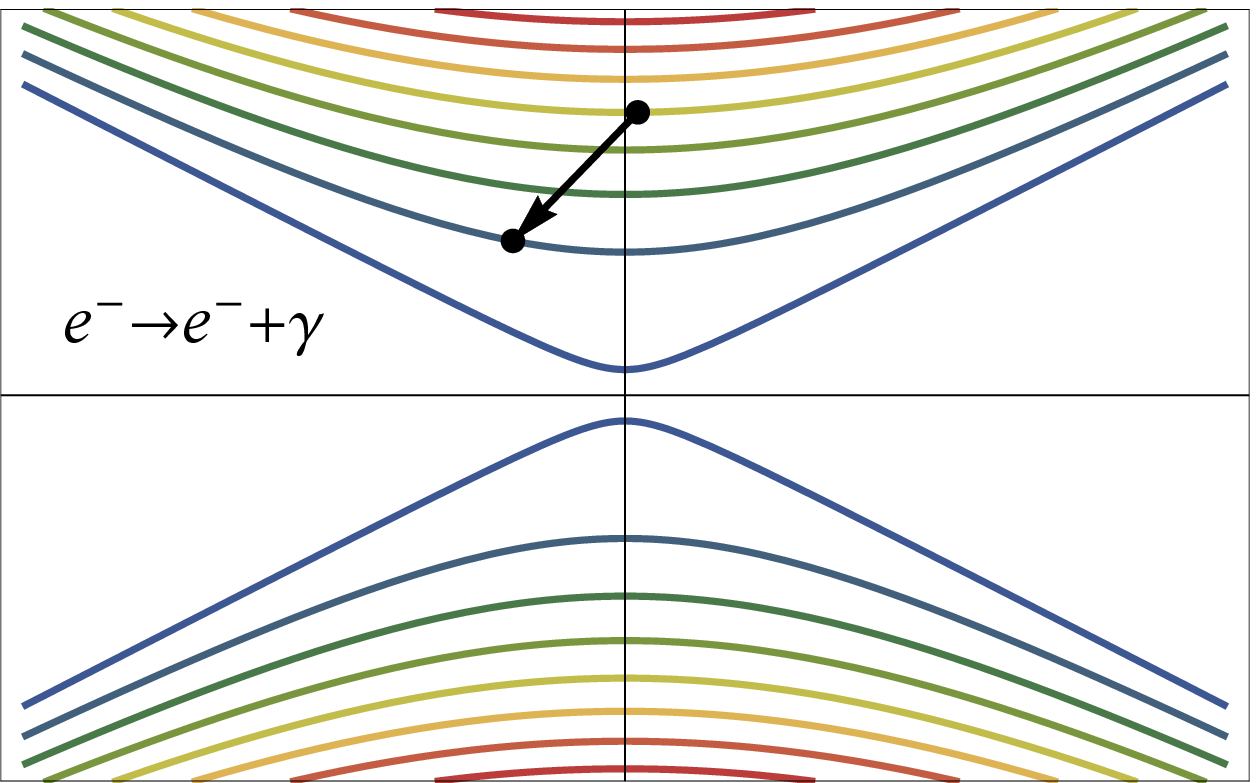}}
  \hspace{0.1\textwidth}
  \subfigure[]{\includegraphics[width=0.25\textwidth]{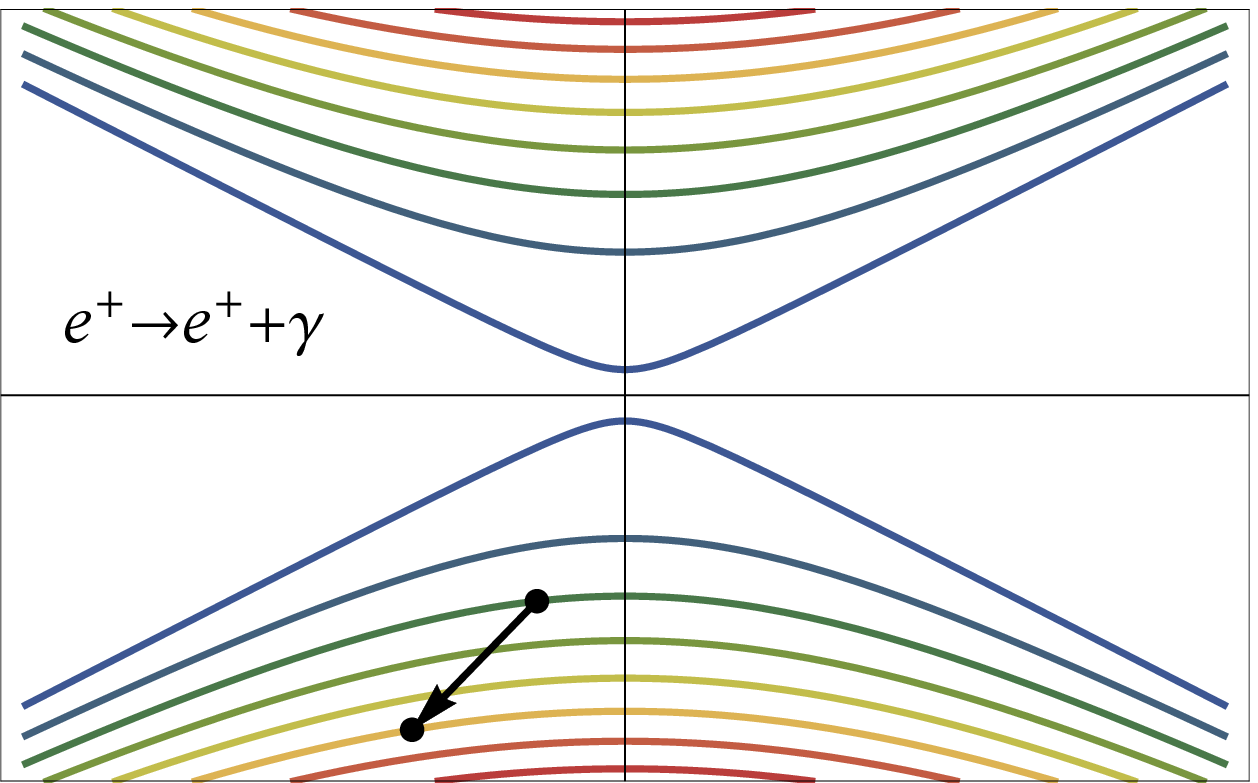}}
  \hspace{0.1\textwidth}
  \subfigure[]{\includegraphics[width=0.25\textwidth]{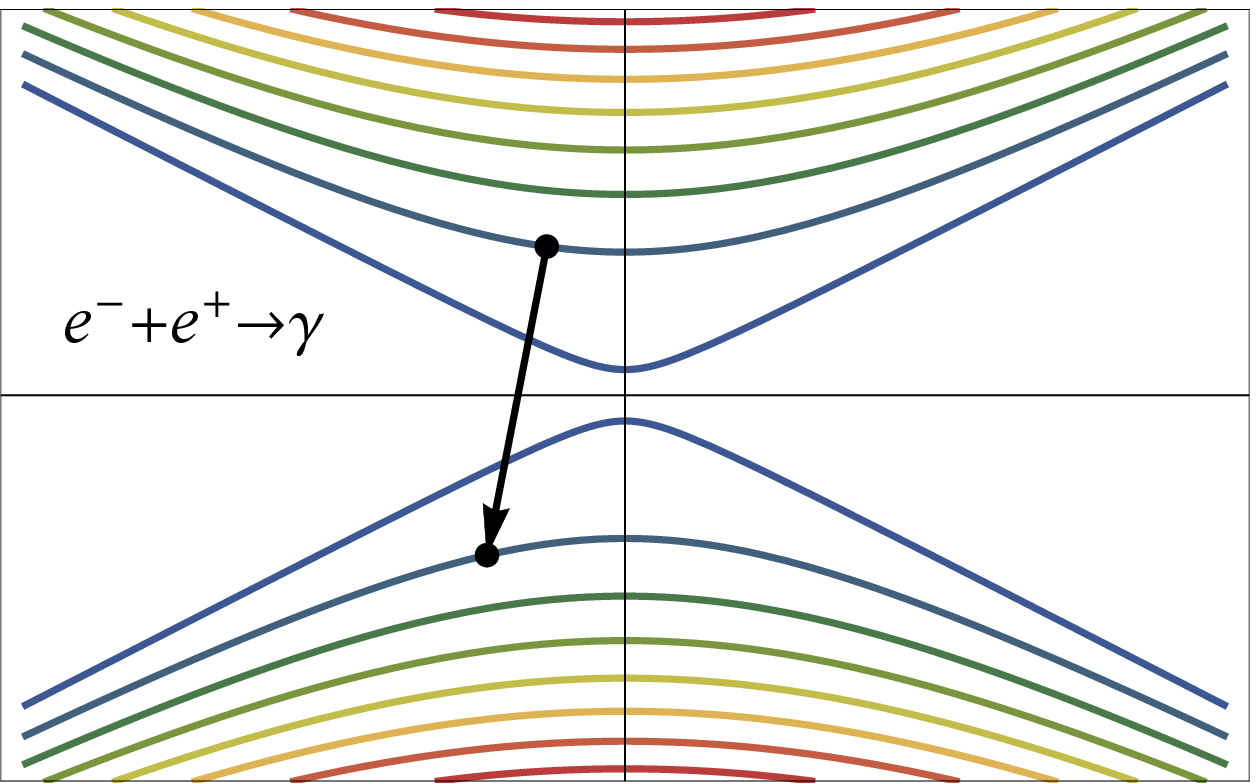}}
\caption{Three types of processes involving fermion states with the Landau level indices $n$ and $n^{\prime}$:
(a) $e^{-} \to e^{-}+\gamma$, (b) $e^{+}\to e^{+}+\gamma$, (c) $e^{-} + e^{+} \to \gamma$. The corresponding 
Landau level transitions are shown schematically in panels (d), (e) and (f), respectively.}
\label{LLprocesses}
\end{figure}

Depending on the choice of the signs of $\lambda$ and $\eta$, the energy conservation equation (\ref{energy-conservation}) represents one of the three possible processes involving particle and/or antiparticle states with the Landau level indices $n$ and $n^{\prime}$. Schematically, the corresponding processes are shown in Fig.~\ref{LLprocesses}. Two of them are the particle and antiparticle splitting processes processes, $e^{-} \to e^{-} +\gamma$ and $e^{+} \to e^{+} +\gamma$, shown in panels (a) and (b), respectively. They correspond to transitions between Landau levels with the energies of the same signs, as shown schematically in panels (d) and (e), respectively. Note that these processes represent the quantum version of the synchrotron radiation. Unlike the quasiclassical version, the current description in terms of the Landau level transitions captures all quantum effects of the synchrotron radiation. The latter includes the quantization of photon energies and the fermion recoil effects due to photon emission.

The third process type is the particle-antiparticle annihilation, shown diagrammatically in panel (c) of Fig.~\ref{LLprocesses}. In terms of Landau levels, it corresponds to transitions between the states with energies of the opposite signs, shown in panel (f).

By examining the energy conservation equation (\ref{energy-conservation}), 
it is easy to check that the two splitting processes are described by setting $\lambda =+1$. The sign of $\eta$
further specifies whether the splitting process is $e^{-} \to e^{-} +\gamma$ ($\eta =-1$) or $e^{+} \to e^{+} +\gamma$ 
($\eta =+1$). In these two cases, real solutions for $p_z$ exist only when the following necessary conditions are
satisfied: 
\begin{eqnarray}
\sqrt{\Omega^2-k_z^2} \leq k_{-}
& \mbox{and}& n>n^{\prime} ,
\label{qqg-cond}
\end{eqnarray}
for $e^{-} \to e^{-} +\gamma$ ($\lambda =+1$, $\eta =-1$), and  
\begin{eqnarray}
\sqrt{\Omega^2-k_z^2} \leq k_{-}
& \mbox{and}&  n<n^{\prime} ,
\quad 
\label{bqbqg-cond}
\end{eqnarray}
for $e^{+} \to e^{+} +\gamma$ ($\lambda =+1$, $\eta =+1$). 
In Eqs.~(\ref{qqg-cond}) and (\ref{bqbqg-cond}), we introduced the shorthand notation for the transverse momentum threshold 
\begin{equation}
k_{-}=\left|\sqrt{m^2+2n|qB|} -\sqrt{m^2+2n^{\prime}|qB|}\right| ,
\label{k-minus}
\end{equation}
which depends on the Landau level indices $n$ and $n^{\prime}$ of the initial and final quantum states, respectively.

The annihilation process $e^{-} + e^{+} \to \gamma$ is realized when $\lambda =-1$ and $\eta = - 1$. 
In this case, the necessary condition for the existence of real-valued solutions for $p_z$ is 
\begin{equation}
\sqrt{\Omega^2-k_z^2} \geq k_{+}, 
\label{qbqg-cond}
\end{equation}
where, by definition, the transverse momentum threshold is
\begin{equation}
k_{+}=\left|\sqrt{m^2+2n|qB|} +\sqrt{m^2+2n^{\prime}|qB|}\right| .
\label{k-plus}
\end{equation}
When $\Omega >0$, there are no physical processes that correspond to $\lambda = -1$ and $\eta = +1$. Accordingly, there are no real solutions to the energy conservation equation (\ref{energy-conservation}) in this case.

In all cases, when real solutions exist, they are given by the following explicit expressions:
\begin{eqnarray}
p_{z}^{(\pm)}&=& \frac{k_z}{2}\left[1+ \frac{2(n-n^{\prime})|qB|}{\Omega^2-k_z^2} 
\pm \frac{\Omega}{|k_z|}  \sqrt{ \left( 1-\frac{k_{-}^2}{\Omega^2-k_z^2} \right)
\left( 1-\frac{k_{+}^2}{\Omega^2-k_z^2}\right)} \right].
\label{pz-solution}
\end{eqnarray}
Note that there are two solutions labeled by the plus and minus signs in the superscript.
By using this result and the energy conservation equation, written in the form $(E_{n,p_z}+\eta\Omega)^2=E_{n^{\prime},p_z-k_z}^2$, it is straightforward to derive the following explicit expressions for the fermions energies:
\begin{eqnarray}
 \left. E_{n,p_z}\right|_{p_z = p_{z}^{(\pm)} } &=& - \frac{\eta \Omega}{2} \left[
1+\frac{2(n-n^{\prime})|qB|}{\Omega^2-k_z^2}\pm \frac{|k_z|}{\Omega}\sqrt{ \left(1-\frac{k_{-}^2}{\Omega^2-k_z^2} \right)\left( 1-\frac{k_{+}^2}{\Omega^2-k_z^2}\right)} 
\right], \label{E1-solution}\\
 \left. E_{n^{\prime},p_z-k_z} \right|_{p_z = p_{z}^{(\pm)} } &=& \frac{\lambda \eta \Omega}{2} \left[
1-\frac{2(n-n^{\prime})|qB|}{\Omega^2-k_z^2}\mp \frac{|k_z|}{\Omega}\sqrt{ \left(1-\frac{k_{-}^2}{\Omega^2-k_z^2} \right)\left( 1-\frac{k_{+}^2}{\Omega^2-k_z^2}\right)} 
\right].
\label{E2-solution}
\end{eqnarray}
As expected, these expressions define positive definite energies when the necessary conditions in Eqs.~(\ref{qqg-cond}), (\ref{bqbqg-cond}), and (\ref{qbqg-cond}) are satisfied. 
 
By making use of the explicit solutions, the $\delta$ function in Eq.~(\ref{Im-Pol-fun-general}) can be rewritten in a much 
simpler form, i.e., 
\begin{eqnarray}
\delta\left(E_{n,p_z}-\lambda E_{n^{\prime},p_z-k_z}+\eta \Omega\right)
=\sum_{s=\pm}\frac{ 2 E_{n,p_z} E_{n^{\prime},p_z-k_z} \delta\left(p_z - p_{z}^{(s)}\right)}{\sqrt{ \left( \Omega^2-k_z^2-k_{-}^2 \right) \left( \Omega^2-k_z^2-k_{+}^2\right)} }.
\label{delta-solution}
\end{eqnarray}
Indeed, this can be derived by using the following result:
\begin{eqnarray}
\left|\frac{\partial \left( E_{n,p_z}-\lambda E_{n^{\prime},p_z-k_z}\right)}{\partial p_z}\right|_{p_z=p_{z}^{(\pm)}}
=\frac{\left| k_z E_{n,p_z} +\eta p_{z}^{(\pm)} \Omega  \right| }{E_{n,p_z} E_{n^{\prime},p_z-k_z}}
\Bigg|_{p_z=p_{z}^{(\pm)}}
= \frac{\sqrt{ \left( \Omega^2-k_z^2-k_{-}^2 \right) \left( \Omega^2-k_z^2-k_{+}^2\right)}}{2E_{n,p_z} E_{n^{\prime},p_z-k_z}}.
\end{eqnarray}
To obtain the last form of the expression, we used Eqs.~(\ref{pz-solution}) and (\ref{E1-solution}). 

By substituting Eq.~(\ref{delta-solution}) into Eq.~(\ref{Im-Pol-fun-general}), we finally derive the expression for the imaginary part of the polarization tensor:
\begin{equation}
\mbox{Im} \left[\Pi_R^{\mu\nu}(\Omega;\mathbf{k}) \right] =
  \frac{\alpha N_{f}}{4\pi \ell^4} \sum_{n,n^\prime=0}^{\infty} 
\sum_{\lambda,\eta=\pm 1}\sum_{s=\pm 1} \Theta_{\lambda, \eta}^{n,n^{\prime}}(\Omega,k_z)
\frac{n_F(E_{n,p_z})-n_F(\lambda E_{n^{\prime},p_z-k_z}) }{\eta\lambda  \sqrt{ \left( \Omega^2-k_z^2-k_{-}^2 \right) \left( \Omega^2-k_z^2-k_{+}^2\right)} }\sum_{i=1}^{4}I_{i}^{\mu\nu} \Bigg|_{p_z = p_{z}^{(s)}} .
\label{Im-Pol-fun}
\end{equation} 
where threshold function $\Theta_{\lambda, \eta}^{n,n^{\prime}}(\Omega,k_z)$ is defined as follows:
\begin{equation}
\Theta_{\lambda, \eta}^{n,n^{\prime}}(\Omega,k_z) = \left\{
\begin{array}{lll}
  \theta\left( k_{-}^2 + k_z^2 - \Omega^2 \right) & \mbox{for}& \lambda=1, \eta=-1, n>n^{\prime},\\
  \theta\left( k_{-}^2 + k_z^2 - \Omega^2 \right) & \mbox{for}& \lambda=1, \eta=1, n<n^{\prime},\\
  \theta\left( \Omega^2 - k_z^2 - k_{+}^2 \right) & \mbox{for}& \lambda=-1, \eta=-1,
  \end{array}
\right.
\end{equation} 
and $\Theta_{\lambda, \eta}^{n,n^{\prime}}(\Omega,k_z) =0$ otherwise. This function defines the windows of the parameter space where the necessary conditions (\ref{qqg-cond}), (\ref{bqbqg-cond}), and (\ref{qbqg-cond}) are satisfied. 

This analytical expression for $\mbox{Im} \left[\Pi_R^{\mu\nu}(\Omega;\mathbf{k}) \right] $ is one of the main results of the paper. We will use it in the next two sections to calculate the photon emission rate and the optical conductivity of a strongly magnetized hot relativistic plasma. Before proceeding to the applications, however, it is instructive to verify that the polarization tensor is transverse (i.e., $k_{\mu} \Pi^{\mu\nu}  =0$ and $\Pi^{\mu\nu} k_{\nu} =0$), as required by the gauge invariance of the theory. 

Since both thermal bath and background magnetic field break the Lorentz symmetry, the final tensor structure of the polarization tensor is rather complicated. Its explicit form is determined by tensors $I_{i}^{\mu\nu}$, defined in Appendix~\ref{AP-tr}. As is clear from Eq.~(\ref{Pi_deriv-1}) and the explicit expressions for $I_{i}^{\mu\nu}$ in Appendix~\ref{AP-tr}, the general tensor structure is same for both real and imaginary parts of the polarization function. Thus, by analyzing all similar terms of the absorptive part of $\Pi_R^{\mu\nu}(\Omega;\mathbf{k}) $ in Appendix~\ref{app:Tensor-structure}, we find that the polarization tensor takes following form:
\begin{eqnarray}
\Pi_R^{\mu\nu}(\Omega;\mathbf{k}) &=&
 \left( \frac{k_\parallel^\mu k_\parallel^\nu }{k_\parallel^2} -  g_{\parallel}^{\mu\nu} \right)  \Pi_{1}
+ \left(g_{\perp}^{\mu\nu}+ \frac{k_\perp^\mu k_\perp^\nu}{k_\perp^2} \right) \Pi_{2}
+ \left( \frac{k_\parallel^\mu \tilde{k}_\parallel^\nu +\tilde{k}_\parallel^\mu k_\parallel^\nu}{k_\parallel^2}  
+ \frac{\tilde{k}_\parallel^\mu k_\perp^\nu +k_\perp^\mu \tilde{k}_\parallel^\nu}{k_\perp^2}\right) \Pi_{3}
\nonumber\\
&+&\left(\frac{k_\parallel^\mu k_\perp^\nu +k_\perp^\mu k_\parallel^\nu}{k_\parallel^2 } + \frac{k_\perp^2}{k_\parallel^2 }  g_{\parallel}^{\mu\nu}  - g_{\perp}^{\mu\nu} \right) \Pi_{4}
+ \left( \frac{F^{\mu\nu} }{B}
+ \frac{k_\parallel^\mu \tilde{k}_{\perp}^\nu -\tilde{k}_{\perp}^\mu k_\parallel^\nu}{k_\parallel^2} \right) \tilde{\Pi}_{5}
+ \frac{\tilde{k}_\parallel^\mu \tilde{k}_{\perp}^\nu -\tilde{k}_{\perp}^\mu \tilde{k}_\parallel^\nu}{k_\parallel^2} \tilde{\Pi}_{6} ,
\label{ImPi-tensor-txt1}
\end{eqnarray}
where we used the following notation:
\begin{equation}
\begin{array}{lll}
g_{\parallel}^{\mu\nu} = \mbox{diag}(1,0,0,-1), \quad & 
k_\parallel^\mu = g_{\parallel}^{\mu\nu}k_{\nu}=k_0\delta^\mu_{0}+k_z\delta^\mu_{3},  \quad & 
\tilde{k}_\parallel^\mu  = - \varepsilon^{\mu 1 2 \nu}k_{\nu}= k_z \delta^{\mu}_{0} +k_0 \delta^{\mu}_{3}, \\
g_{\perp}^{\mu\nu} = \mbox{diag}(0,-1,-1,0),  \quad &  
k_\perp^\mu = g_{\perp}^{\mu\nu}k_{\nu}= k_x\delta^\mu_{1}+k_y\delta^\mu_{2}, \quad & 
\tilde{k}_{\perp}^{\mu}= -\varepsilon^{0 \mu \nu 3}k_{\nu}= k_{y} \delta^{\mu}_{1} -k_x \delta^{\mu}_{2}.\\
\end{array}
\label{ggkkkk}
\end{equation}
Note that $\tilde{k}_{\perp,\mu} \tilde{k}_{\perp}^{\mu} = k_{\perp,\mu} k_{\perp}^{\mu} = -k_\perp^2$,  
$\tilde{k}_{\parallel,\mu} \tilde{k}_\parallel^\mu = -k_{\parallel,\mu} k_\parallel^\mu= - k_\parallel^2$, and
$k_\mu \tilde{k}_{\perp}^{\mu} =k_{\mu} \tilde{k}_\parallel^\mu =0$. 
As we see the polarization tensor in Eq.~(\ref{ImPi-tensor-txt1}) contains four {\em symmetric} tensor structures and two {\em antisymmetric} ones. 
Naively, it may appear that another symmetric tensor, i.e., $\tilde{k}_\parallel^\mu \tilde{k}_\parallel^\nu /k_\parallel^2$, can be constructed from $\tilde{k}_\parallel^\mu$. As one can verify, however, it is exactly the same as $k_\parallel^\mu k_\parallel^\nu/k_\parallel^2 -  g_{\parallel}^{\mu\nu}$. Note that the antisymmetric contributions, defined by the component functions $\tilde{\Pi}_{5}$ and $\tilde{\Pi}_{6}$, appear because the time-reversal symmetry is broken by a nonzero magnetic field.

By using the definitions in Eq.~(\ref{ggkkkk}), one can verify that all six tensor structures in Eq.~(\ref{ImPi-tensor-txt1}) are transverse. Therefore, the polarization tensor is transverse as well, i.e., $k_{\mu} \Pi^{\mu\nu}  =0$ and $\Pi^{\mu\nu} k_{\nu} =0$, which is consistent with the gauge invariance. The explicit expressions for the (absorptive part of) component functions are given in Appendix~\ref{app:Tensor-structure}.

\section{Photon emission from QED plasma}
\label{sec:emission-QED}

Here we will use the general result for $\mbox{Im} [\Pi_R^{\mu\nu}(\Omega;\mathbf{k}) ] $ which is obtained in the previous section to study the photon emission from a strongly magnetized hot QED plasma. The corresponding emission is directly observable when the plasma is optically thin, i.e., its size is small compared to the photon mean free path in the plasma. For optically thick systems, the photon emission will come only from the surface layer of a depth comparable to the mean free path. 

\subsection{Photon emission: Analytical results}

In quantum field theory, the photon production rate from a thermally equilibrated charged plasma can be expressed in terms of the imaginary part of the retarded polarization tensor $ \Pi^{\mu\nu}_{R}(\Omega,\mathbf{k})$ as follows \cite{Kapusta:2006pm}:
\begin{equation}
\Omega\frac{d^3R}{d^3\mathbf{k}}=-\frac{1}{(2\pi)^3}\frac{\mbox{Im}\left[\Pi^{\mu}_{R,\mu}(\Omega,\mathbf{k})\right]}{\exp\left(\frac{\Omega}{T}\right)-1},
\label{diff-rate-definition}
\end{equation}
where $\Omega$ and $\mathbf{k}$ are the photon frequency and momentum, respectively, and $T$ is the temperature of the plasma. 

Because of the rotation symmetry about the $z$ axis, it is convenient to use the spherical coordinates and write $d^3\mathbf{k} = k^2 dk \sin\theta d\theta d\phi$. Note that the production rate is independent of the polar angle $\phi$. Thus, without loss of generality, the differential rate is fully characterized by 
 \begin{equation}
\frac{d^2 R}{k \, d k\, d(\cos \theta)}=-\frac{1}{(2\pi)^2}\frac{\mbox{Im}\left[\Pi^{\mu}_{R,\mu}(k,\mathbf{k})\right]}{\exp\left(\frac{k}{T}\right)-1},
\label{diff-rate-cylindrical}
\end{equation}
where we substituted $\Omega=k$ for the on-shell photons. The photon momentum is given by $k=\sqrt{k_\perp^2+k_z^2}$, where $k_{\perp} = k\sin\theta$ and $k_z = k \cos\theta$ are the components perpendicular and parallel to the magnetic field, respectively.

By performing the Lorentz contraction in expression (\ref{Im-Pol-fun}), we derive the following result:
\begin{eqnarray}
\mbox{Im} \left[\Pi^{\mu}_{R,\mu}(\Omega,\mathbf{k})\right] &=&
  \frac{\alpha N_{f}}{4\pi  \ell^4} \sum_{n>n^\prime}^{\infty}  \theta\left( k_{-}^2 +k_z^2 - \Omega^2 \right)
\sum_{s=\pm 1} 
\frac{n_F(E_{n^{\prime},p_z-k_z}) -n_F(E_{n,p_z})}{\sqrt{ \left( k_{-}^2 +k_z^2 - \Omega^2 \right)\left( k_{+}^2 +k_z^2 - \Omega^2 \right) } } 
\sum_{i=1}^{4} \mathcal{F}_{i}
\Bigg|_{p_z = p_{z}^{(s)} , \lambda =1, \eta =-1}
\nonumber\\
&+&  \frac{\alpha N_{f}}{4\pi  \ell^4} \sum_{n<n^\prime}^{\infty}  \theta\left( k_{-}^2 +k_z^2 - \Omega^2 \right)
\sum_{s=\pm 1} 
\frac{n_F(E_{n,p_z})-n_F(E_{n^{\prime},p_z-k_z}) }{\sqrt{ \left( k_{-}^2 +k_z^2 - \Omega^2 \right)\left( k_{+}^2 +k_z^2 - \Omega^2 \right) } }
\sum_{i=1}^{4} \mathcal{F}_{i}
\Bigg|_{p_z = p_{z}^{(s)}  , \lambda =\eta =1}
\nonumber\\
&+&  \frac{\alpha N_{f}}{4\pi  \ell^4} \sum_{n,n^\prime=0}^{\infty} \theta\left( \Omega^2-k_z^2-k_{+}^2\right)
\sum_{s=\pm 1} 
\frac{n_F(E_{n,p_z}) +n_F(E_{n^{\prime},p_z-k_z}) -1}{\sqrt{ \left(  \Omega^2-k_z^2-k_{-}^2 \right) \left( \Omega^2-k_z^2-k_{+}^2 \right)  } }
\sum_{i=1}^{4} \mathcal{F}_{i}
\Bigg|_{p_z = p_{z}^{(s)} , \lambda =\eta =-1} ,
\label{Im-Pi-mumu-three}
\end{eqnarray}
where, by definition,  $\mathcal{F}_i =g_{\mu\nu} I_{i}^{\mu\nu}$  (with $i=1,2,3,4$). As shown in Appendix~\ref{AP-tr}, 
$\mathcal{F}_2 =\mathcal{F}_3 =0$, while $\mathcal{F}_1$ and $\mathcal{F}_4$ are nonzero. The explicit expressions for the latter two functions are given by 
\begin{eqnarray}
\mathcal{F}_1&=& 4\pi \left[\left(\Omega^2-k_z^2\right)\ell^2 -2 (n+n^\prime) \right]
\left( \mathcal{I}_{0}^{n-1,n^{\prime}}(\xi) +\mathcal{I}_{0}^{n,n^{\prime}-1}(\xi)  \right)
+8\pi m^2 \ell^2 \left( \mathcal{I}_{0}^{n,n^{\prime}}(\xi) +\mathcal{I}_{0}^{n-1,n^{\prime}-1}(\xi)  \right),\\
\mathcal{F}_4&=& 16 \pi \, \mathcal{I}_{2}^{n-1,n^{\prime}-1}(\xi),
\end{eqnarray}
where $\xi = k_{\perp}^2\ell^{2}/2$. 
For the definition of functions  $\mathcal{I}_{i}^{n,n^{\prime}}(\xi)$ in terms of the Laguerre polynomials, see Appendix~\ref{prop-use}.

\begin{figure}[t]
\centering
\subfigure[]{\includegraphics[width=0.32\textwidth]{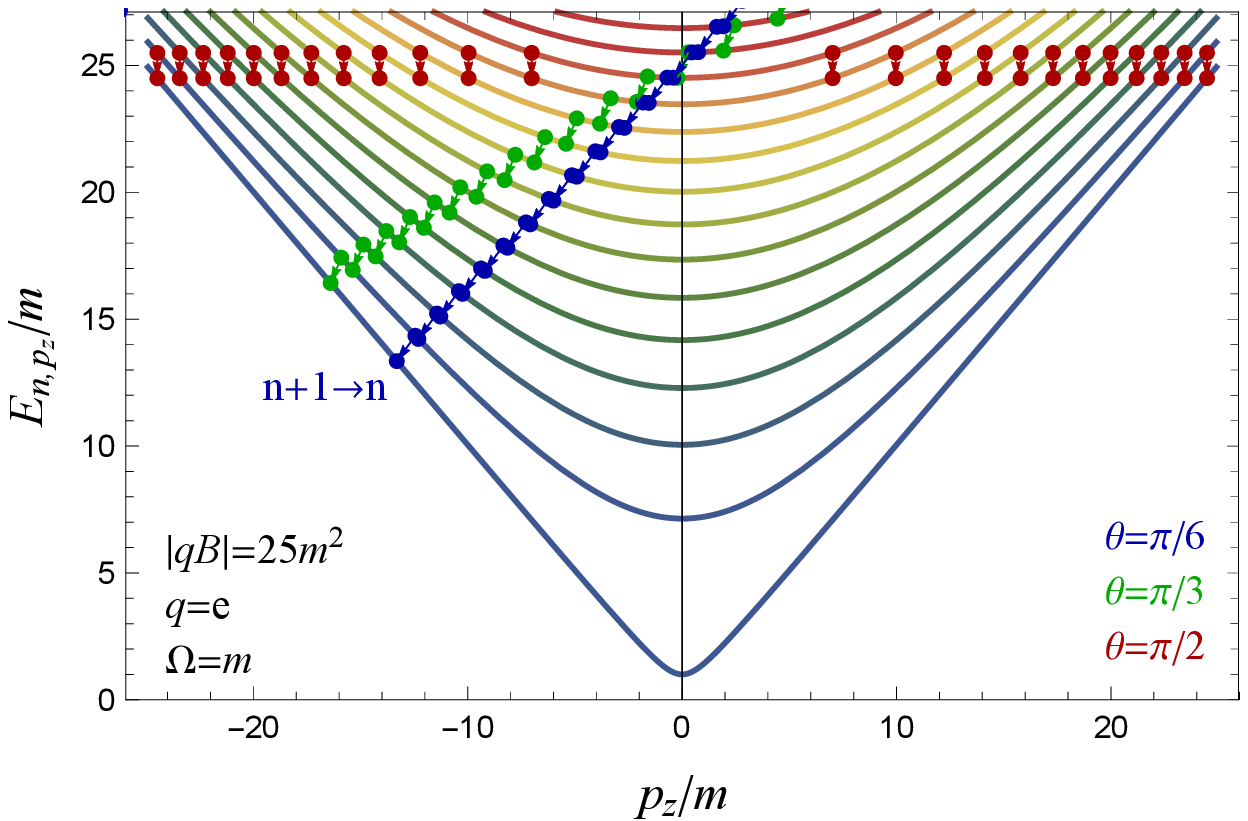}}
\subfigure[]{\includegraphics[width=0.32\textwidth]{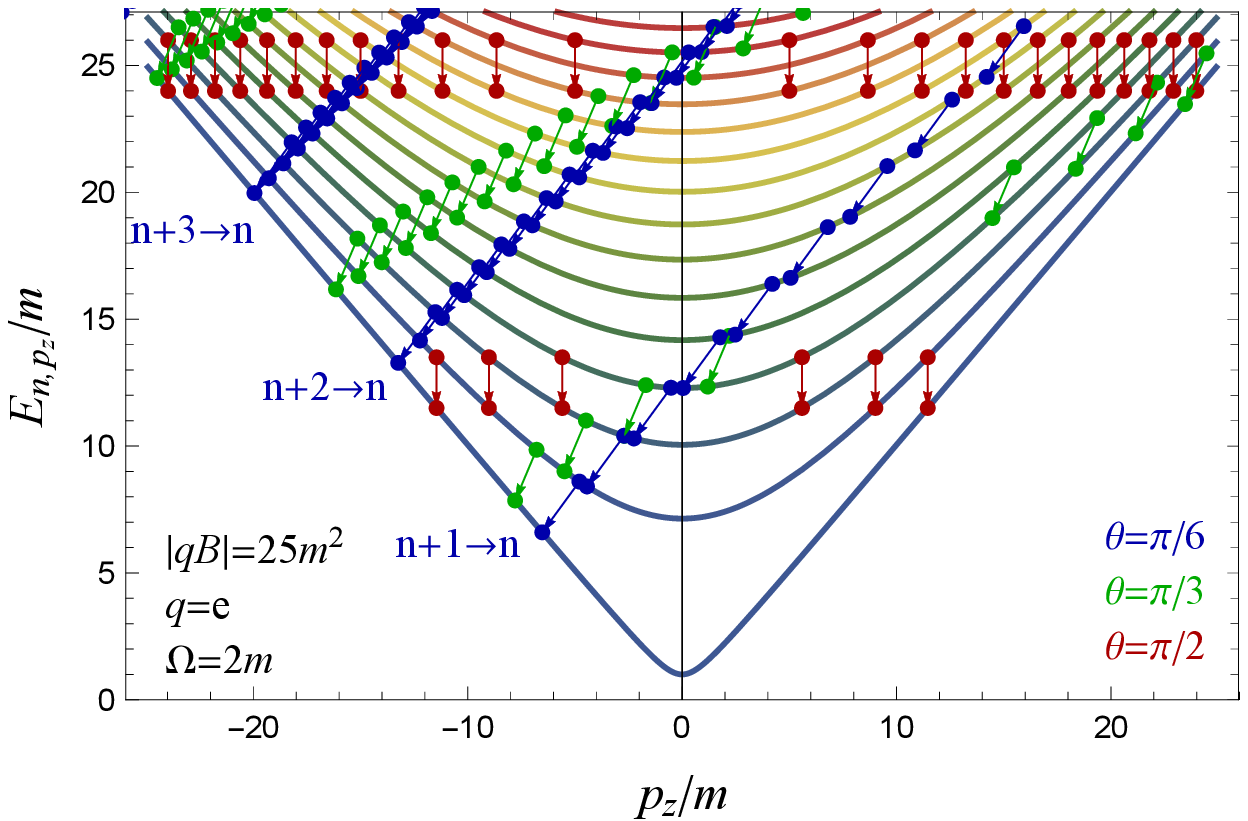}}
\subfigure[]{\includegraphics[width=0.32\textwidth]{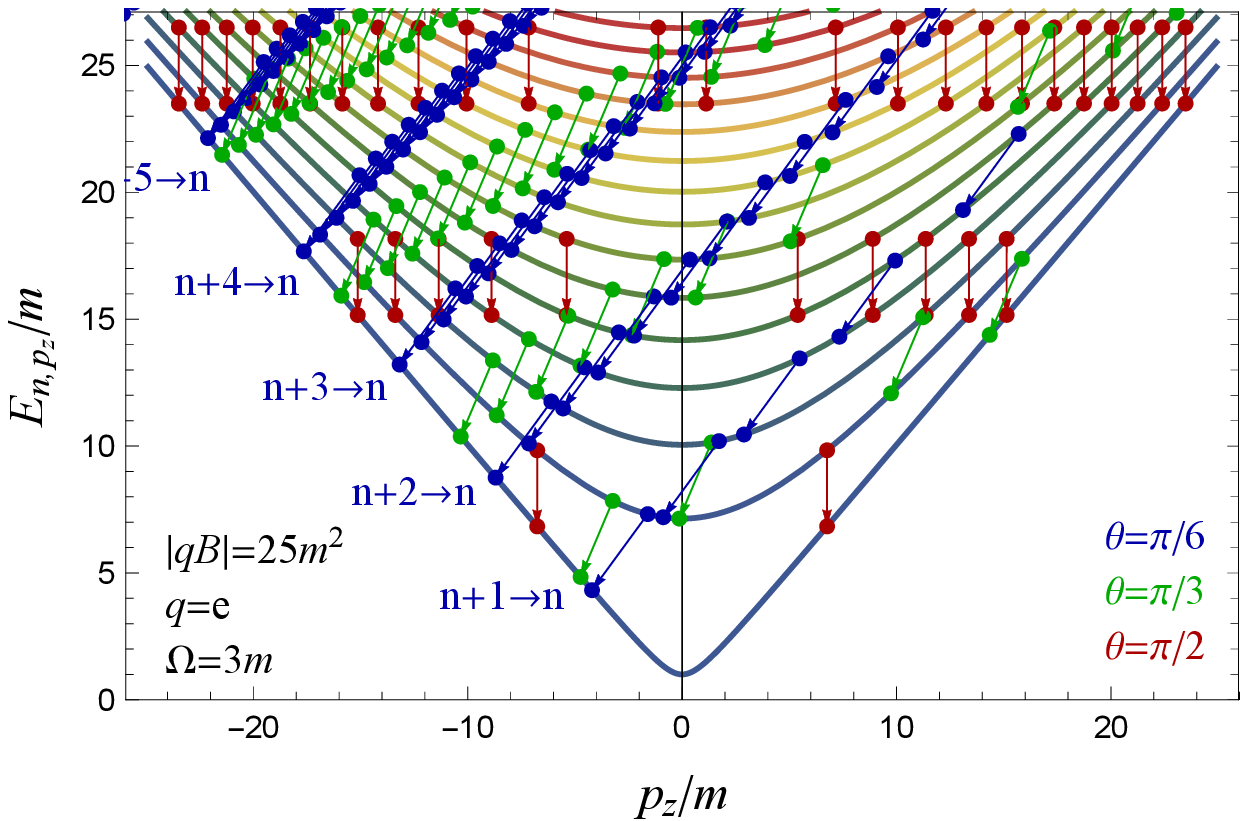}}
\caption{Schematic representation of Landau level transitions for the splitting processes $e^{-} \to e^{-}+\gamma$ with a fixed direction of the photon emission, i.e., $\theta=0$ (blue), $\theta=\pi/3$ (green), and $\theta=\pi/2$ (red), allowed by the energy conservation. Note that, at $\theta = 0$, only the transitions between the adjacent Landau levels ($n+1\to n$) contribute nontrivially. The individual panels show the results for different photon energies: 
(a) $\Omega=m$,  (b) $\Omega=2m$,  (c) $\Omega=3m$.}
\label{fig:transitions}
\end{figure}

\begin{figure}[t]
\centering
\subfigure[]{\includegraphics[width=0.32\textwidth]{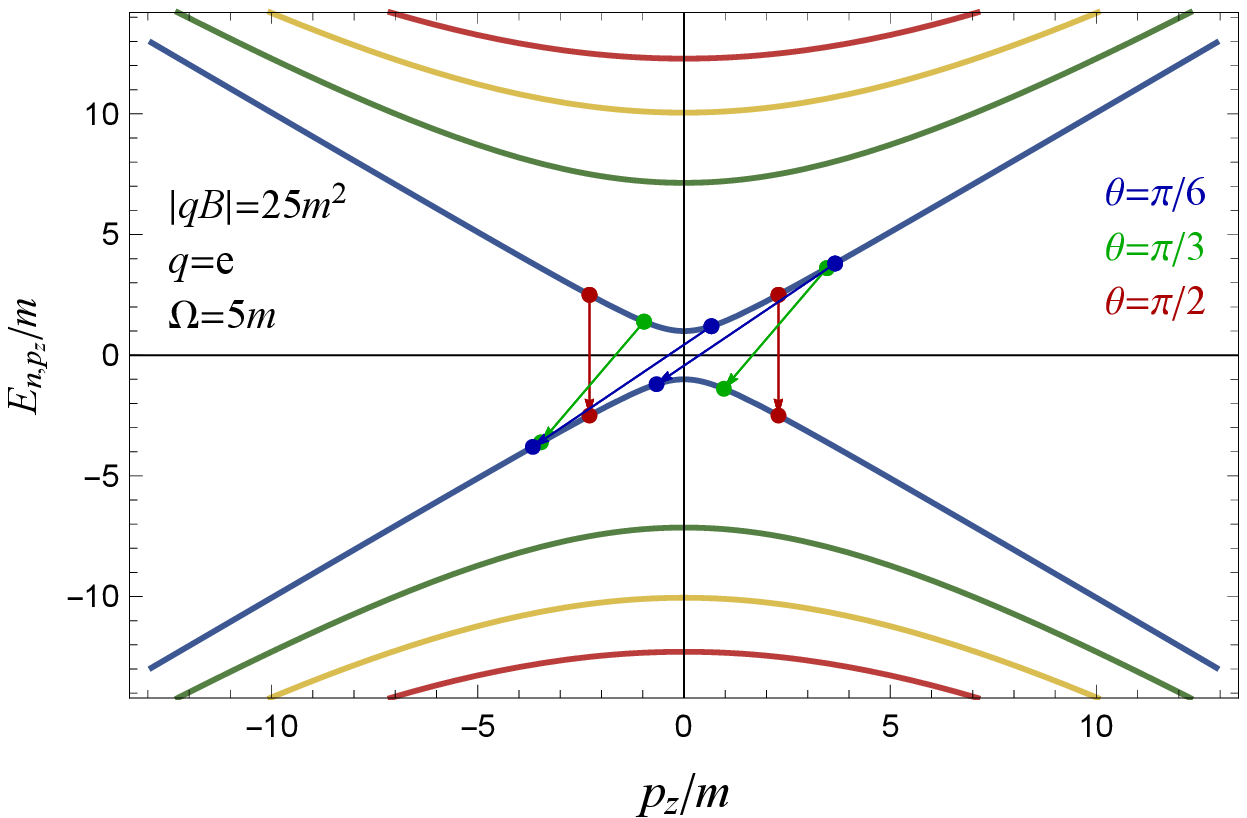}}
\subfigure[]{\includegraphics[width=0.32\textwidth]{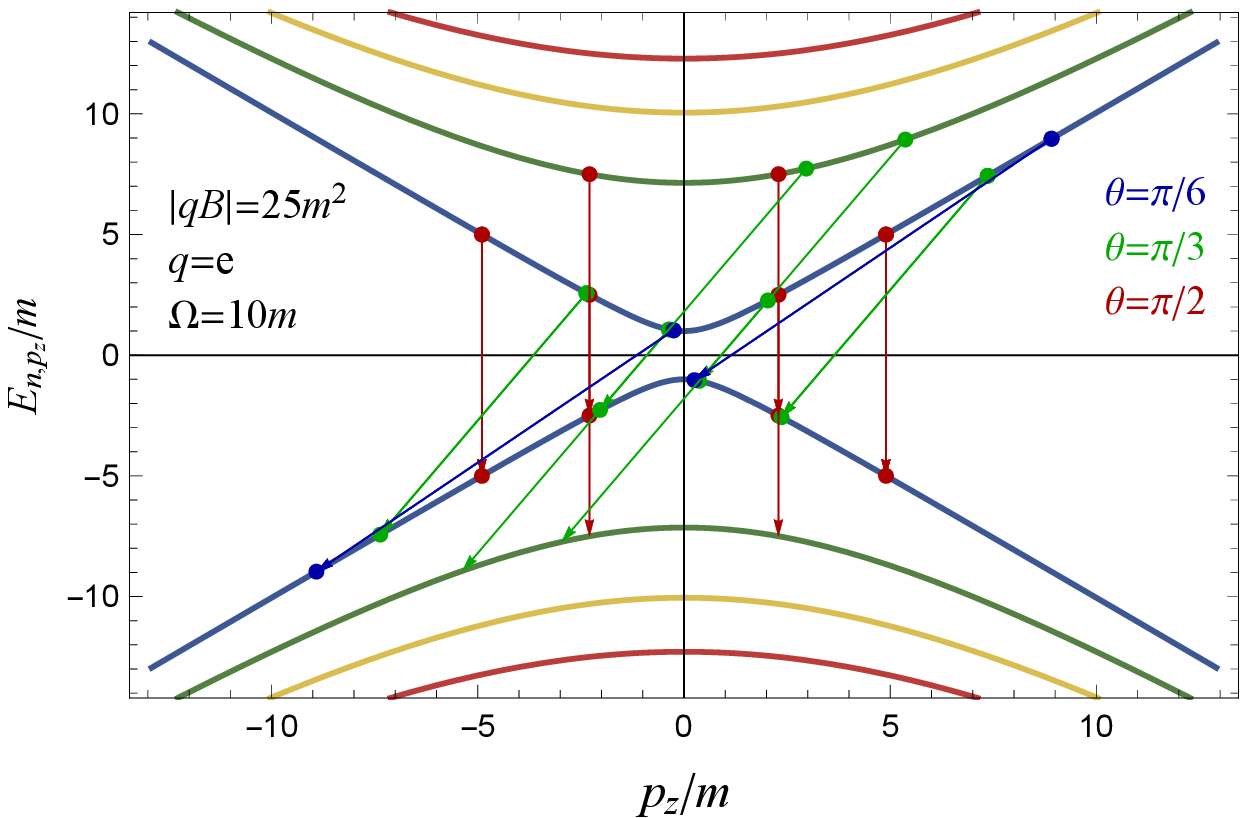}}
\subfigure[]{\includegraphics[width=0.32\textwidth]{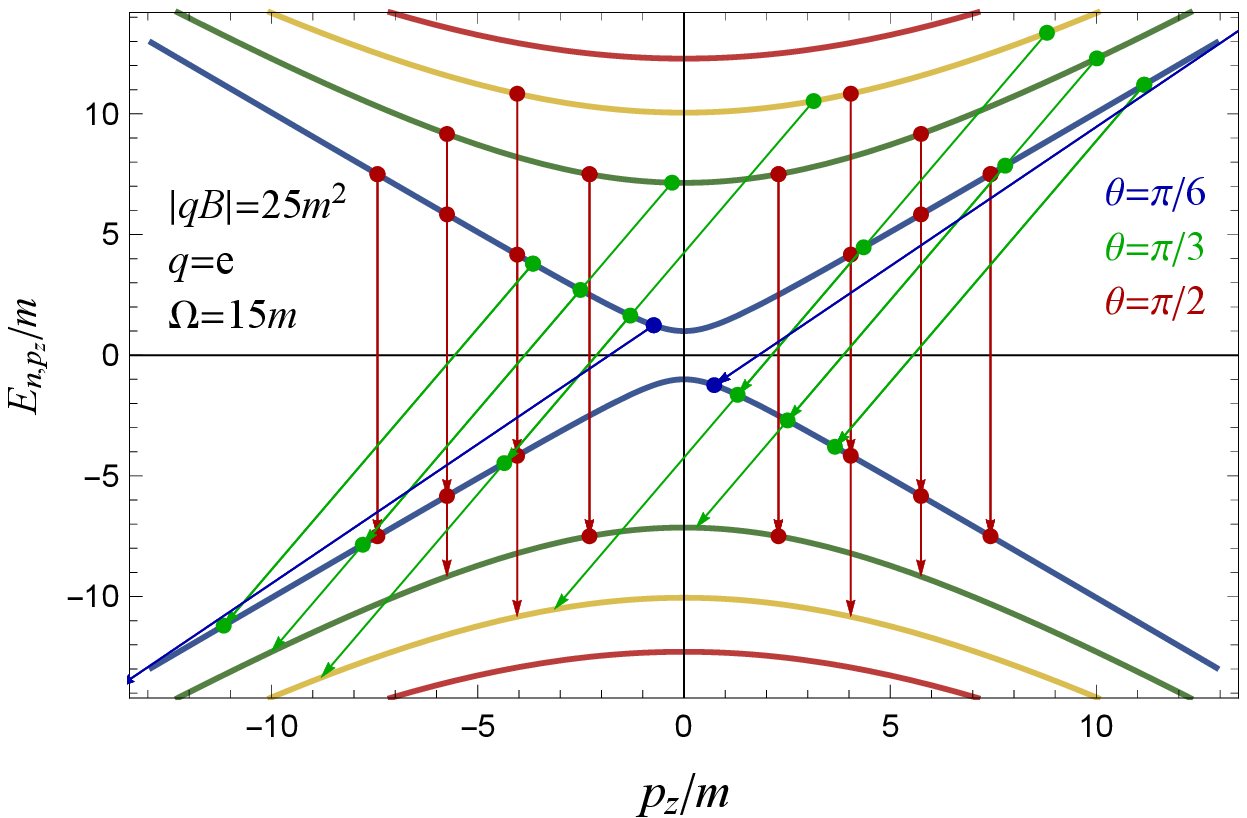}}
\caption{Schematic representation of Landau level transitions for the annihilation processes $e^{-} + e^{+}\to \gamma$ with a fixed direction of the photon emission, i.e., $\theta=\pi/6$ (blue), $\theta=\pi/3$ (green), and $\theta=\pi/2$ (red),  allowed by the energy conservation. Note that, the emission is forbidden at $\theta = 0$. The individual panels show the results for different photon energies: 
(a) $\Omega=5m$,  (b) $\Omega=10m$,  (c) $\Omega=15m$.}
\label{fig:transitions-annih}
\end{figure}

By making use of the explicit expressions for the fermion energies in Eqs.~(\ref{E1-solution}) and (\ref{E2-solution}), it is straightforward to show that the first two contributions in Eq.~(\ref{Im-Pi-mumu-three}), which describe the photon emission due to the particle splitting $e^{-} \to e^{-} +\gamma$ ($\lambda =+1$, $\eta =-1$) and the antiparticles splitting  $ e^{+} \to e^{+} +\gamma$ ($\lambda =+1$, $\eta =1$) processes, are equal to each other. This is not surprising, of course, because the two processes are related by the charge conjugation symmetry. (Note, however, that the same is not true in a plasma at a nonzero chemical potential \cite{Wang:2021eud}.) By taking this symmetry into account and substituting the result into Eq.~(\ref{diff-rate-cylindrical}), we obtain the final expression for the rate:
\begin{eqnarray}
\frac{d^2 R}{k \, d k\, d(\cos \theta)}&=&\frac{\alpha N_f }{(2\pi)^3\ell^4\left[\exp\left(\frac{k}{T}\right)-1\right]}
\sum_{n>n^\prime}^{\infty}  
\frac{g(n, n^{\prime}) 
\left[\theta\left(\Omega^2-k_z^2-k_{+}^2\right) 
-\theta\left(k_{-}^2+k_z^2-\Omega^2\right)
\right]
 }{\sqrt{ \left( k_{-}^2+k_z^2-\Omega^2 \right)\left( k_{+}^2+k_z^2-\Omega^2\right) } } \left(
\mathcal{F}_1+\mathcal{F}_4 \right)
\nonumber\\
&+&  \frac{\alpha N_{f}}{2(2\pi)^3 \ell^4\left[\exp\left(\frac{k}{T}\right)-1\right]} \sum_{n=0}^{\infty} 
\frac{g_0(n)\theta\left(\Omega^2-k_z^2-k_{+}^2\right)}{\sqrt{ \left(\Omega^2-k_z^2\right)\left( \Omega^2-k_z^2-k_{+}^2 \right)} }\left(
\mathcal{F}_1+\mathcal{F}_4 \right)  ,
\label{photon-rate-sum}
\end{eqnarray}
where
\begin{eqnarray}
g(n, n^{\prime}) &=& 2-\sum_{s_1,s_2=\pm}
n_F\left(\frac{\Omega}{2}  +s_1 \frac{\Omega(n-n^{\prime})|qB|}{\Omega^2-k_z^2}+s_2 \frac{|k_z|}{2(\Omega^2-k_z^2)}\sqrt{  \left(\Omega^2-k_z^2-k_{-}^2 \right) \left( \Omega^2-k_z^2-k_{+}^2\right)} 
\right), \label{gnn-11} \\
g_0(n) &=& g(n, n)  =2 - 2\sum_{s=\pm}
n_F\left(\frac{\Omega}{2} +s \frac{|k_z|}{2}\sqrt{1- \frac{4(m^2+2n|q B|)}{ \Omega^2-k_z^2}}
\right) .  \label{gnn-00}
\end{eqnarray}
While the analytical expression in Eq.~(\ref{photon-rate-sum}) for the photon production rate is relatively simple, its predictions are not obvious. As is clear, each term in the sum has a simple interpretation as a partial contribution from a splitting or annihilation process that involves a quantum transition between Landau levels $n$ and $n^\prime$. Schematically, the relevant low-energy transitions for the two types of processes ($e^{-} \to e^{-}+\gamma$ and $e^{-} +e^{+}\to \gamma$, respectively) are shown in Figs.~\ref{fig:transitions} and \ref{fig:transitions-annih} for several fixed values of the photon energy and several fixed directions of emission. Because of the superposition of a large (infinite) set of relevant transitions, the final energy and angular dependence of the differential rate is hard to grasp from Eq.~(\ref{photon-rate-sum}). Therefore, in the next subsection, we use numerical methods to study the photon production rate in more detail. 

Let us note in passing that the general result in Eq.~(\ref{photon-rate-sum}), derived in the strong-field case, should also have a well-defined zero-field limit. To obtain such a limit, one has to perform an infinite sum over the Landau levels analytically. While such a task is beyond the scope of this study, it may not be hopeless if the resumming technique of Ref.~\cite{Sokolov:1986nk} is utilized.

\subsection{Photon emission: Numerical results}

To get an insight into the underlying physics of the photon emission from a magnetized relativistic plasma, here we study the corresponding differential rate by numerical methods. To stay as general as possible, we will express all dimensionful parameters (e.g., the magnetic field strength and temperature) in units of the fermion mass $m$. In particular, we will consider the following three values of temperature: $T=5m$, $T=15m$, and $T=50 m$, which cover a range of different regimes (from moderately relativistic to ultrarelativistic). To explore the interplay of the magnetic field and thermal effects in each regime, we will calculate the photon production rate for two different values of the magnetic field, i.e., $\sqrt{|qB|} = 5m$ and $\sqrt{|qB|} = 15m$.

For the QED plasma, the mass of particles (electrons) is $m=m_e=0.511~\mbox{MeV}$ and the charge is $q=-e$. In this case, as is easy to check, the two choices of the magnetic field $\sqrt{|qB|} = 5m_{e}$ and $\sqrt{|qB|} = 15m_{e}$ correspond to $B\approx 1.1\times 10^{15}~\mbox{G}$ and $B \approx 9.9\times 10^{15}~\mbox{G}$, respectively. Note that the approximate conversion formula reads $B \approx 4.414\times 10^{13}(\sqrt{|qB|}/m_e)^2~\mbox{G}$, where both $\sqrt{|qB|}$ and $m_e$ are given in the energy units. On the other hand, if the charged particles are quarks with charge $q$ and mass $m=5~\mbox{MeV}$, the same numerical results, rescaled by a factor of $(q/e)^2$, will also describe the partial contribution of photon emission from the given quark species in a magnetized quark-gluon plasma at $B \approx 1.1\times 10^{17}|e/q|~\mbox{G}$ and $B \approx 9.5\times 10^{17}|e/q|~\mbox{G}$, respectively. 

When calculating the rate in Eq.~(\ref{photon-rate-sum}) numerically, the sum over Landau levels has to be truncated at some finite value of $n_{\rm max}$. Qualitatively, such truncation is equivalent to setting an ultraviolet energy cutoff at about $\Lambda_{UV}=\sqrt{2n_{\rm max} |qB|}$. In application to the photon emission, this implies that the photon energy should be limited from above by the cutoff, i.e., $\Omega \lesssim \Lambda_{UV}$. Indeed, the photon emission is meaningful within this framework only when the energies of both initial and final Landau-level states lie below $\Lambda_{UV}$. In general, it is expected that the emission rate in the plasma is a rapidly decreasing function of the photon energy at large $\Omega$. However, its value can be calculated reliably from Eq.~(\ref{photon-rate-sum}) only if sufficiently many Landau levels are included in the sum, i.e., $n_{\rm max}\gtrsim \Omega^2/(2|qB|)$. 

As the numerical analysis will show, the photon emission is dominated by the particle and antiparticle splitting processes ($e^{-} \to e^{-}+\gamma$ and $e^{+} \to e^{+}+\gamma $), provided the temperature is comparable to or higher than the magnetic energy scale $\sqrt{|qB|}$ and the photon energy $\Omega$ is not much larger than $\sqrt{|qB|}$. Nevertheless, the particle-antiparticle annihilation processes ($e^{-} + e^{+}\to \gamma$) also contribute, especially at small temperatures and large $\Omega$. For a fixed value of $\Omega$, we find that the maximum Landau-level indices are constrained by Eq.~(\ref{qbqg-cond}), namely, $\mbox{max}(n,n^\prime)\leq n^{*}_{\Omega} = \left[\Omega(\Omega-2m)/(2|qB|)\right]$. Therefore, the corresponding sum of the annihilation contributions can be limited to $n_{\rm max} = n^{*}_{\Omega}$. Note that $n^{*}_{\Omega}$ grows quadratically with the photon energy $\Omega$ and decreases with the magnetic field as the inverse power law $1/B$. Numerically, one finds that $n^{*}_{\Omega} = 196$ for $\Omega=100m$ and $|qB|=25m^2$. On the other hand, $n^{*}_{\Omega} = 21$ for $\Omega=100m$ and $|qB|=225m^2$.

As we will discuss later in more detail, the low-energy quantization of Landau levels also sets a limit for the photon energy from below, namely $\Omega \gtrsim \Lambda_{IR}$, where $\Lambda_{IR}=\sqrt{|qB|} /\sqrt{2n_{\rm max}}$. The restriction comes from the fact that the separation between the fermionic Landau levels is of the order of $|qB|$ at low energies. As a result, the emission of photons with small energies, $\Omega\ll\sqrt{|qB|}$, can come only from the quantum transitions between the states with sufficiently large longitudinal momenta and/or Landau-level indices, where the vertical separation between the levels is equal to or smaller than $\Omega$, see Fig.~\ref{fig:transitions}(a). From a physics viewpoint, this implies that the rate is strongly suppressed when $\Omega\ll\sqrt{|qB|}$. On a technical side, however, this means that a reliable calculation requires the inclusion of a sufficiently large number of Landau levels, i.e., $n_{\rm max}\gtrsim |qB|/(2\Omega^2)$. In the numerical calculation below, we will set $n_{\rm max}\simeq 450$. Even with such large $n_{\rm max}$, one finds that the range of photon energies $\Omega$ is strongly constrained, i.e., $\sqrt{|qB|} /30 \lesssim \Omega \lesssim 30 \sqrt{|qB|}$. To remain in the range of validity, therefore, we will consider the following range of photon energies: $m\leq \Omega\leq 100m$, which lies within the required limits for both choices of the magnetic field. 

\begin{figure}[th]
\centering
\subfigure[]{\includegraphics[width=0.45\textwidth]{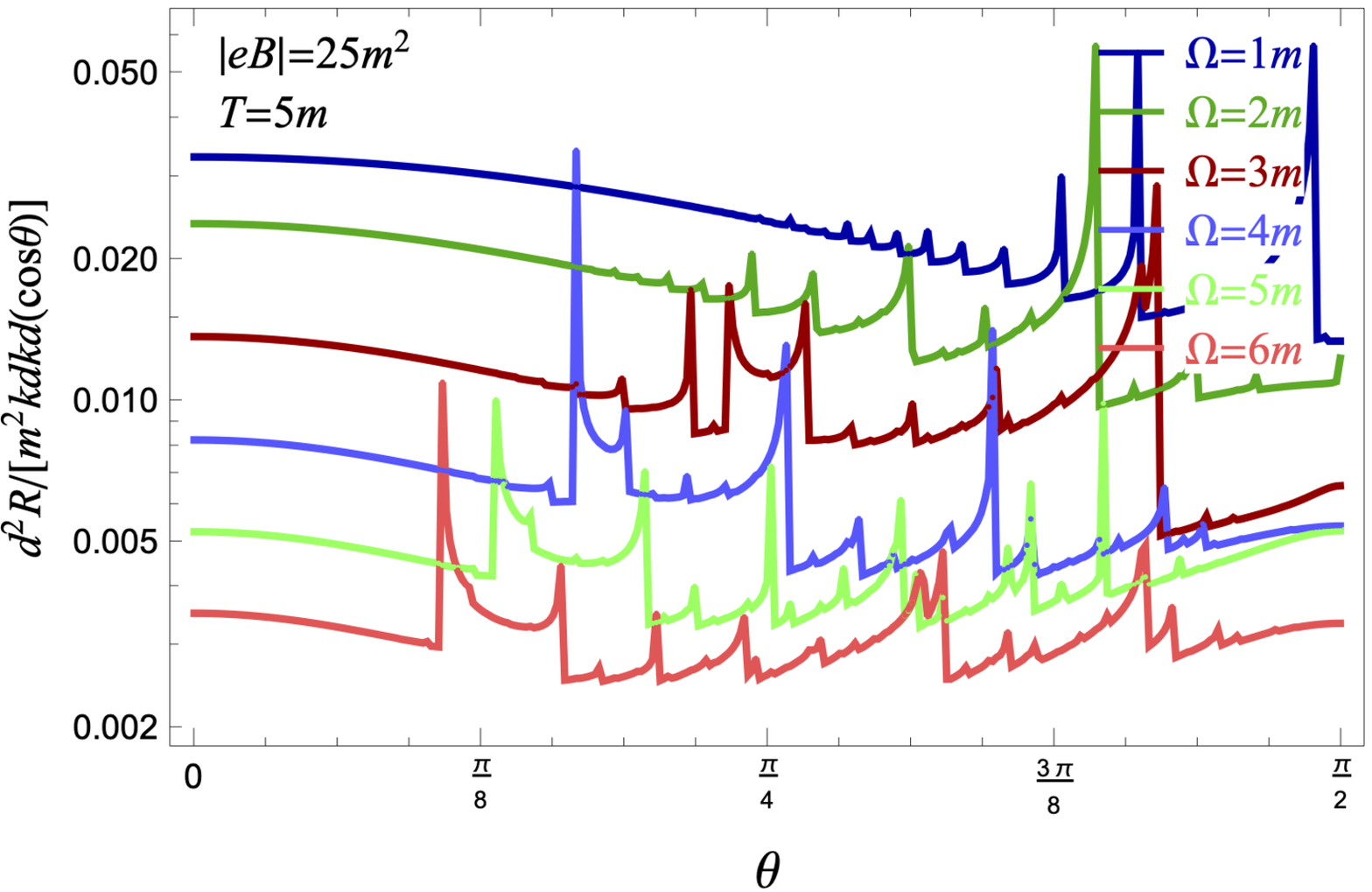}}
  \hspace{0.05\textwidth}
\subfigure[]{\includegraphics[width=0.45\textwidth]{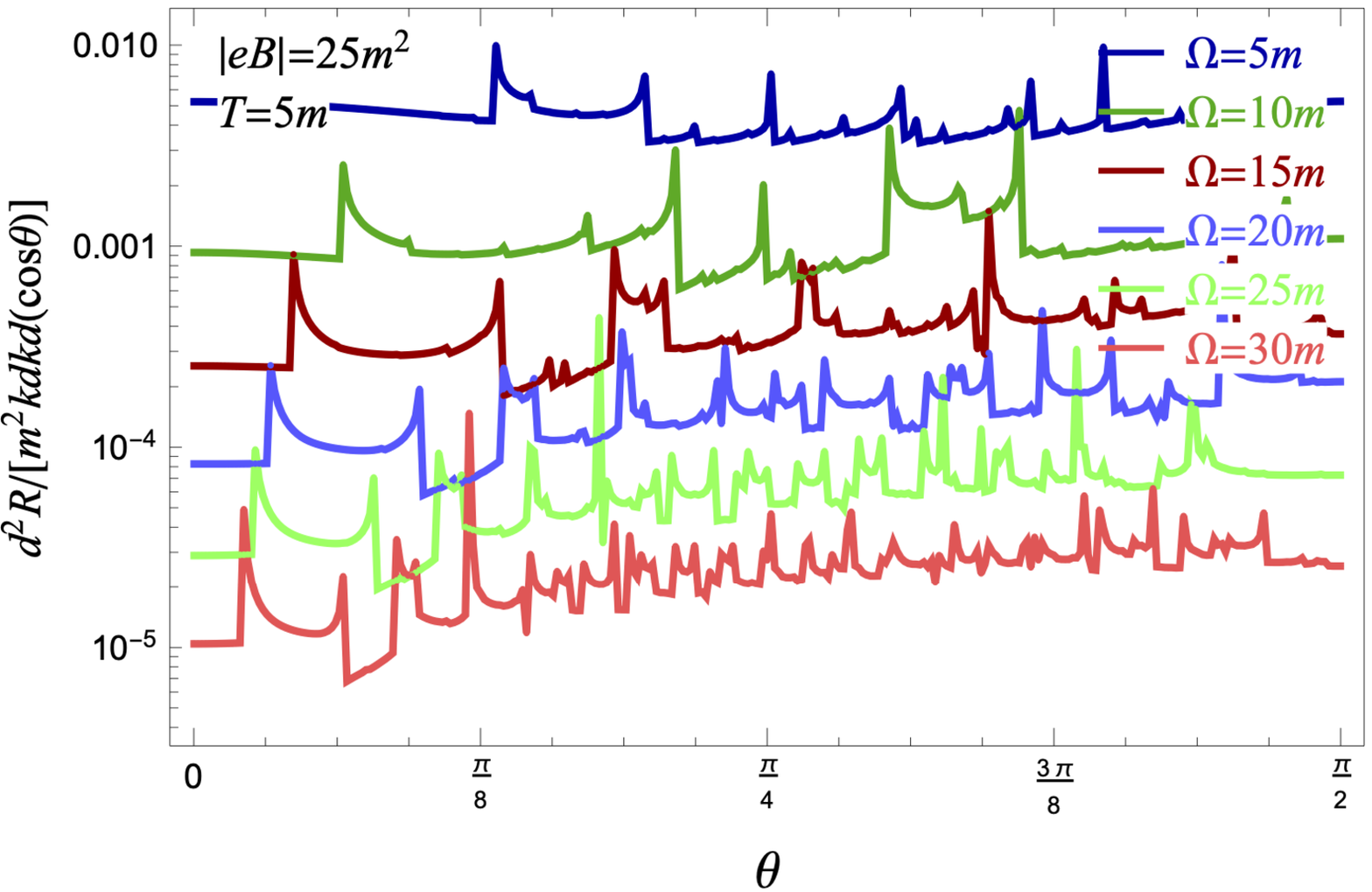}}\\
\subfigure[]{\includegraphics[width=0.3\textwidth]{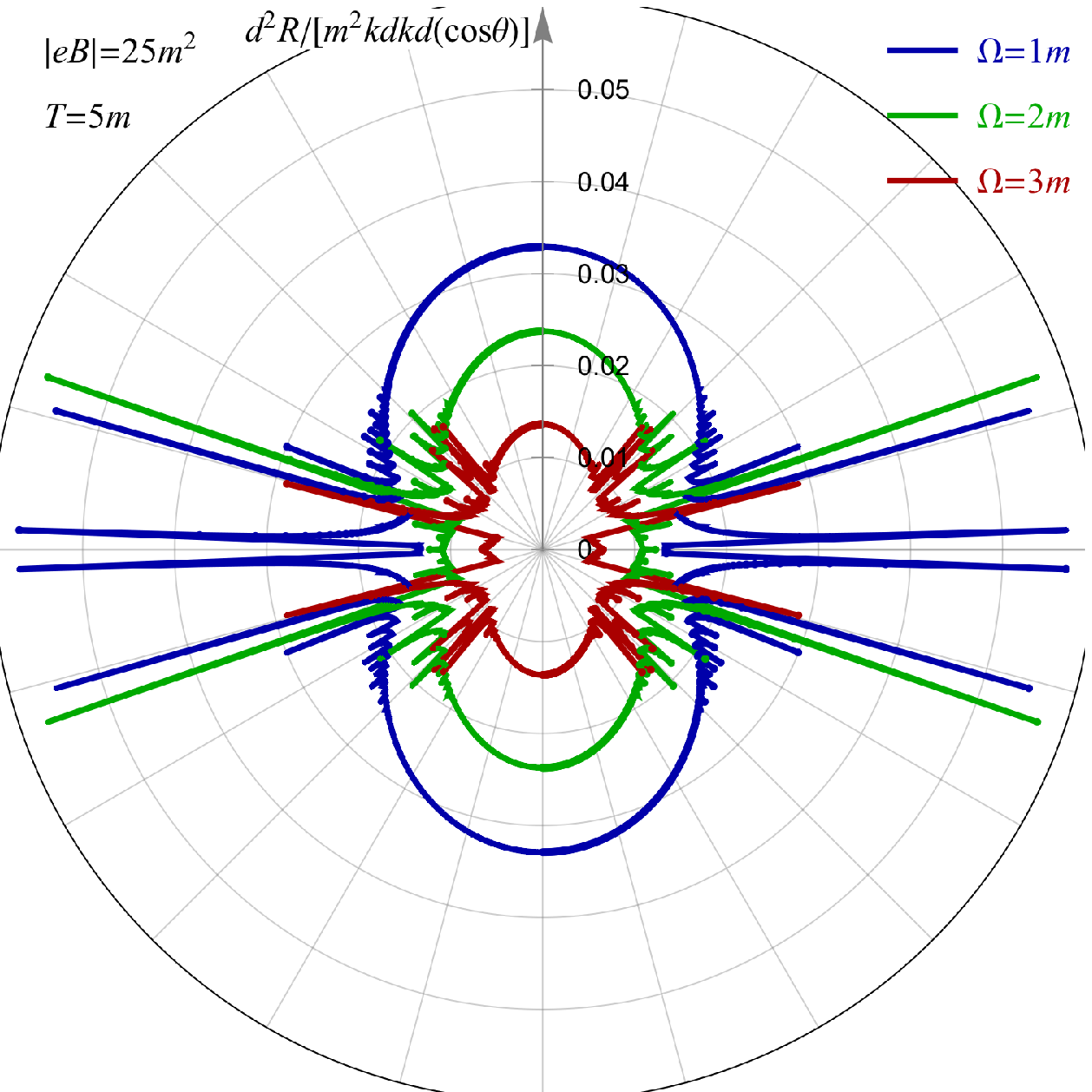}} 
  \hspace{0.02\textwidth}
\subfigure[]{\includegraphics[width=0.3\textwidth]{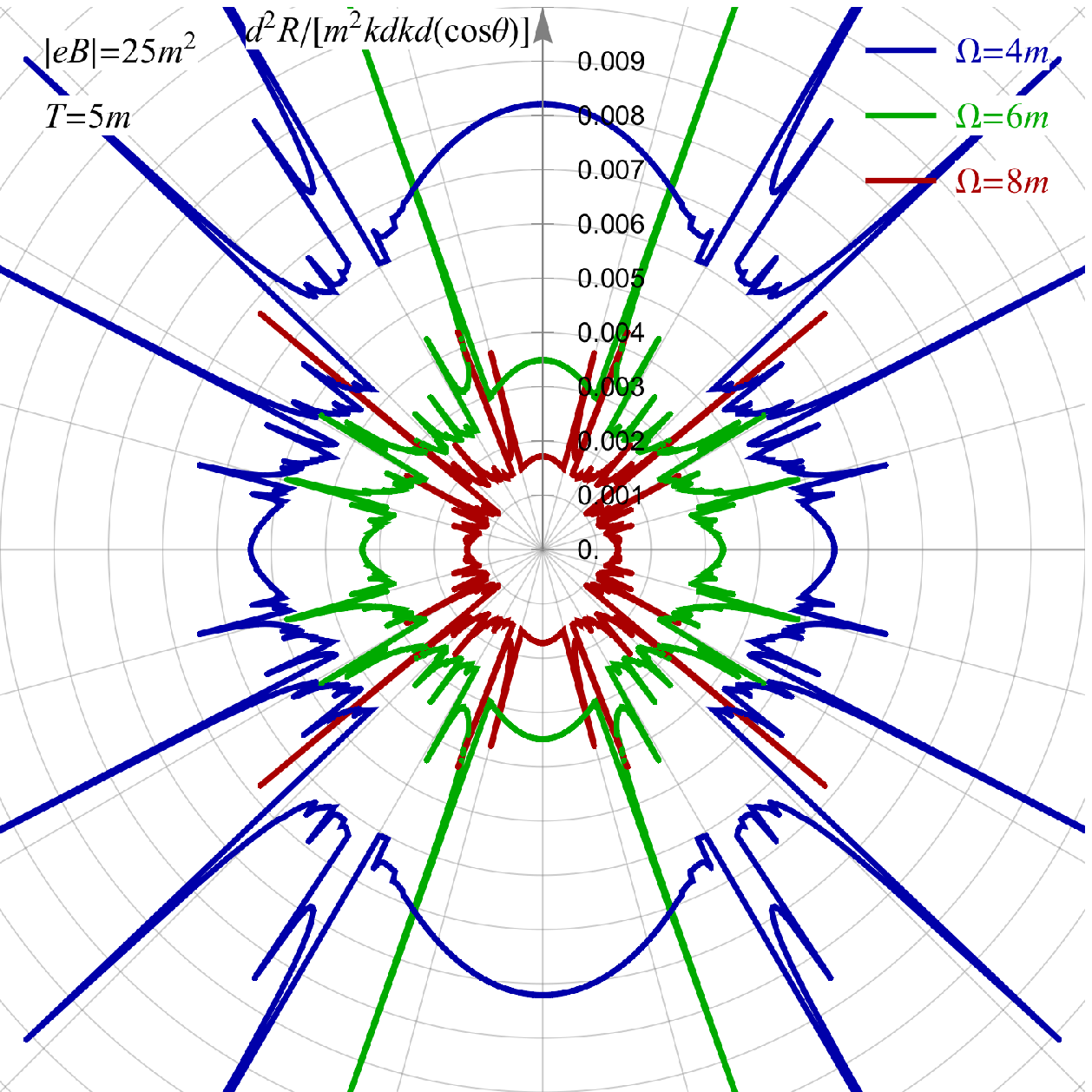}} 
  \hspace{0.02\textwidth}
\subfigure[]{\includegraphics[width=0.3\textwidth]{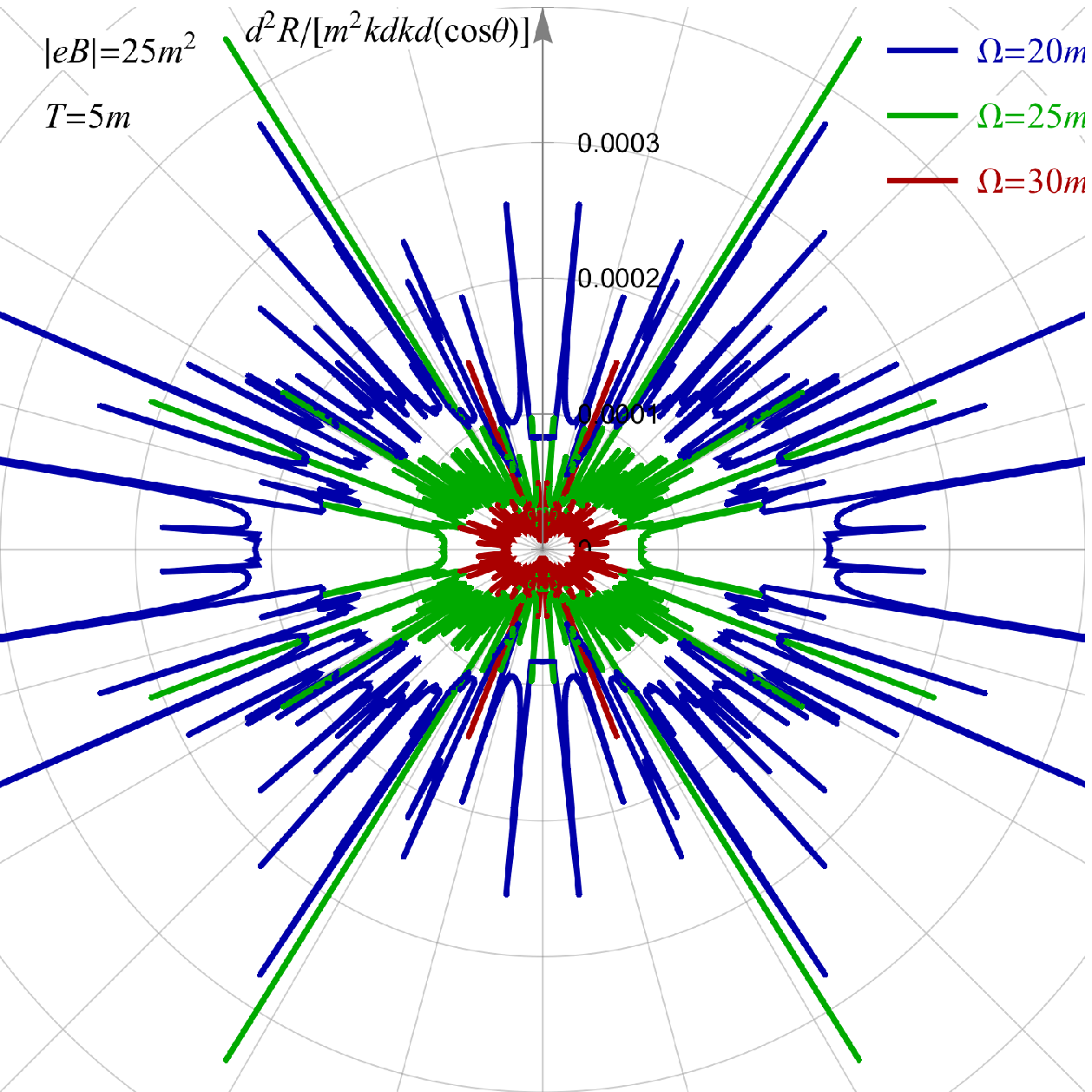}} 
\caption{The differential photon production rate as a function of the angle $\theta$ for $|qB|=25m^2$ and $T=5m$.}
\label{fig:resultsB25T5}
\end{figure}
\begin{figure}[th]
\centering
\subfigure[]{\includegraphics[width=0.45\textwidth]{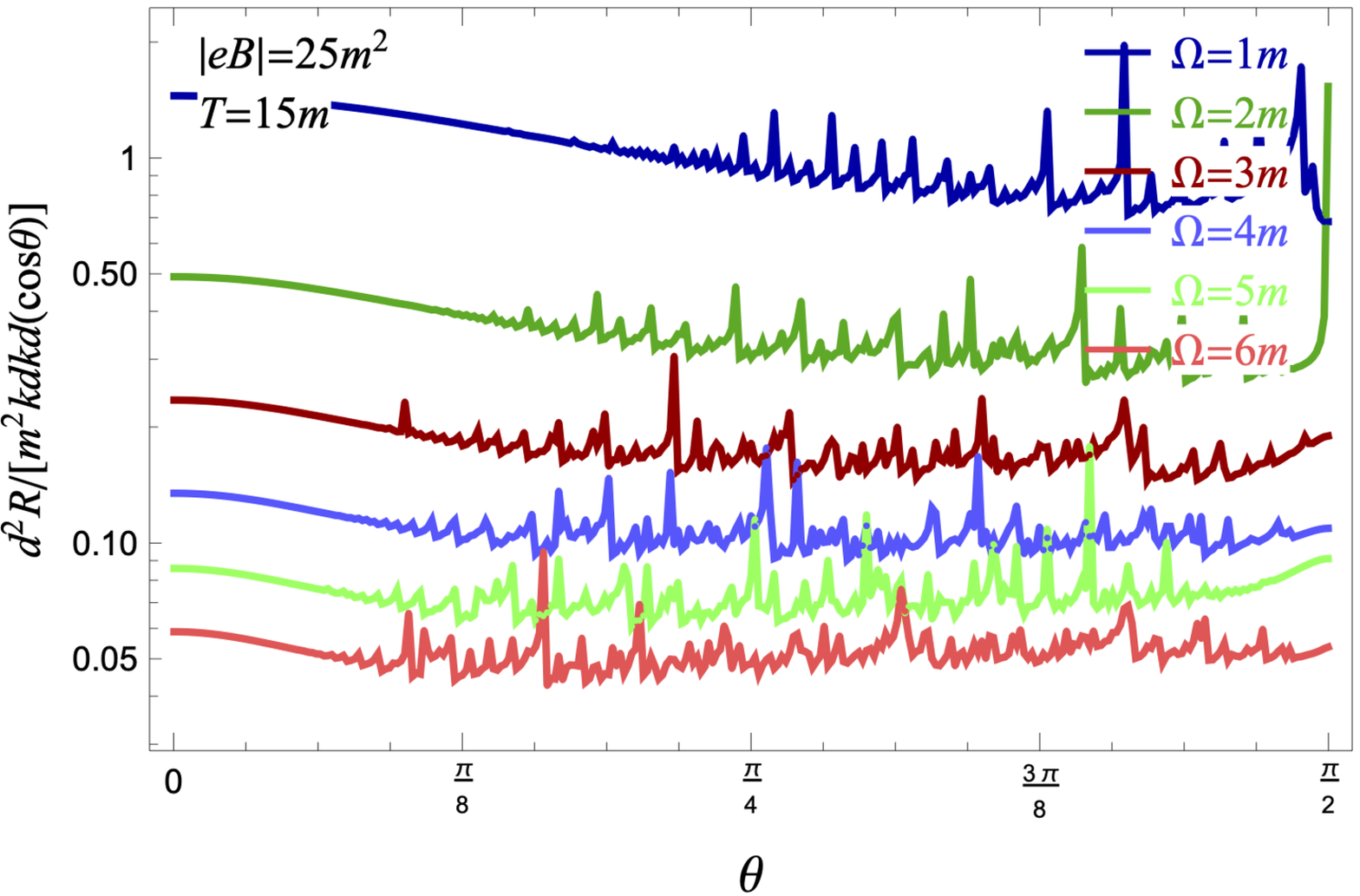}}
  \hspace{0.05\textwidth}
\subfigure[]{\includegraphics[width=0.45\textwidth]{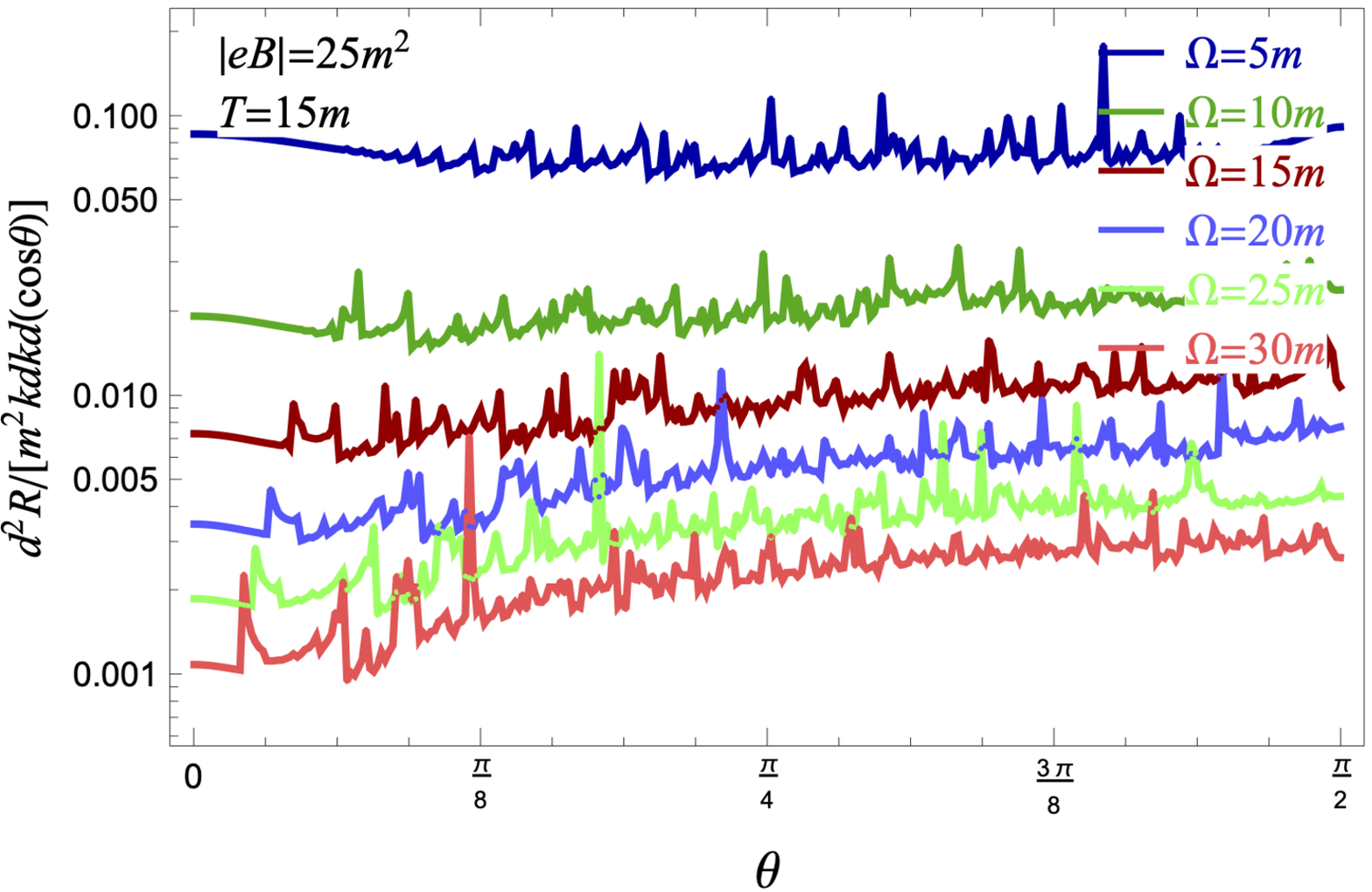}}\\
\subfigure[]{\includegraphics[width=0.3\textwidth]{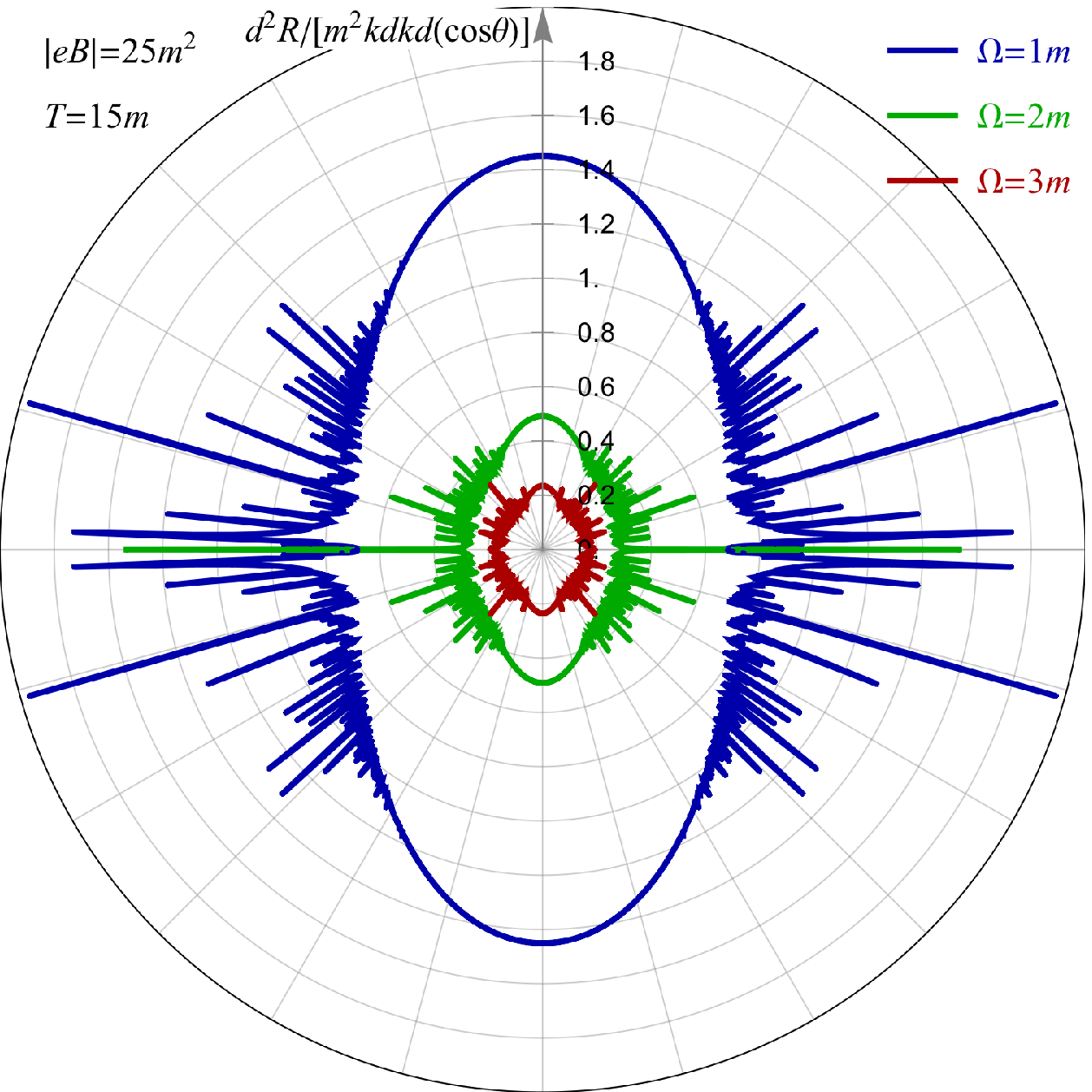}} 
  \hspace{0.02\textwidth}
\subfigure[]{\includegraphics[width=0.3\textwidth]{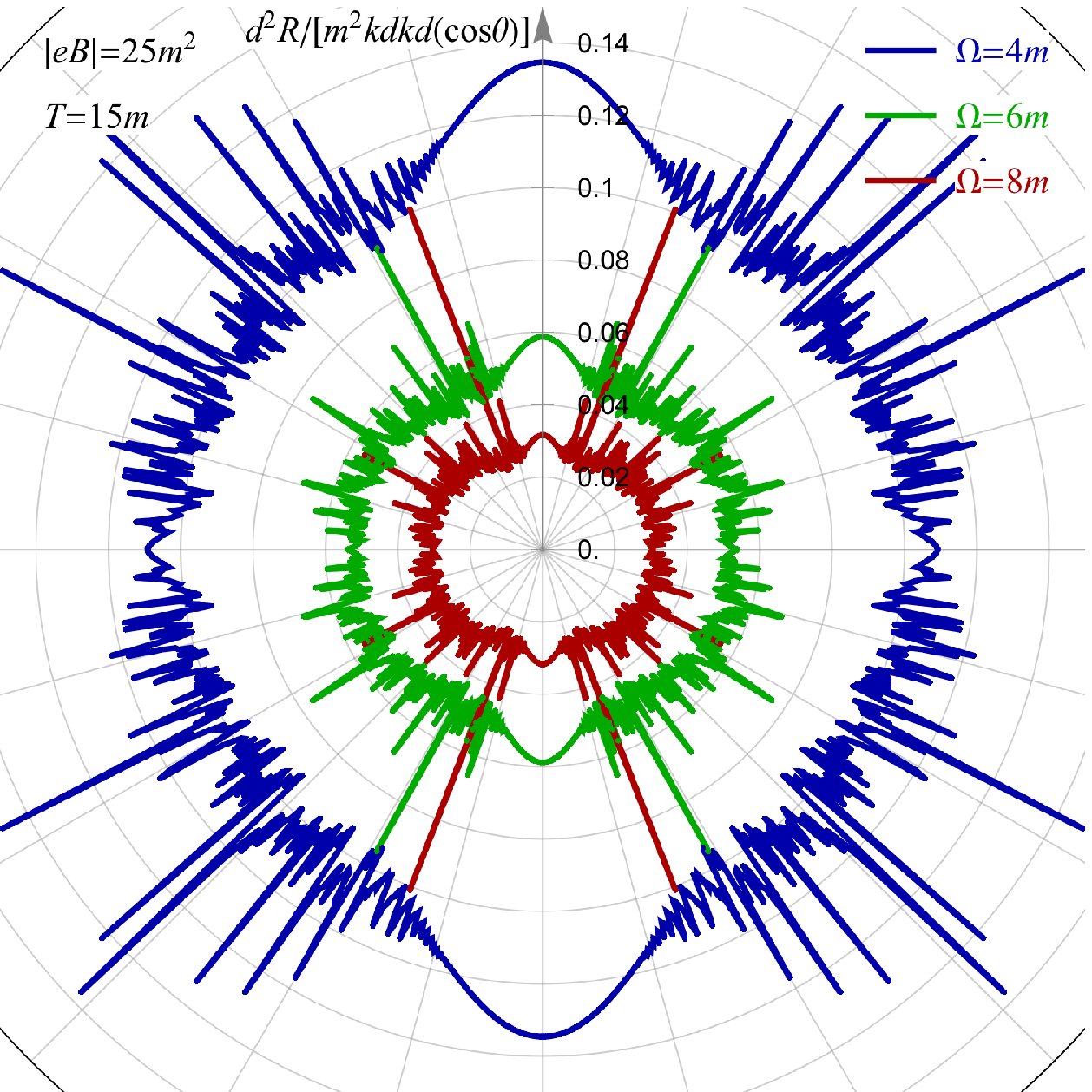}} 
  \hspace{0.02\textwidth}
\subfigure[]{\includegraphics[width=0.3\textwidth]{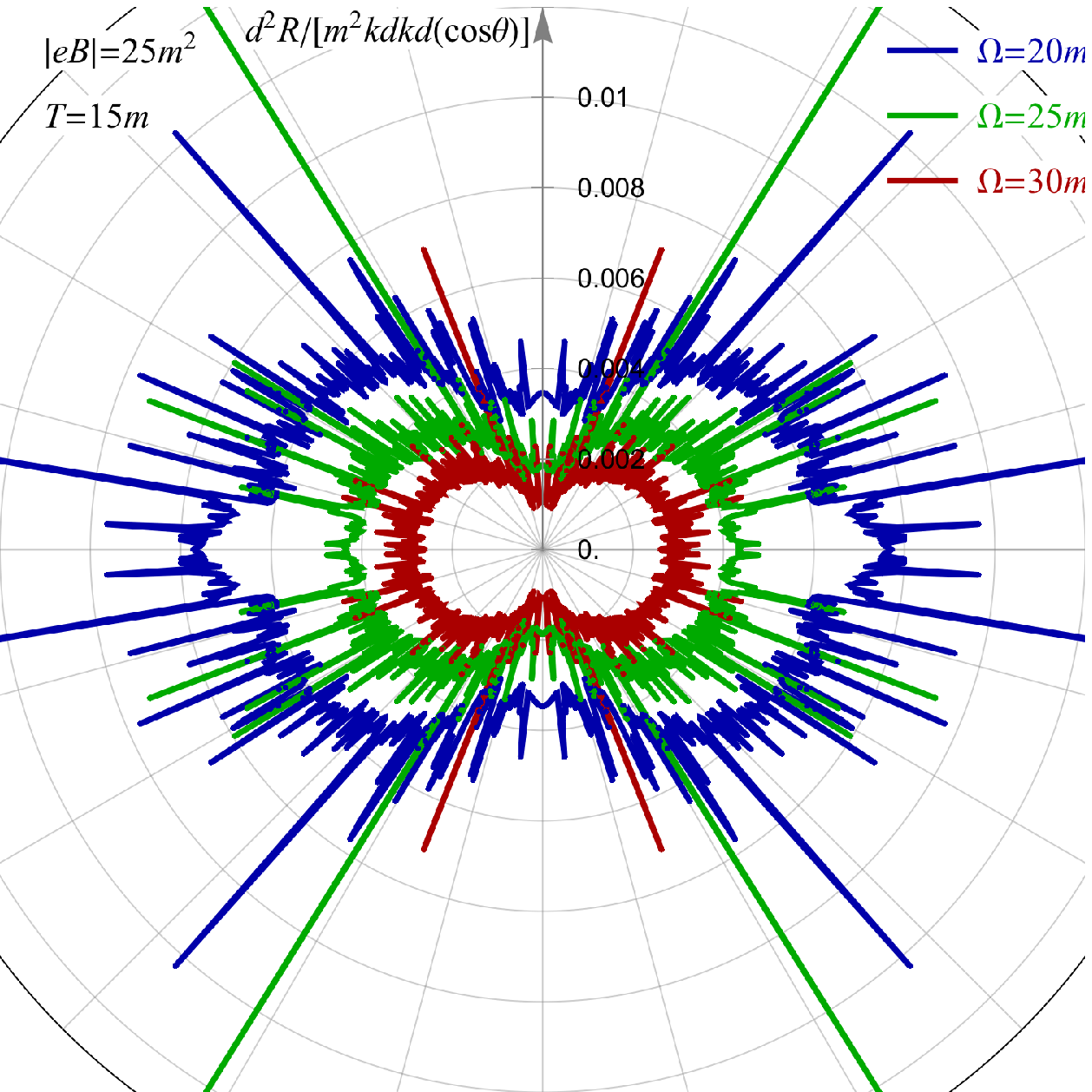}} 
\caption{The differential photon production rate as a function of the angle $\theta$ for $|qB|=25m^2$ and $T=15m$.}
\label{fig:resultsB25T15}
\end{figure}
\begin{figure}[th]
\centering
\subfigure[]{\includegraphics[width=0.45\textwidth]{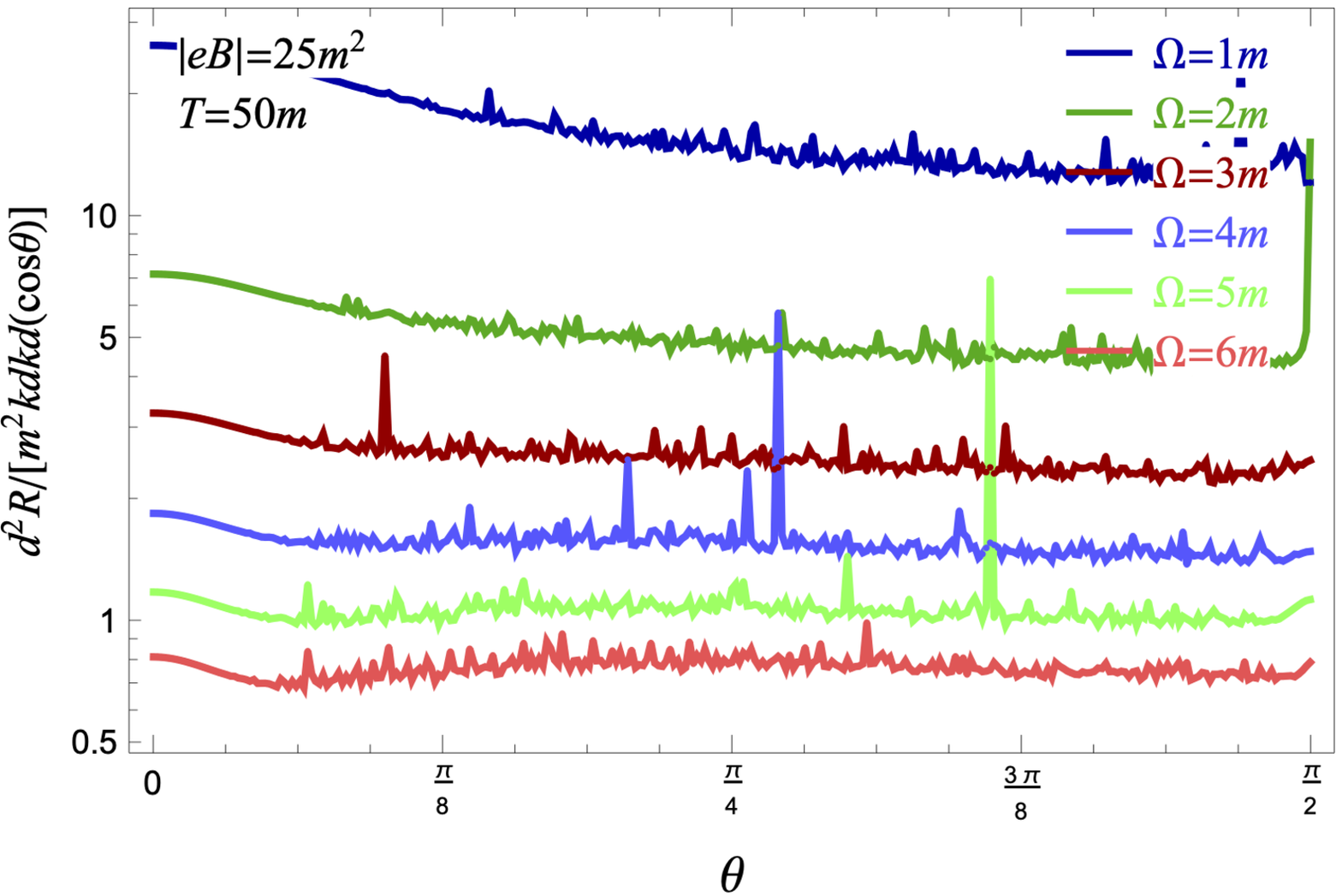}}
  \hspace{0.05\textwidth}
\subfigure[]{\includegraphics[width=0.45\textwidth]{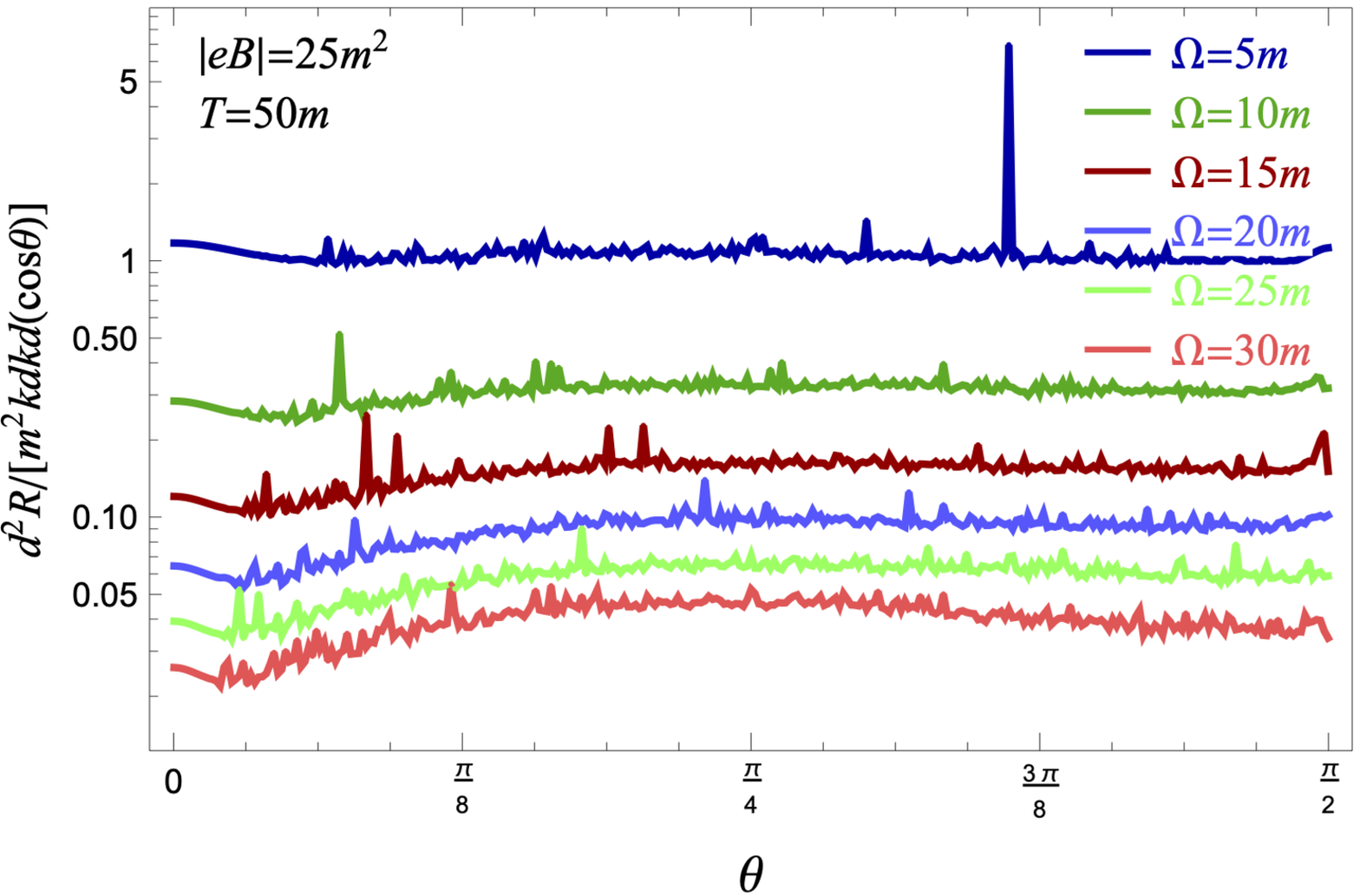}}\\
\subfigure[]{\includegraphics[width=0.3\textwidth]{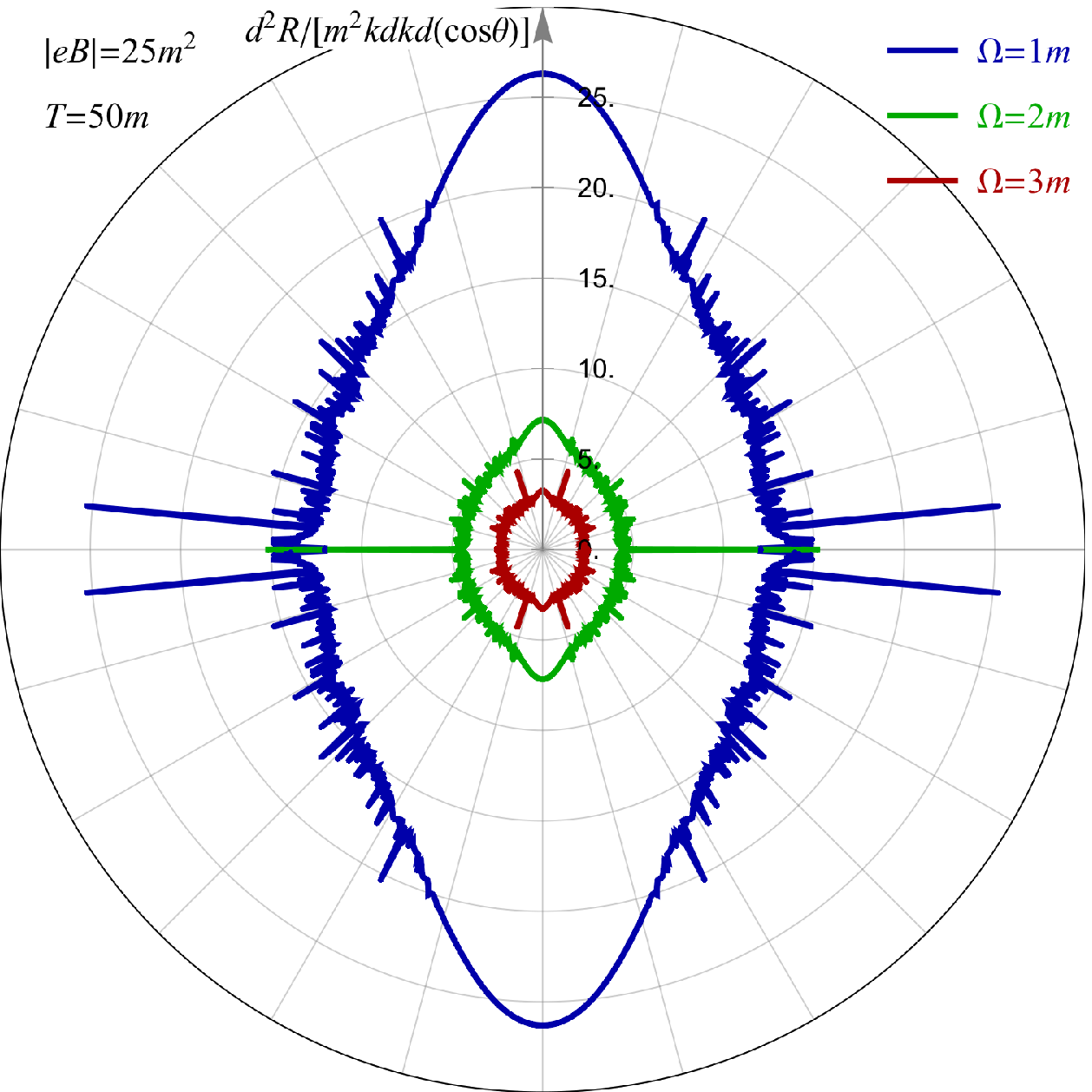}} 
  \hspace{0.02\textwidth}
\subfigure[]{\includegraphics[width=0.3\textwidth]{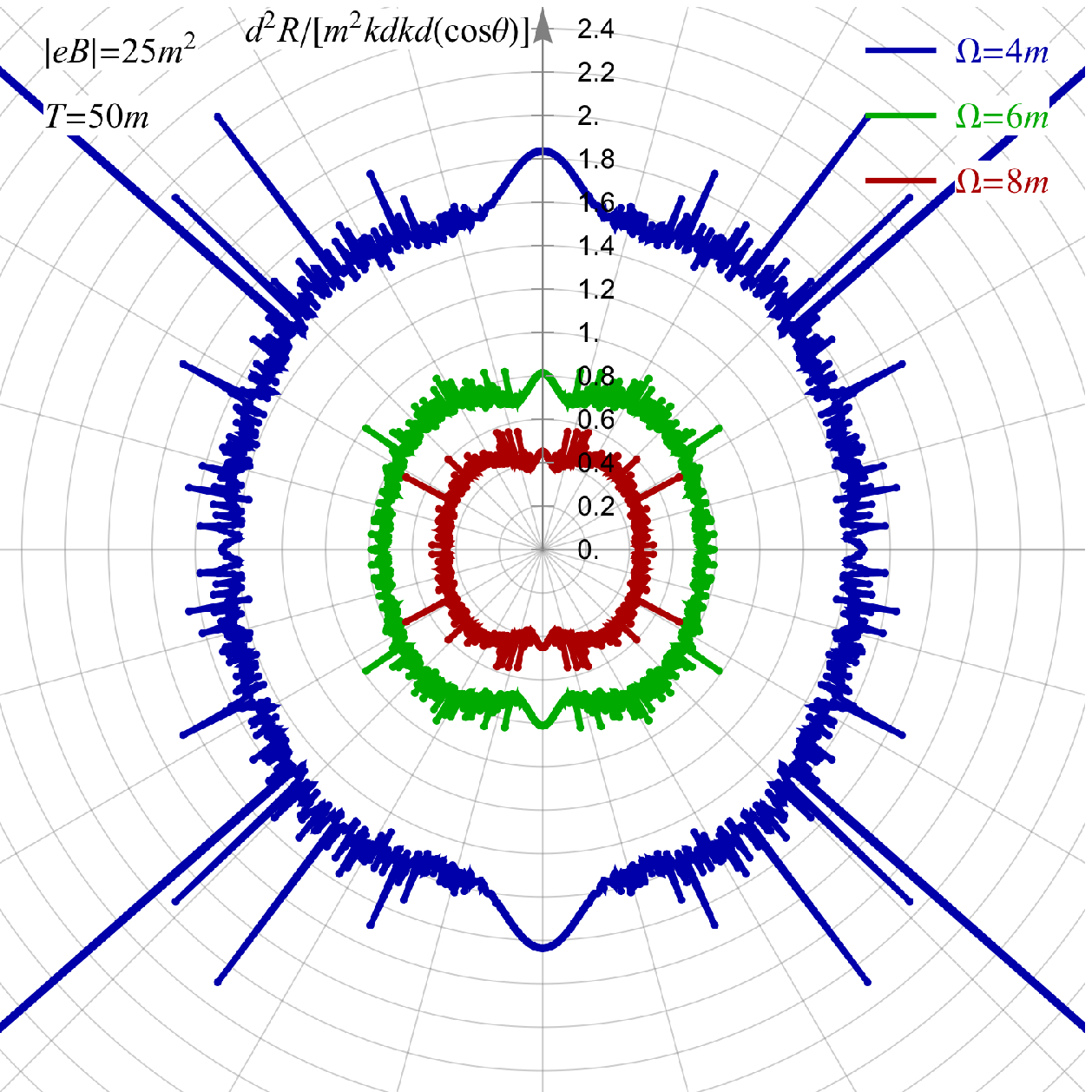}} 
  \hspace{0.02\textwidth}
\subfigure[]{\includegraphics[width=0.3\textwidth]{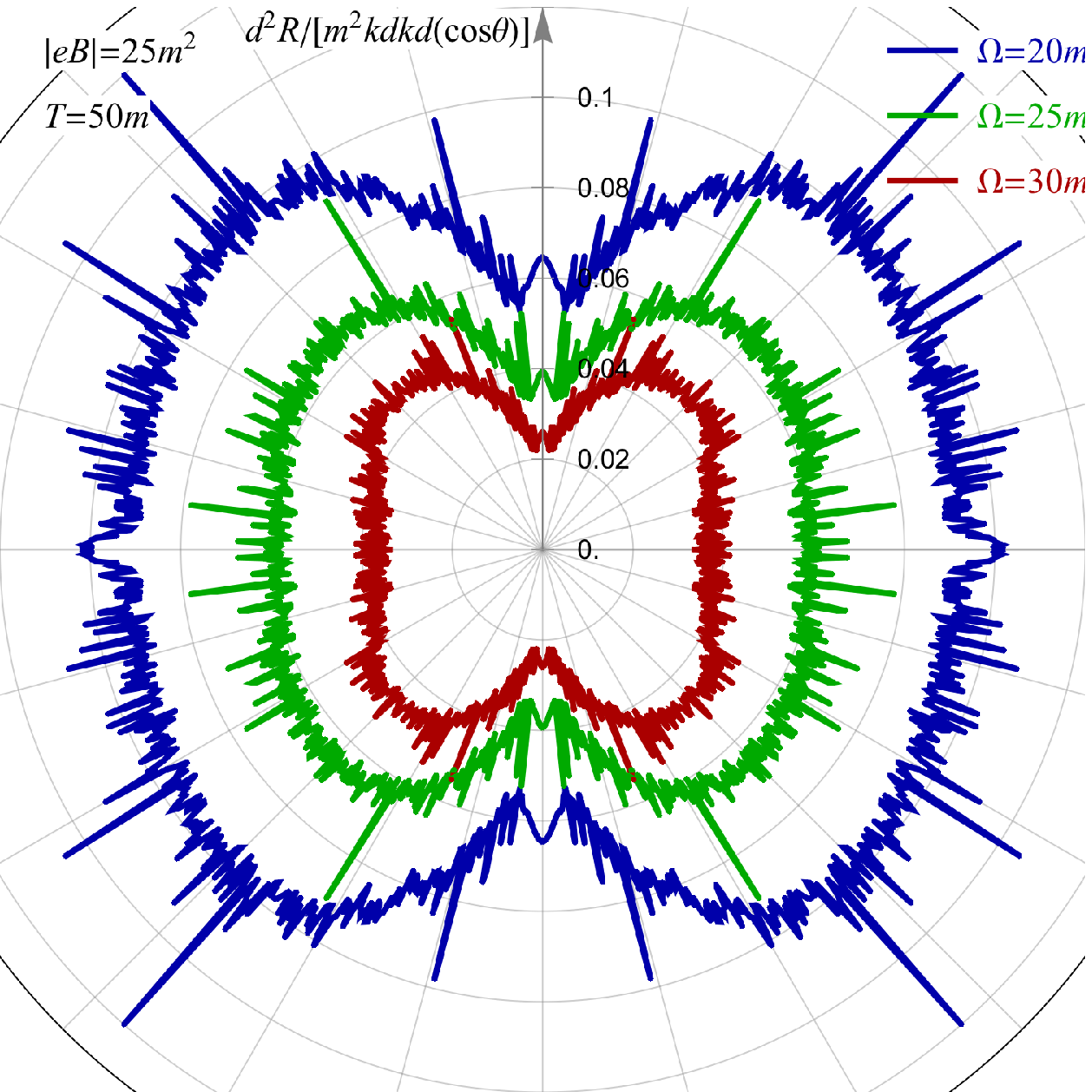}} 
\caption{The differential photon production rate as a function of the angle $\theta$ for $|qB|=25m^2$ and $T=50m$.}
\label{fig:resultsB25T50}
\end{figure}

\begin{figure}[th]
\centering
\subfigure[]{\includegraphics[width=0.45\textwidth]{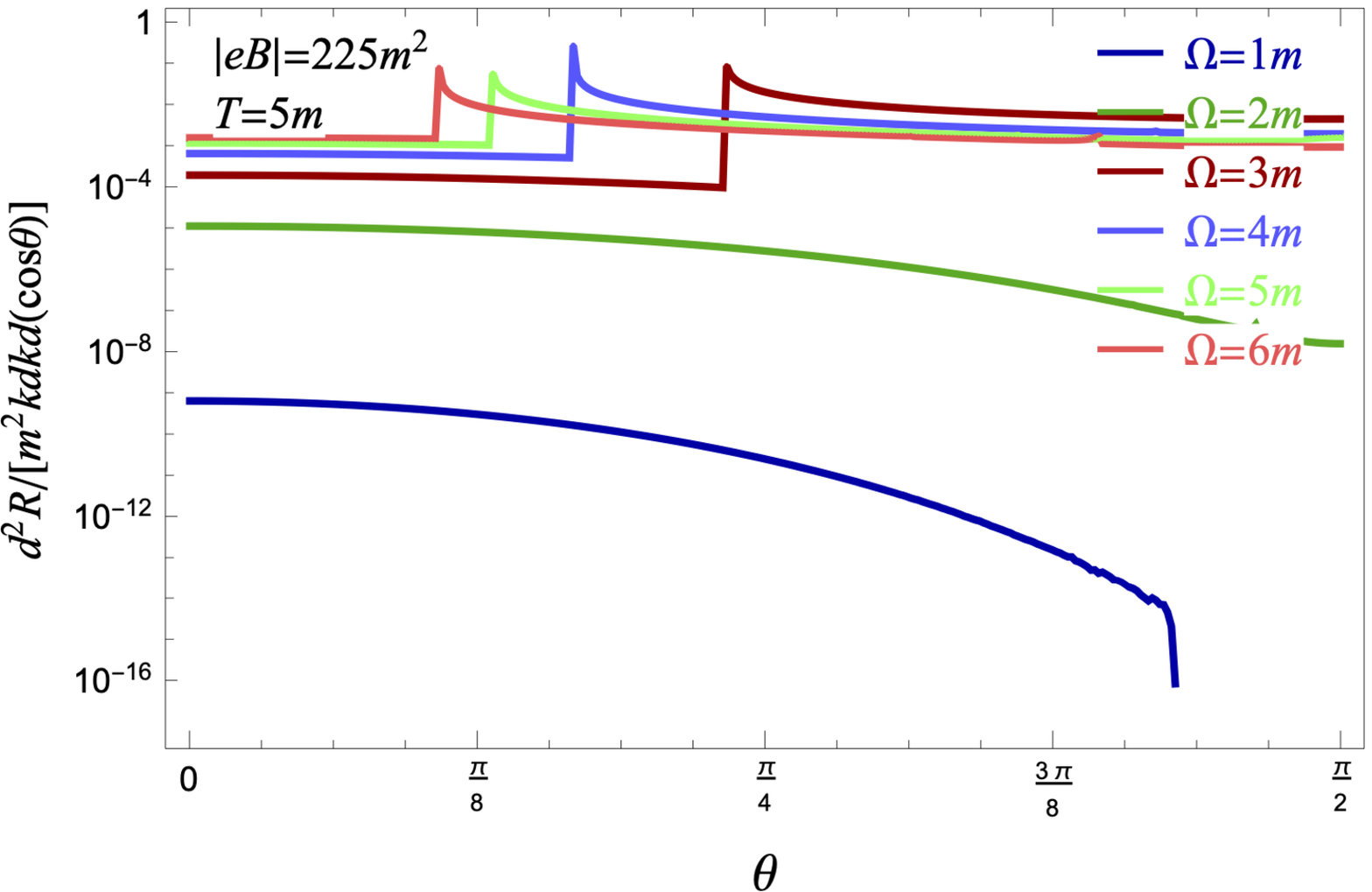}}
  \hspace{0.05\textwidth}
\subfigure[]{\includegraphics[width=0.45\textwidth]{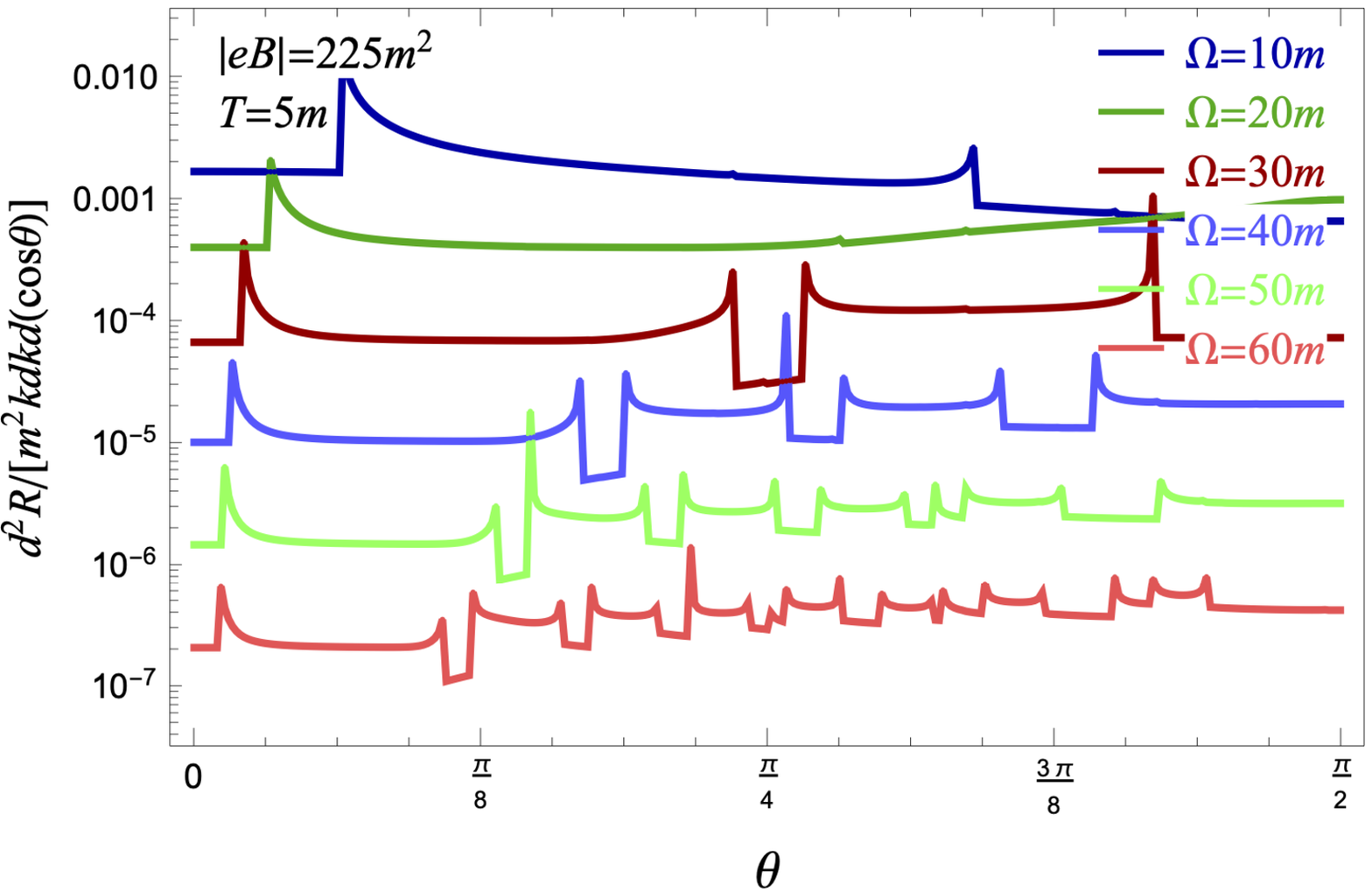}}\\
\subfigure[]{\includegraphics[width=0.3\textwidth]{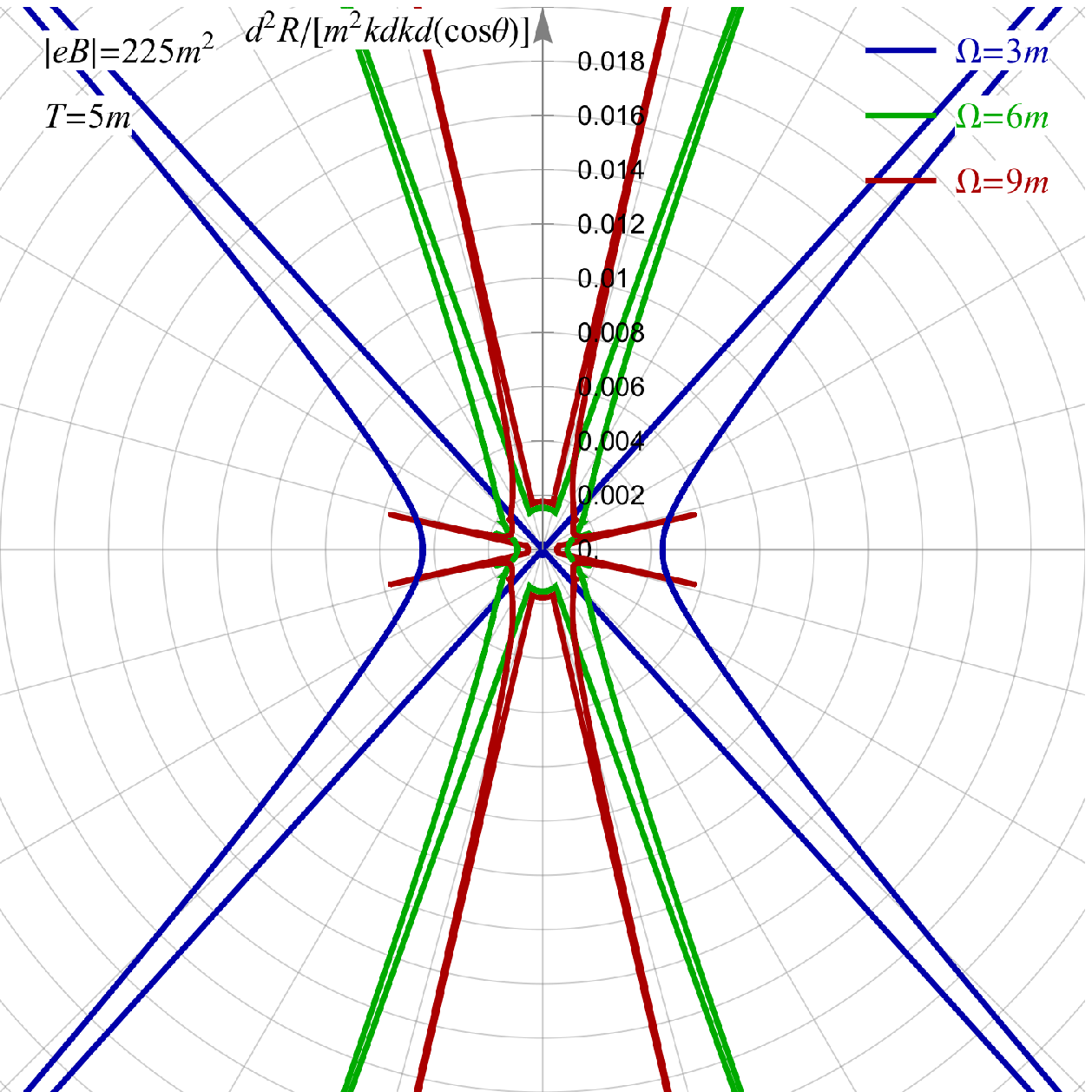}} 
  \hspace{0.02\textwidth}
\subfigure[]{\includegraphics[width=0.3\textwidth]{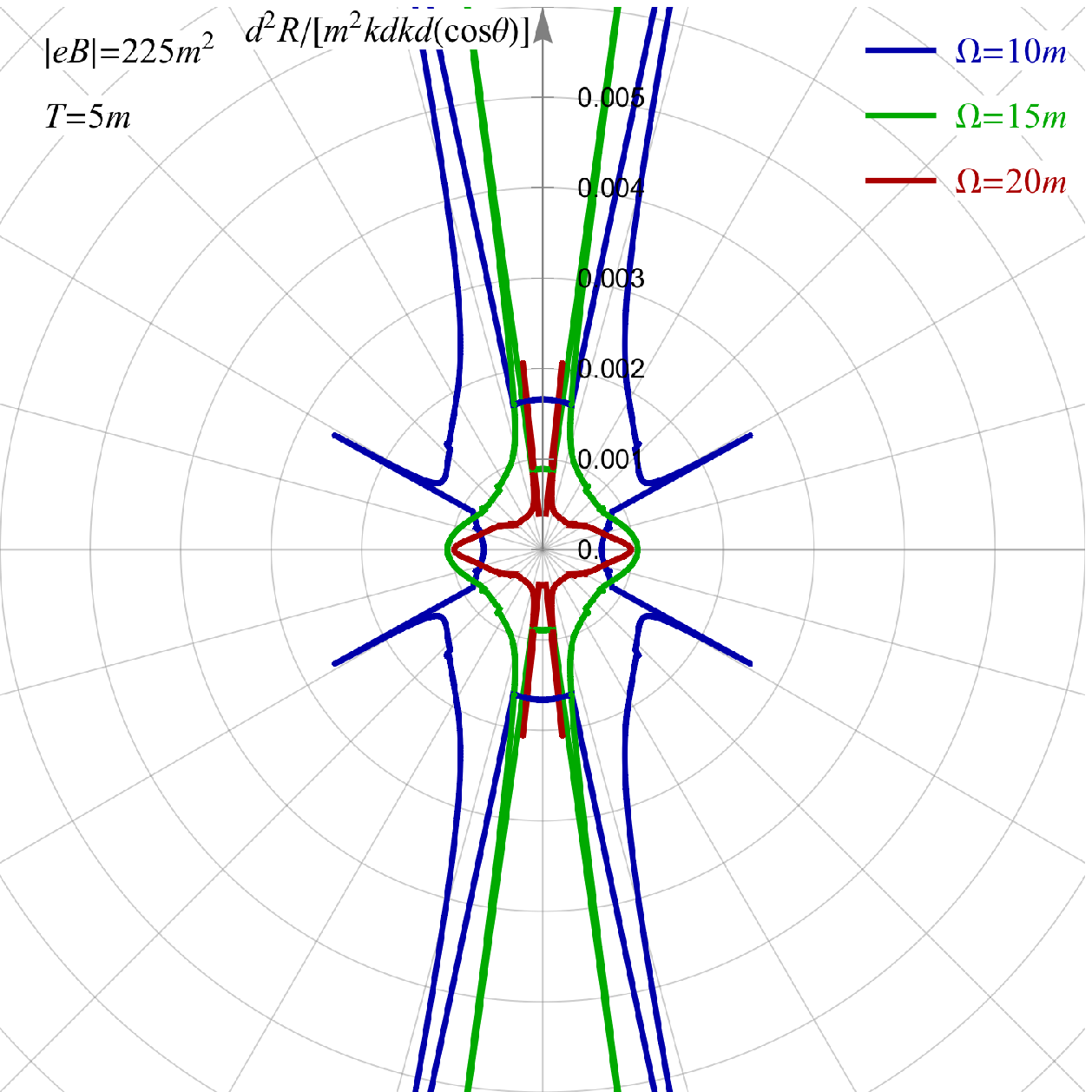}} 
  \hspace{0.02\textwidth}
\subfigure[]{\includegraphics[width=0.3\textwidth]{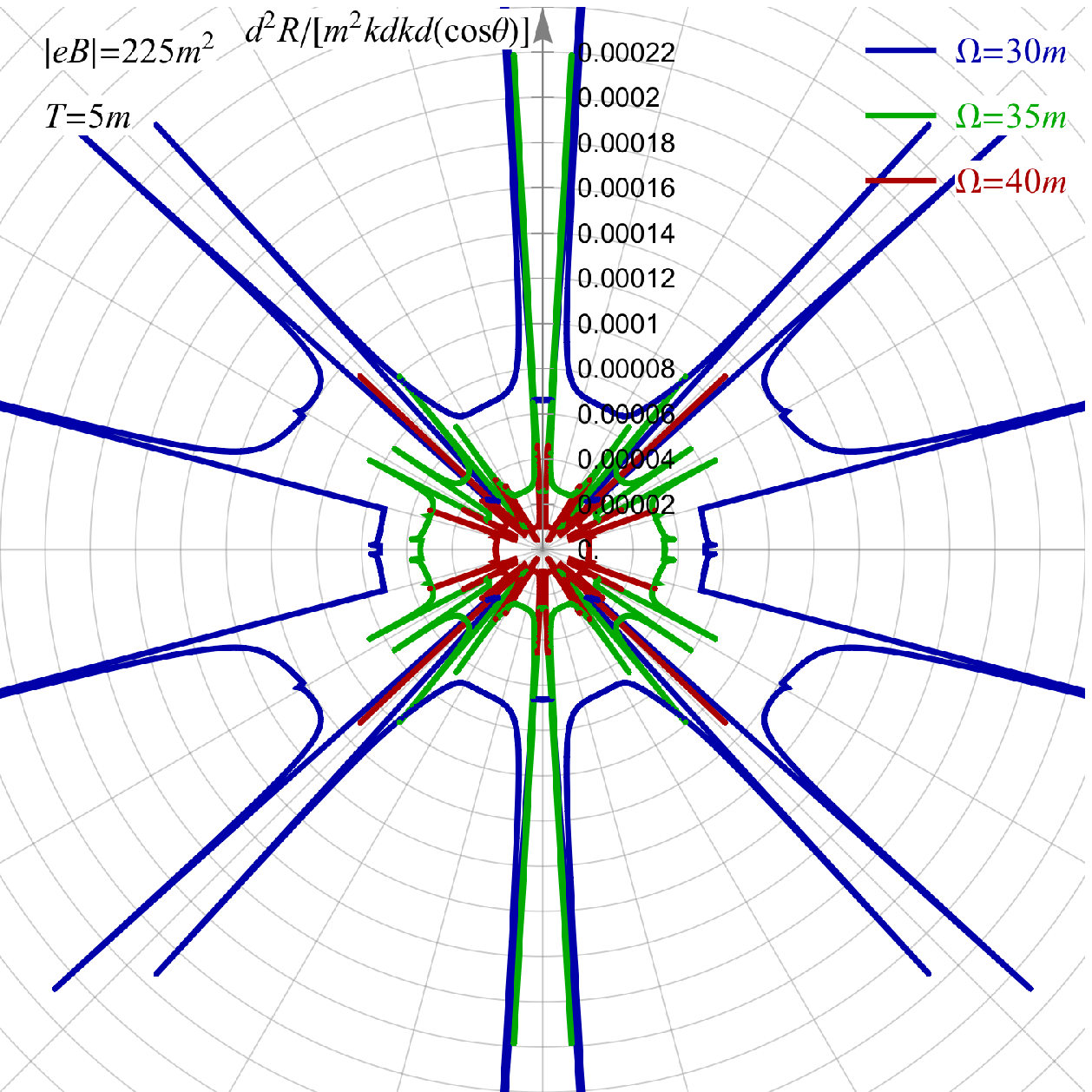}} 
\caption{The differential photon production rate as a function of the angle $\theta$ for $|qB|=225m^2$ and $T=5m$.}
\label{fig:resultsB225T5}
\end{figure}
\begin{figure}[th]
\centering
\subfigure[]{\includegraphics[width=0.44\textwidth]{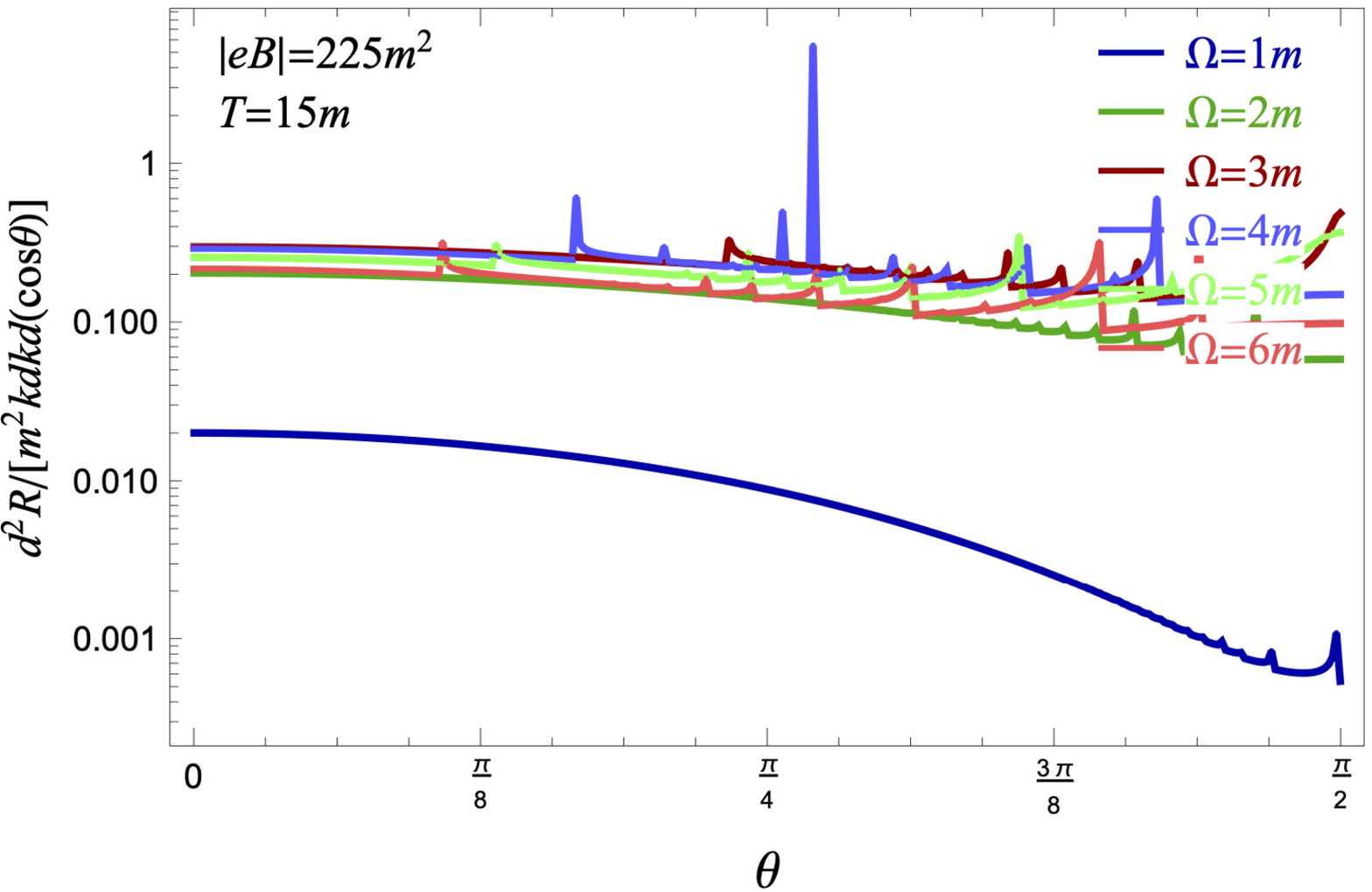}}
  \hspace{0.05\textwidth}
\subfigure[]{\includegraphics[width=0.45\textwidth]{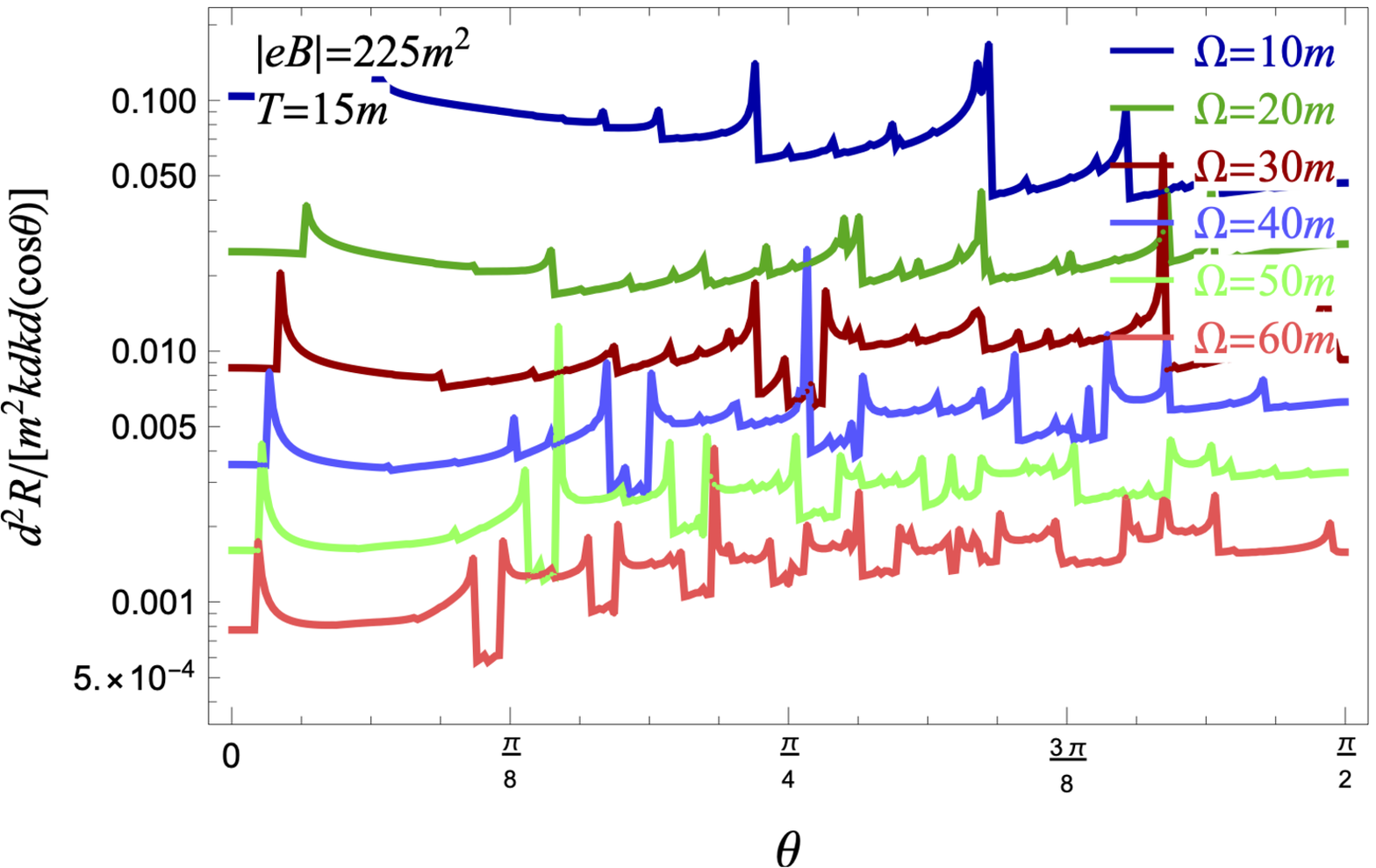}}\\
\subfigure[]{\includegraphics[width=0.3\textwidth]{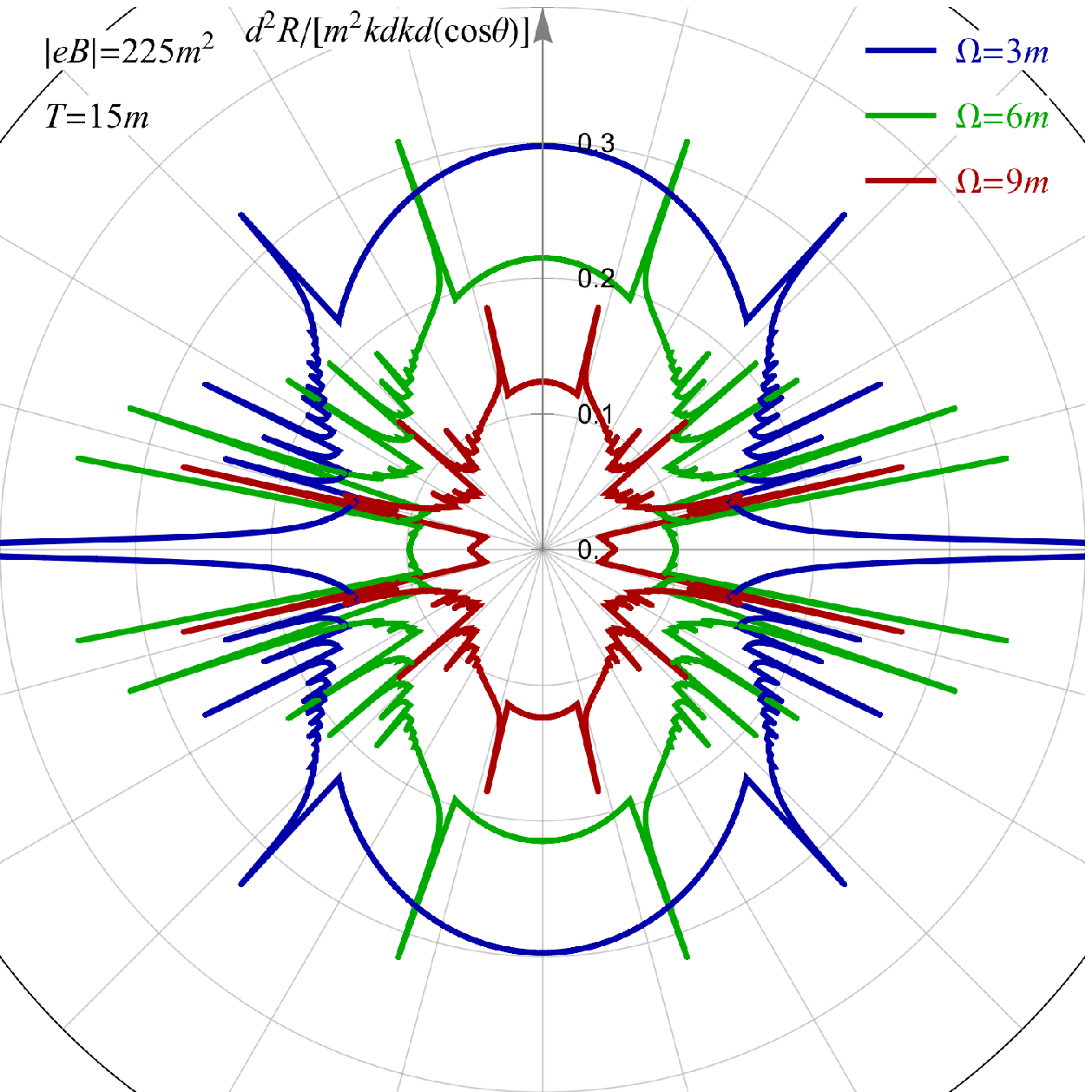}} 
  \hspace{0.02\textwidth}
\subfigure[]{\includegraphics[width=0.3\textwidth]{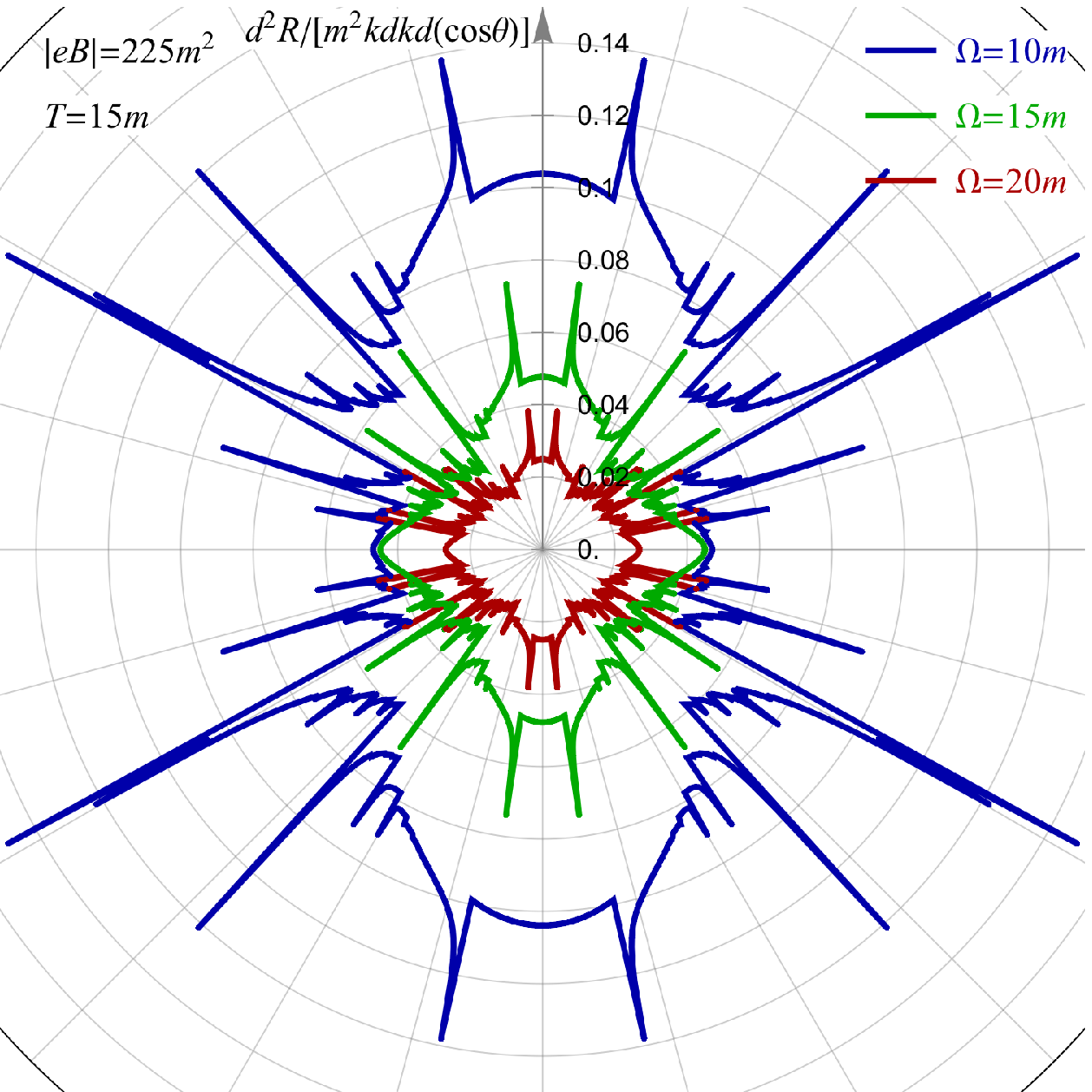}} 
  \hspace{0.02\textwidth}
\subfigure[]{\includegraphics[width=0.3\textwidth]{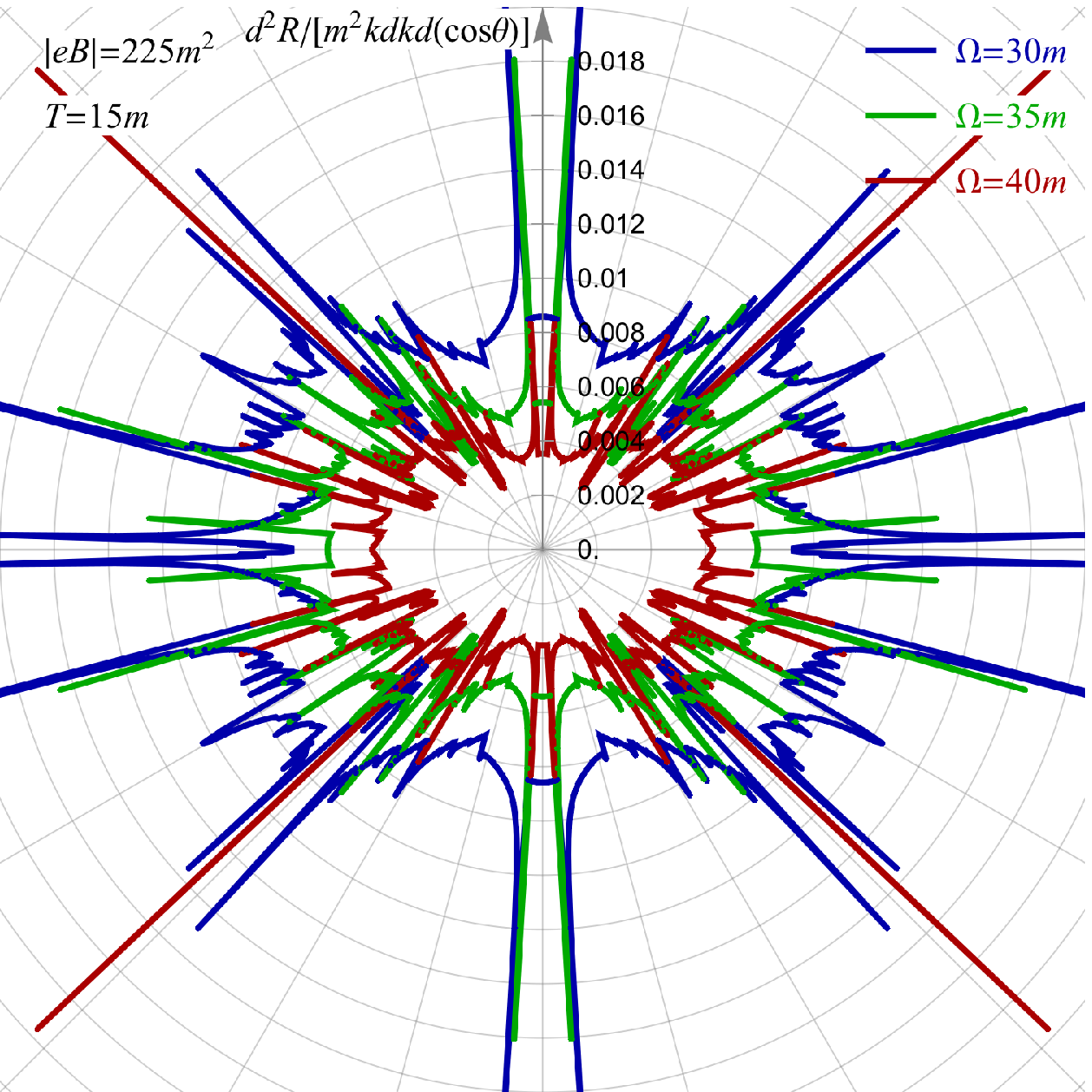}} 
\caption{The differential photon production rate as a function of the angle $\theta$ for $|qB|=225m^2$ and $T=15m$.}
\label{fig:resultsB225T15}
\end{figure}
\begin{figure}[th]
\centering
\subfigure[]{\includegraphics[width=0.44\textwidth]{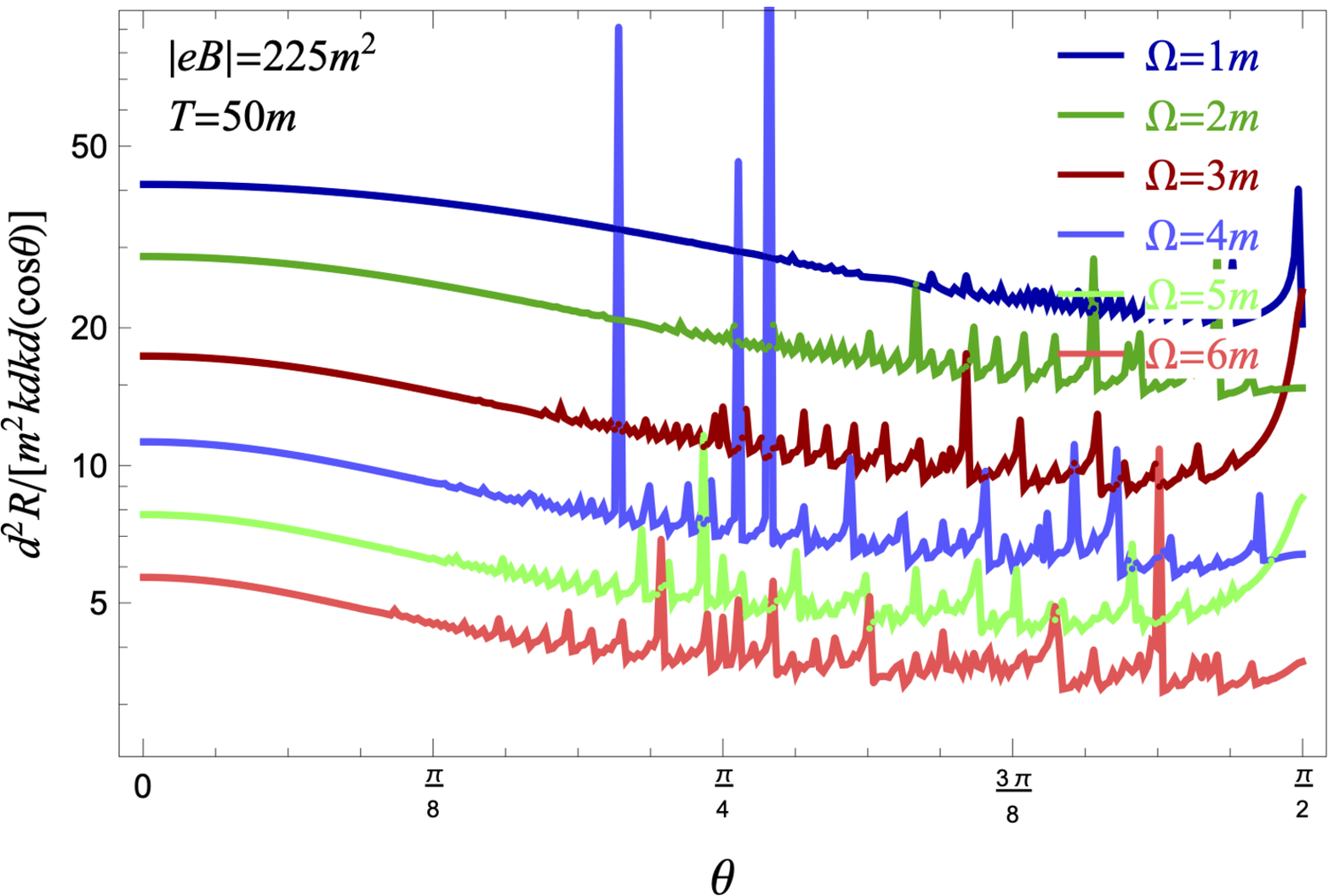}}
  \hspace{0.05\textwidth}
\subfigure[]{\includegraphics[width=0.45\textwidth]{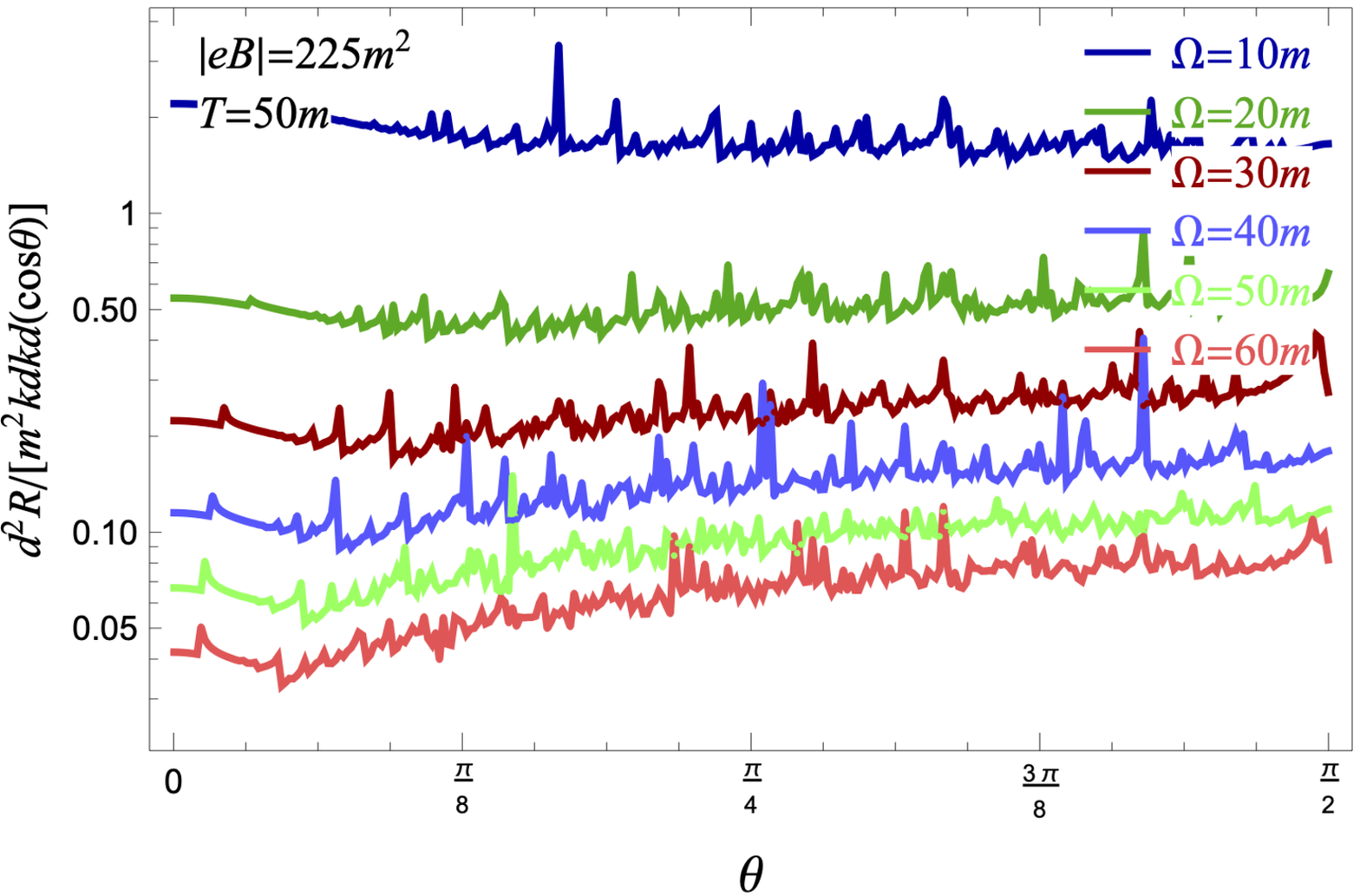}}\\
\subfigure[]{\includegraphics[width=0.3\textwidth]{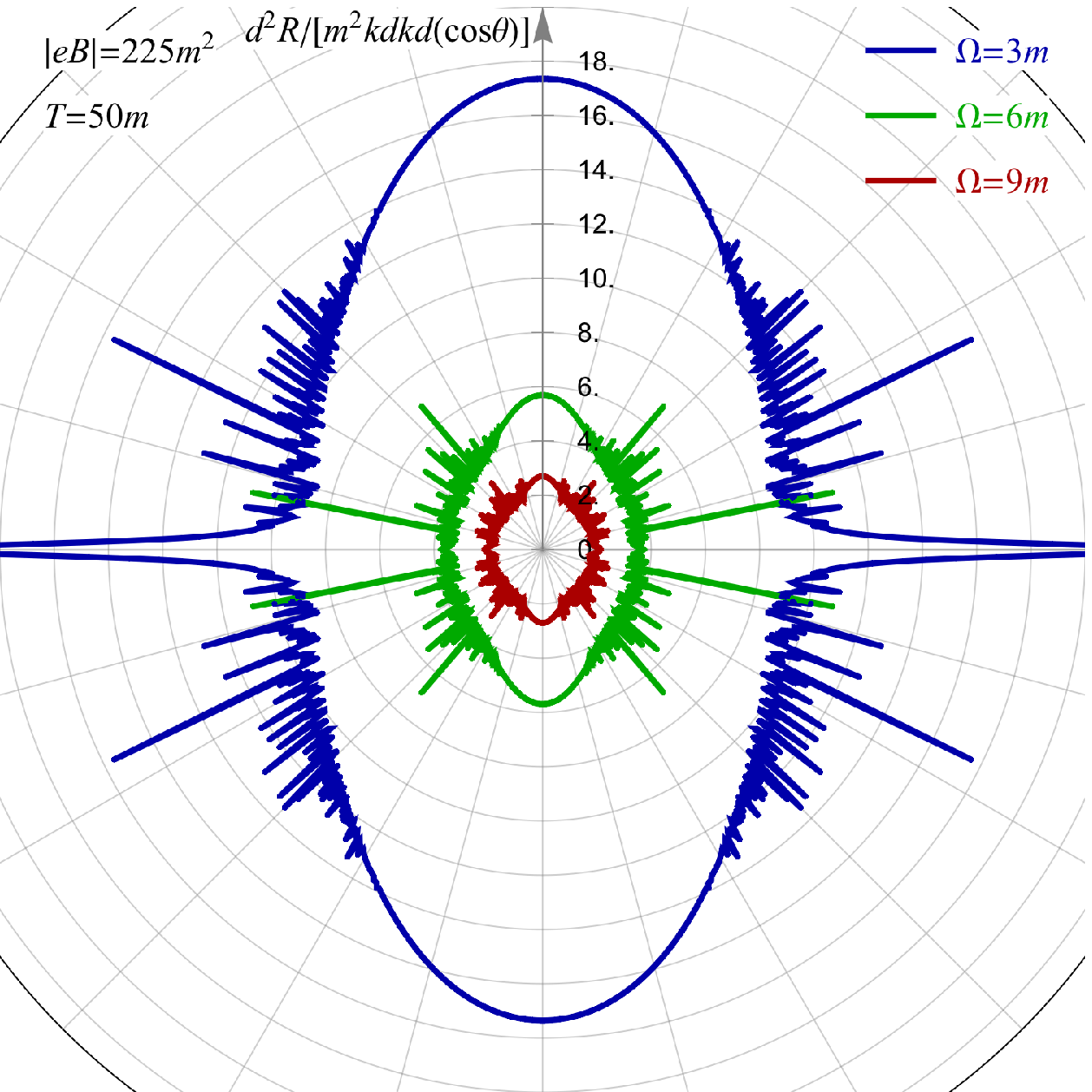}} 
  \hspace{0.02\textwidth}
\subfigure[]{\includegraphics[width=0.3\textwidth]{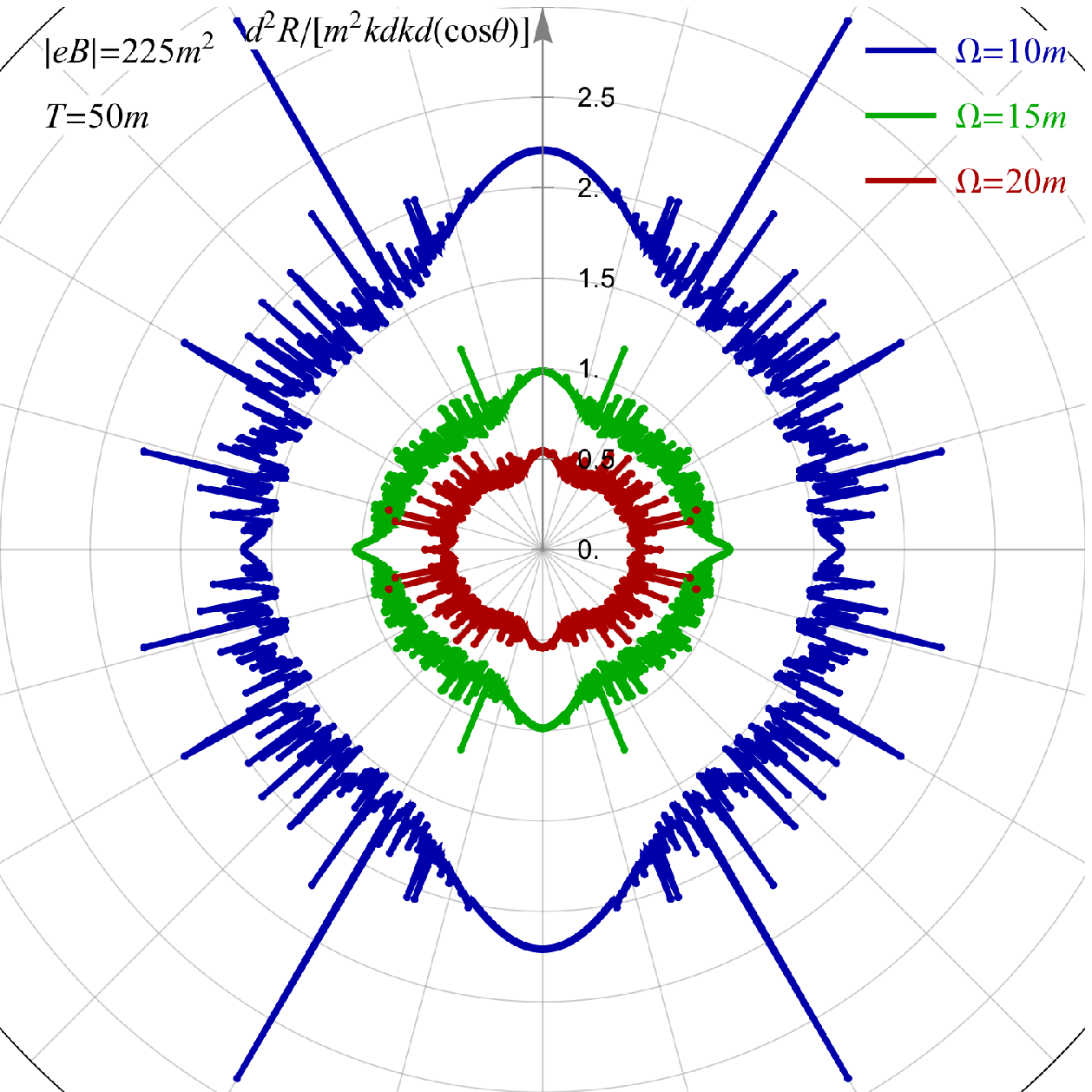}} 
  \hspace{0.02\textwidth}
\subfigure[]{\includegraphics[width=0.3\textwidth]{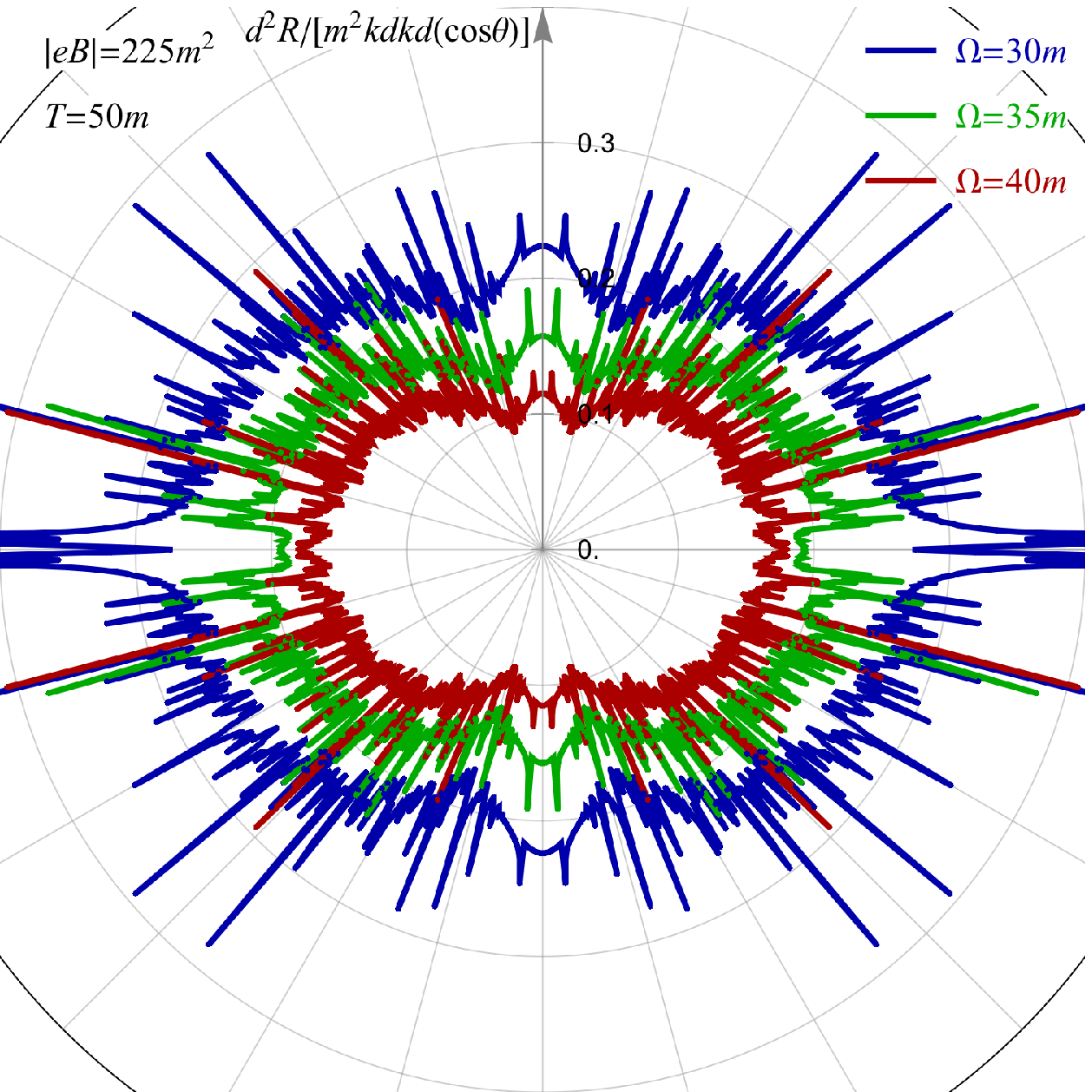}} 
\caption{The differential photon production rate as a function of the angle $\theta$ for $|qB|=225m^2$ and $T=50m$.}
\label{fig:resultsB225T50}
\end{figure}

\subsection{Photon emission at $|qB| =(5m)^2$}
\label{sec:Photon-emission-qB25}

Let us start by presenting the numerical results for the differential photon emission rate in the case the magnetic field $|qB| =(5m)^2$ and temperature $T=5m$. The corresponding rates as functions of the angular coordinate $\theta$ are shown in Fig.~\ref{fig:resultsB25T5} for several representative sets of photon energies. The results for six smaller energies ($\Omega \lesssim \sqrt{|qB|}$) are shown in panel (a), while the results for six larger energies ($\Omega \gtrsim \sqrt{|qB|}$) are shown in panel (b). By comparing the results, we see that the differential rate has a qualitatively different overall behavior in the two regimes. While at small photon energies the emission tends to be strongest along the magnetic field direction ($\theta\approx 0$), at large energies the emission tends to be strongest in the directions perpendicular to the magnetic field ($\theta\approx \pi/2$). In other words, the photon emission has a nonzero ellipticity \cite{Wang:2020dsr}, with the overall profile changing from a prolate shape at $\Omega \lesssim \sqrt{|qB|}$ to an oblate shape at $\Omega \gtrsim \sqrt{|qB|}$. The corresponding qualitative change of the emission profile is visualized by panels (c), (d), and (e) in Fig.~\ref{fig:resultsB25T5}, where the polar plots of the differential rates are presented. As one can see from panel (a), the overall photon emission has a well-pronounced prolate shape at small energies. The profile gradually changes from prolate to oblate at intermediate energies, see panel (b). Finally, as seen from panel (c), it becomes oblate at high photon energies. 

As is evident from Fig.~\ref{fig:resultsB25T5}, the photon emission rate is not a smooth function of the angular coordinate $\theta$. It has numerous spikes that come from the Landau-level quantization in the spectrum of charged fermions in a magnetic field. In essence, the spikes result from a near-threshold photon production when $k_\perp=\Omega\sin\theta\approx k_{\pm}$. As one can see from Eq.~(\ref{photon-rate-sum}), the rate near the Landau-level thresholds is described by a singular inverse-square-root dependence on the energy (at fixed $\theta$) or the angular coordinate $\theta$ (at fixed energy). Of course, the actual singularity is a consequence of the clean limit approximation that neglects interaction effects of charged particles. In a more refined approximation, as we will demonstrate in Sec.~\ref{interaction-effects}, interactions lead to a nonzero quasiparticle width that smoothes out the dependence of the rate on the energy and the angular coordinate $\theta$. For simplicity, however, we will ignore the interaction effects since they are not expected to change the overall qualitative behavior of the emission rate. Moreover, since the inverse-square-root singularities are integrable, they have little effect on the net integrated rate (see also Sec.~\ref{interaction-effects}).  

As is clear from Eq.~(\ref{photon-rate-sum}), there are two types of thresholds in the photon production. One of them is associated with the splitting  (i.e., $e^{-}\rightarrow e^{-}+\gamma$ and $e^{+}\rightarrow e^{+}+\gamma$) processes, while the other with the  annihilation (i.e., $e^{-}+e^{+}\rightarrow \gamma$) processes. They are characterized by a qualitatively different dependence of the rate on the angular coordinate $\theta$ in Fig.~\ref{fig:resultsB25T5}. For the particle splitting processes, the rate grows gradually as  $\theta$ approaches the threshold value from below and then drops suddenly at the threshold.  For the annihilation processes, on the other hand, the rate first jumps to a large value (infinite value in the clean limit) at the threshold and then decreases gradually with $\theta$. Both types could be easily identified in the differential rates shown in Fig.~\ref{fig:resultsB25T5}. 

For example, one can readily identify the first annihilation peak at $\sin\theta=2m/\Omega$ (provided $\Omega\geq 2m$), which is connected with the threshold for transitions between the lowest Landau-level states with positive and negative energies. For example, the rates for the photon energies $\Omega=3m$, $\Omega=4m$, $\Omega=5m$, and $\Omega=6m$, presented in Fig.~\ref{fig:resultsB25T5}(a), have the corresponding peaks at $\theta\approx 0.232 \pi$, $\theta= \pi/6$, $\theta\approx 0.131 \pi$, and $\theta\approx 0.108 \pi$, respectively. 
 
Similarly, one can identify some thresholds for the splitting processes. The thresholds for the transitions between the $n=1$ and $n=0$ Landau levels appear when $\sin\theta=(\sqrt{m^2+2|qB|}-m)/\Omega$ (provided $\Omega\gtrsim 6m$). For example, the rates for the photon energies $\Omega=10m$, $\Omega=15m$, $\Omega=20m$, $\Omega=25m$, and $\Omega=30m$, presented in Fig.~\ref{fig:resultsB25T5}(b), have the corresponding peaks at $\theta\approx 0.210 \pi$, $\theta= 0.134\pi$, $\theta\approx 0.099 \pi$, $\theta\approx 0.079 \pi$, and $\theta\approx 0.066 \pi$, respectively. In the rates at some smaller photon energies, shown in Fig.~\ref{fig:resultsB25T5}(a), one can also identify the thresholds for the transitions between the $n=2$ and $n=1$ Landau levels which occur when $\sin\theta=(\sqrt{m^2+4|qB|}-\sqrt{m^2+2|qB|})/\Omega$ (provided $\Omega\gtrsim 3m$). For $\Omega=3m$, $\Omega=4m$, $\Omega=5m$, and $\Omega=6m$, the corresponding peaks appear at $\theta\approx 0.421\pi$, $\theta=0.259\pi$, $\theta\approx 0.198 \pi$, and $\theta\approx 0.161 \pi$, respectively.

By comparing the results at different photon energies $\Omega$ presented in Fig.~\ref{fig:resultsB25T5}, we see that the rates tend to decrease overall with increasing $\Omega$. This qualitative conclusion is not surprising and will be reconfirmed more rigorously later by analyzing the dependence of the integrated rates on the photon energy. It is explained largely by the strong energy dependence of the Fermi-Dirac distribution functions that appear in the definition of the polarization tensor, see Eq.~(\ref{Im-Pi-mumu-three}). From a physics viewpoint, the emission of high-energy photons is possible only when there exist occupied Landau-level states with sufficiently large positive energies or empty Landau-level states with sufficiently large negative energies. In a thermal plasma, however, the availability of such states rapidly decreases with increasing the energy. 

For the most part, the same qualitative behavior of the differential photon emission rate is obtained at higher temperatures, $T=15m$ and $T=50m$. The corresponding numerical results are presented in Figs.~\ref{fig:resultsB25T15} and \ref{fig:resultsB25T50}, respectively. Naturally, the photon rate grows as a whole with increasing the temperature of plasma. Also, as one might expect, it tends to become a smoother function of the angular coordinate $\theta$ at higher temperatures. Most importantly, we find that, with increasing the temperature, the ellipticity of photon emission remains qualitatively the same. In particular, the emission profile starts from a prolate shape at low energies ($\Omega\lesssim \sqrt{|qB|}$) and gradually evolves to an oblate shape at high energies ($\Omega\gtrsim \sqrt{|qB|}$). As one might check, even the degree of the ellipticity at high energies remains about the same (i.e., $\sim 20\%$) .

\subsection{Photon emission at $|qB| =(15m)^2$}
\label{sec:Photon-emission-qB225}

The results for the stronger magnetic field $qB=(15m)^2$ are shown in Figs.~\ref{fig:resultsB225T5}, \ref{fig:resultsB225T15} and \ref{fig:resultsB225T50} for the same three default choices of temperature, $T=5m$, $T=15m$, and $T=50m$, respectively. Qualitatively, the functional dependence of the differential rate has many features similar to those at the smaller magnetic field. However, some differences can be clearly seen as well. For example, the quantization effects at the lowest temperature $T=5m$ are much more pronounced. While the overall rates are small in the low-temperature regime, the partial contribution of the annihilation processes is substantial. For $\Omega\leq 6m$, as one can see from Fig.~\ref{fig:resultsB225T5}(a), there is only one well pronounced singularity in the angular dependence and it is associated with the first singularity at $\sin\theta=2m/\Omega$, provided the photon energy exceeds the threshold, i.e., $\Omega>2m$. Below the threshold energy ($\Omega < 2m$), on the other hand, the rate is strongly suppressed. The latter is explained by the fact that the charge particles of a low-temperature plasma ($T\ll \sqrt{|qB|}$) reside primarily in the lowest Landau level. Since the occupation numbers of the higher levels are negligible, the phase space for the quark and antiquark splitting processes is highly restricted and, in turn, the corresponding contributions to the rate are negligible. 

From the polar representation of the differential rates, shown in panels (c) through (e) of Fig.~\ref{fig:resultsB225T5}, the ellipticity of the emission profile may not be obvious at the lowest temperature, $T=5m$. The main complication comes from a very nonsmooth dependence of the rate on the angular coordinate $\theta$. By utilizing a Fourier transform, however, one can still verify that the profile evolves from a largely prolate shape at small energies ($\Omega\lesssim \sqrt{|qB|}$) to an oblate shape at large energies ($\Omega\gtrsim \sqrt{|qB|}$). 

By comparing the results in Figs.~\ref{fig:resultsB225T5}, \ref{fig:resultsB225T15} and \ref{fig:resultsB225T50}, we see that the differential rates not only grow with increasing the temperature, but also their angular dependence gets smoother. One can also verify that the relative contribution of the splitting (annihilation) processes gets larger (smaller) with at temperatures. Because of the smoother rates at $T=15m$ and  $T=50m$, the ellipticity of the emission profiles can be easily identified from the polar representation of the differential rates, see (c) through (e) of Figs.~\ref{fig:resultsB225T15} and \ref{fig:resultsB225T50}. They are qualitatively the same as in the case of the weaker $qB=(5m)^2$.

\subsection{Integrated photon production rate}
\label{sec:Photon-Integrated-rate}

\begin{figure}[t]
\centering
\subfigure[]{\includegraphics[width=0.48\textwidth]{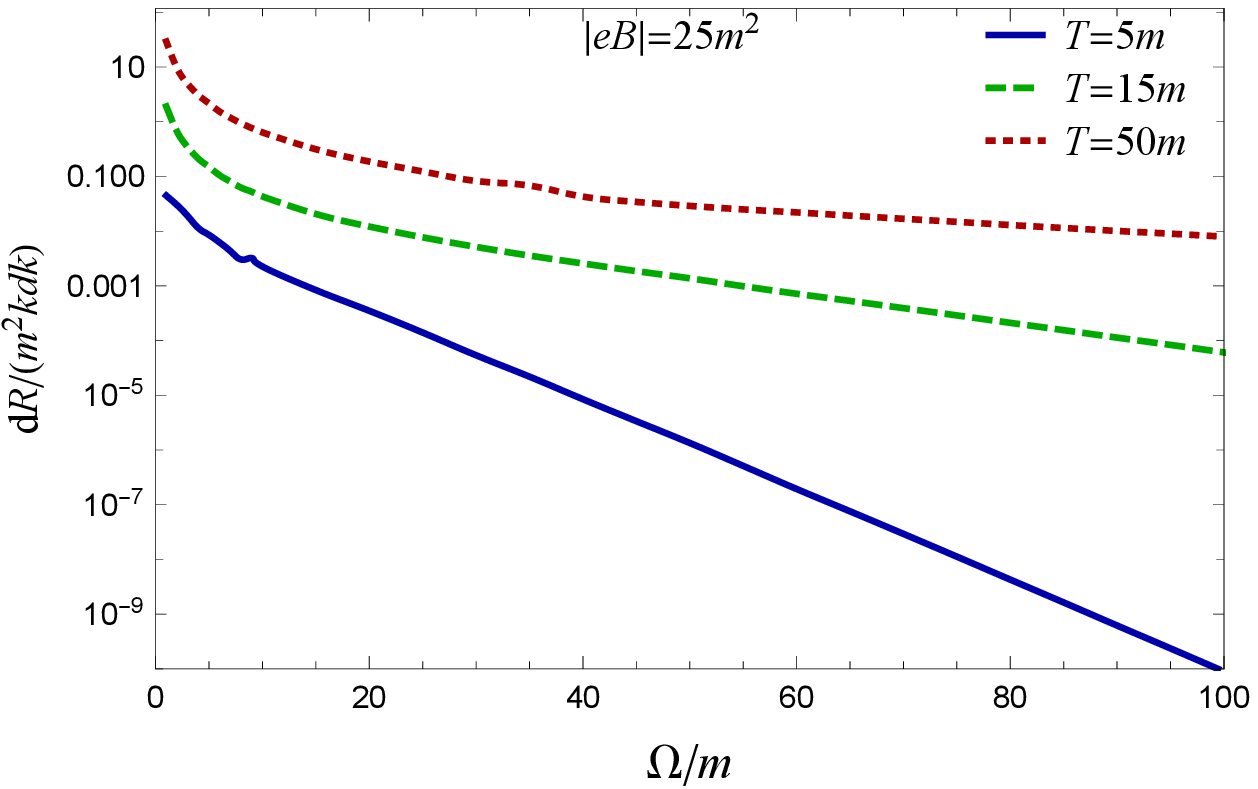}}
  \hspace{0.01\textwidth}
\subfigure[]{\includegraphics[width=0.48\textwidth]{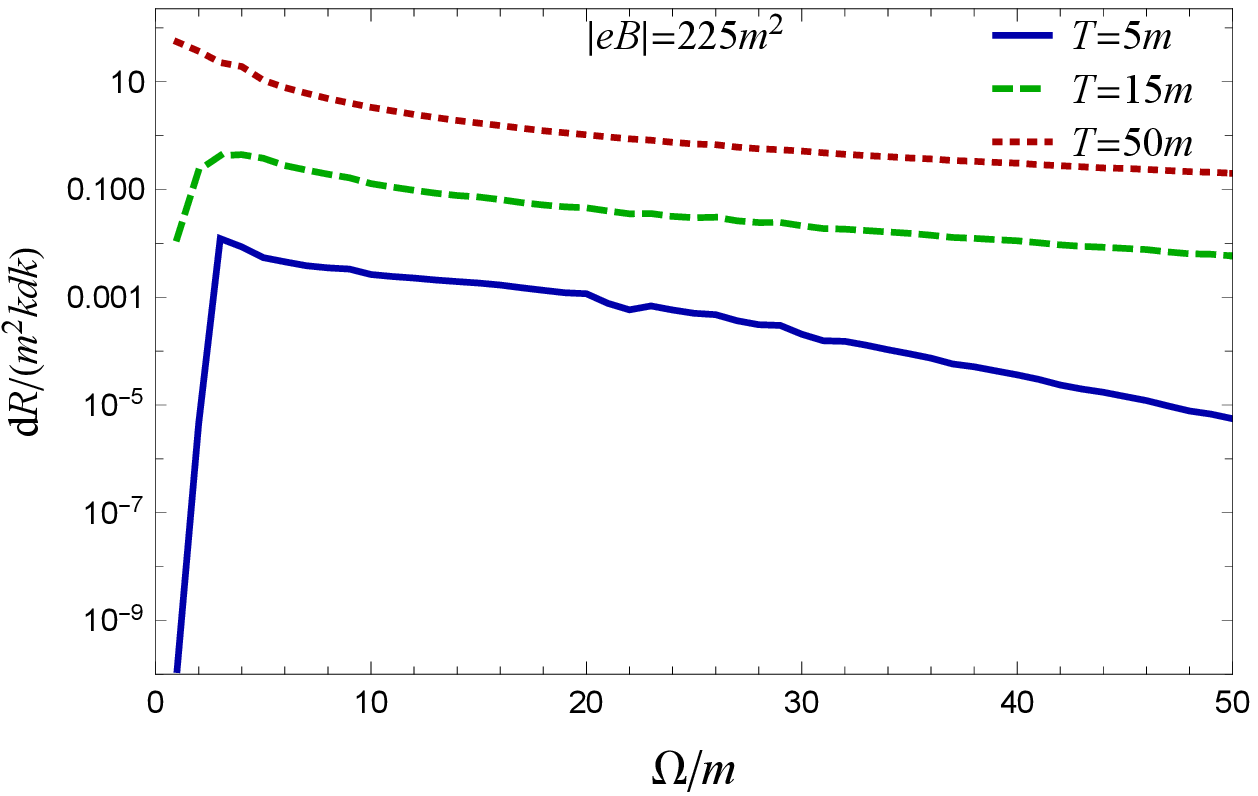}}
\caption{Panel (a): The energy dependence of the integrated photon production for $|qB|=25m^2$
and three different temperatures $T=5m$ (blue line) and $T=15m$ (green line),  and $T=50m$ (red line).
Panel (b): Same as panel (a) but for $|qB|=225m^2$.}
\label{fig:ProdRate-Integrated}
\end{figure}

For both values of the magnetic field, $qB=(5m)^2$ and $qB=(15m)^2$, we see that the differential rates tend to grow overall with  increasing the temperature. In order to see this clearly, it is instructive to calculate the total integrated photon production rate, i.e.,
\begin{equation}
\frac{dR}{k \, d k }=\int_{0}^{\pi}\frac{d^2 R}{k \, d k\, d(\cos \theta)} \sin\theta d\theta =-\frac{1}{(2\pi)^2}\int_{0}^{\pi}\frac{\mbox{Im}\left[\Pi^{\mu}_{R,\mu}(k,\mathbf{k})\right]}{\exp\left(\frac{k}{T}\right)-1}\sin\theta d\theta.
\label{integrated-rate}
\end{equation}
The corresponding numerical results for the integrated rates as a function of the photon energy are presented in Fig.~\ref{fig:ProdRate-Integrated} for each of the three fixed values of temperature, $T=5m$, $T=15m$, and $T=50m$. The two separate panels contains the results for $qB=(5m)^2$ and $qB=(15m)^2$, respectively. Overall, as we see, the photon rates grow with the temperature. This is not surprising since hotter plasmas should shine brighter.  

The energy dependence in both panels of Fig.~\ref{fig:ProdRate-Integrated} shows that the integrated rates fall rapidly at large values of $\Omega$. This is expected, of course, since the emission of high-energy photons is possible only when there are some occupied (empty) charged particle (antiparticle) states with sufficiently high positive (negative) energies. In a thermal plasma at fixed temperature $T$, however, the number densities of such states are controlled by the Fermi-Dirac distributions, which drop quickly with increasing particle energies. 

The dependence of the integrated rates in the region of small photon energies is more subtle. Our analysis reveals that the rate has a local maximum at some value of $\Omega$, determined by the magnetic field strength and the temperature. When $\Omega$ decreases further, the rate gets strongly suppressed. This is seen in the case of the stronger magnetic field $|qB|=225m^2$ in panel (b) of Fig.~\ref{fig:ProdRate-Integrated} for the rates at $T=5 m$ and $T=15m$. As we will explain in Sec.~\ref{sec:small-Omega} below, the same should happen also in the case of the weaker magnetic field, $|qB|=25m^2$, but at smaller energies. The corresponding suppression of the rate at small photon energies is caused by the Landau-level quantization.

\subsection{Landau-level quantization at small $\Omega$}
\label{sec:small-Omega}

As one might expect, the Landau-level quantization has a strong effect on the photon emission at small energies. This is clear already from the illustration of the relevant quantum transitions in Figs.~\ref{fig:transitions} and \ref{fig:transitions-annih}. The corresponding   processes are possible only when the photon energy is greater than or equal to the separation between the neighboring Landau levels. This is a necessary condition imposed by the energy conservation.  

From general considerations, the Landau quantization has the strongest effect when $\Omega\ll \sqrt{|qB|}$. In this case, transitions between particle states are highly restricted because the typical separation between the low-lying Landau levels is of the order of $\sqrt{|qB|}$, while the photon energy is much smaller. Strictly speaking, this does not apply to the annihilation processes that involve  only the (positive- and negative-energy) states in the lowest Landau level. They still contribute when $\Omega> 2m$, see Appendix~\ref{LLLAL}. Nevertheless, the partial contribution of such annihilation processes to the rate decreases rapidly when $\Omega$ goes down and approaches the threshold. Furthermore, for $\Omega<2m$, it turns off completely. Thus, to understand the universal suppression mechanism for the photon emission rate at small $\Omega$, we will concentrate on the splitting processes and ignore the annihilation channel. 

The kinematic restrictions for the splitting processes at small $\Omega$ can be understood from the visualization of the quantum transitions in Fig.~\ref{fig:transitions}(a). In particular, the transitions between the neighboring Landau levels with small indices $n$ are possible only at large enough longitudinal momenta, where the energy levels are sufficiently close to each other. The quantitative estimate for the momenta can be obtained from the solution of the energy conservation condition in Eq.~(\ref{pz-solution}). In particular, one finds that $| p_z | \sim  |qB|/\left[\Omega (1+|\cos\theta |) \right]$. Since the particle states with such large $| p_z |$ also have large energies, i.e., $E_{p_z} \gtrsim |qB|/\left[\Omega (1+|\cos\theta |) \right]$, their occupation numbers and, in turn, their contributions to the photon emission rate are exponentially suppressed when $\Omega\to 0$. One should add that there are also numerous transitions involving states with large Landau indices (with $n$ up to about $|qB|/\Omega^2$) that contribute to the rate too. However, all relevant transitions occur between states with comparable or larger energies, implying that their contributions are subject to the same exponential suppression by the Fermi-Dirac distributions.

As the value of $\Omega$ grows and becomes comparable to the Landau energy scale, $\Omega\sim \sqrt{|qB|}$, the quantization effects relax gradually. This is seen from the visualization of quantum transitions in panels (b) and (c) of Fig.~\ref{fig:transitions}(a). When the photon energy $\Omega$ grows, the transitions start to involve particle states with lower energies. The larger occupation numbers of such states lead to a higher photon production rate. With further increasing of $\Omega$, however, the role of quantization gradually diminishes. Eventually, when $\Omega\gg \sqrt{|qB|}$, the rates start to decrease again with increasing $\Omega$. Therefore, the photon production rate as a function of the photon energy has the following universal behavior. It starts with a vanishingly small value when $\Omega\simeq 0$, reaches a maximum at an intermediate energy, and then decreases at large $\Omega$. 

As argued in Ref.~\cite{Wang:2020dsr}, the quantization of Landau levels also explains the prolate shape of the photon emission profile in the region of small $\Omega$. The latter follows from the angular dependence of the particle momenta of the lowest energy states participating in the splitting processes, i.e., $|p_z| \sim |qB|/\left[\Omega (1+|\cos\theta |) \right]$. The smallest values of $|p_z|$ (and, thus, the particle energies) are achieved when $\theta = 0$ and $\theta =\pi$ (along the line of the magnetic field). These determine the directions of the strongest photon emission. In contrast, the largest values of $|p_z|$ correspond to $\theta =\pi/2$, implying that the photons emitted perpendicular to the magnetic field have the lowest rate.

\subsection{Interaction effects in the vicinity of Landau-level thresholds}
\label{interaction-effects}

In this study we used the clean limit approximation that ignores all particle interactions except for the interaction with a strong background magnetic field. One of the consequences of such an approximation was the appearance of numerous singular spikes in the differential photon rate as a function of the angular coordinate $\theta$ or energy $\Omega$. The corresponding inverse-square-root singularities were connected with Landau-level thresholds in the three processes responsible for the photon production (i.e., $e^{-}\rightarrow e^{-}+\gamma$, $e^{+}\rightarrow e^{+}+\gamma$, and $e^{-}+e^{+}\rightarrow \gamma$). As we explain below, the singularities will be removed and the rate will have smooth dependence when interaction effects are taken into account. Qualitatively, this can be understood as follows. 

\begin{figure}[th]
\centering
\includegraphics[width=0.45\textwidth]{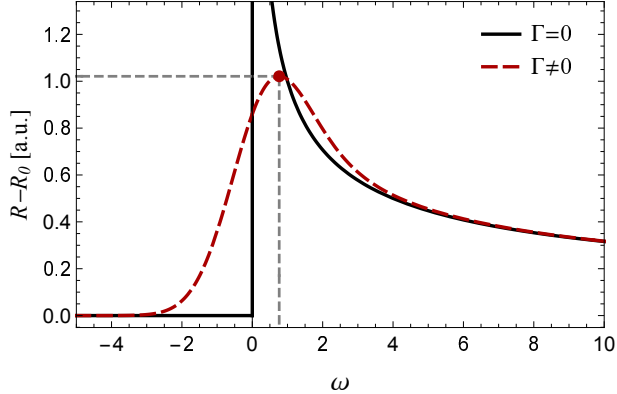} 
\caption{The effect of quasiparticle width $\Gamma$ on the near-threshold behavior of the rate as a function of the photon energy. A singular dependence $R-R_0\propto \theta(\omega)/\sqrt{\omega}$, where $\omega=(\Omega-\Omega_{\rm thr})/\Gamma$ and $R_0$ is the background rate, is replaced by a smooth function defined in Eq.~(\ref{smooth-function}).}
\label{fig:threshold}
\end{figure}

In an interacting plasma, charged particles should have a nonzero quasiparticle width $\Gamma$, which is determined by the imaginary part of the self-energy. When the quasiparticle width is small, its effect can be captured in the calculation of the photon emission rate by replacing the $\delta$ function in the particle spectral weight with a Gaussian profile of width $\Gamma$. The inclusion of $\Gamma$ will modify the on-shell condition for interacting particles and, in turn, relax slightly the energy conservation relation in the processes responsible for the emission. From a technical point of view, one can replace the $\delta$ function, that enforces the energy conservation condition on the right-hand side of Eq.~(\ref{Im-Pol-fun}), with a Gaussian profile, i.e., 
\begin{eqnarray}
\delta\left(E_{n,p_z}-\lambda E_{n^{\prime},p_z-k_z}+\eta \Omega\right) &\to & 
\frac{1}{\sqrt{2\pi} \Gamma}
\int_{-\infty}^{\infty} d\varepsilon e^{-\frac{\varepsilon^2}{2\Gamma^2}} 
\delta\left(E_{n,p_z}-\lambda E_{n^{\prime},p_z-k_z}+\eta \Omega+\varepsilon\right)
\nonumber\\
&=& \frac{1}{\sqrt{2\pi} \Gamma}
\int_{-\infty}^{\infty} d\Omega^\prime e^{-\frac{(\Omega^\prime-\Omega)^2}{2\Gamma^2}}
\delta\left(E_{n,p_z}-\lambda E_{n^{\prime},p_z-k_z}+\eta \Omega^\prime\right),
\end{eqnarray}
where, in the second line, we changed the integration variable from $\varepsilon$ to $\Omega^\prime =\Omega+\eta \varepsilon$.
Such an introduction of the quasiparticle width effects leads to a simple modification of the expression for the imaginary part of the polarization tensor:
\begin{eqnarray}
\mbox{Im} \left[\Pi^{\mu}_{R,\mu} (\Omega,\mathbf{k}) \right]\to  \frac{1}{\sqrt{2\pi} \Gamma}
\int_{-\infty}^{\infty} d\Omega^\prime 
e^{-\frac{(\Omega^\prime-\Omega)^2}{2\Gamma^2}}
\mbox{Im} \left[\Pi^{\mu}_{R,\mu} (\Omega^\prime,\mathbf{k}) \right].
\end{eqnarray}
Note that the result is a convolution of the result obtained in the clean limit and a Gaussian distribution. Let us verify now that the modification has the predicted smearing effect on the inverse-square-root singularities in the near-threshold production of photons. It suffices to consider an isolated singularity at $\Omega=\Omega_{\rm thr}$, e.g.,
\begin{eqnarray}
\frac{\theta(\Omega-\Omega_{\rm thr})}{\sqrt{\Omega-\Omega_{\rm thr}}}f(\Omega)
&\to & \frac{1}{\sqrt{2\pi} \Gamma}
\int_{-\infty}^{\infty} d\Omega^\prime 
e^{-\frac{(\Omega^\prime-\Omega)^2}{2\Gamma^2}}
\frac{\theta(\Omega^\prime-\Omega_{\rm thr})}{\sqrt{\Omega^\prime-\Omega_{\rm thr}}}f(\Omega^\prime)
\nonumber\\
&\approx &\sqrt{\frac{\pi |\omega|}{8\Gamma}} e^{-\frac{\omega^2}{4}} \left\{
\theta(\omega) \left[I_{-1/4}\left(\frac{\omega^2}{4}\right)+I_{1/4}\left(\frac{\omega^2}{4}\right)\right]
+\frac{\sqrt{2}}{\pi}\theta(-\omega)K_{1/4}\left(\frac{\omega^2}{4}\right)
\right\}f(\Omega_{\rm thr}) ,
\label{smooth-function}
\end{eqnarray}
where $\omega= (\Omega-\Omega_{\rm thr})/\Gamma$ is a dimensionless deviation from the threshold energy and $f(\Omega)$ is a smooth function. The effect of smearing is visualized in Fig.~\ref{fig:threshold}, where we show the partial contribution of an isolated threshold singularity over the background value $R_0$. Note that the local peak of the near-threshold rate is shifted from $\Omega_{\rm thr}$ to $\Omega^{*}_{\rm thr} \approx \Omega_{\rm thr}+0.765\Gamma$. The maximum value of the function is approximately $1.021f(\Omega_{\rm thr}) /\sqrt{\Gamma}$.

As we see, the quasiparticle width $\Gamma$ indeed smoothes out the threshold singularities in the photon rate. In the weakly interacting regime, when $\Gamma$ is small, the corresponding modifications occur only in small windows of energies near the thresholds, i.e., $|\Omega-\Omega_{\rm thr}|\lesssim\Gamma$. Thus, the overall features of the photon rate obtained above in the clean limit of a strongly magnetized plasma will remain approximately the same even after the interaction effects are accounted for. Moreover, one can also verify that the integrated rates will not change much either. Indeed, by integrating both singular and smoothed functions in Eq.~(\ref{smooth-function}), one can verify that the relative error is less than $0.84\%$ when the integration range is as small as $|\Omega-\Omega_{\rm thr}|\leq 4\Gamma$. With expanding the integration range further, the error rapidly decreases too.

\section{Magneto-optical conductivity of relativistic plasma}
\label{sec:Numerical-results}

In this section, we demonstrate another use of the general result for the imaginary part of the photon polarization function obtained in Sec.~\ref{sec:Polarization}. In particular, below we calculate the absorptive part of the magneto-optical conductivity of a hot relativistic plasma. The corresponding frequency-dependent transport coefficient is a valuable characteristics in a plasma under time dependent perturbations. The general theoretical predictions could be also applied to pseudorelativistic electron plasmas of Dirac and Weyl semimetals \cite{Ashby:2013aa,Long:2018aa,PhysRevB.102}, where it can be tested experimentally \cite{PRL115.176404}. 

In the linear response theory, the magneto-optical conductivity can be expressed in terms of the imaginary part of the photon polarization tensor as follows: 
\begin{equation}
\sigma^{ij}(\Omega) =\frac{\mbox{Im}\left[\Pi_{R}^{ij}(\Omega;\bm{0})\right]}{\Omega}.
\label{sigma-ij}
\end{equation}
Note that only the diagonal components of the conductivity tensor are nonzero in the case of the vanishing chemical potential, which is studied in this paper. Indeed, the off-diagonal (Hall) conductivity $\sigma_{H}(\Omega)$ is trivial because of the charge conjugation symmetry. 

Compared to the general result for the polarization tensor obtained in Sec.~\ref{sec:Polarization}, the conductivity tensor has 
a much simpler structure. This is the result of taking the limit $\mathbf{k}\to \bm{0}$. Indeed, by making use of the expression in Eq.~(\ref{ImPi-tensor-txt1}), supplemented by the definitions of individual component functions in Appendix~\ref{app:Tensor-structure}, we find that the magneto-optical conductivity takes the following form:
\begin{equation}
\sigma^{ij}(\Omega) =\delta^{i}_{3}\delta^{j}_{3}\sigma_{\parallel}(\Omega) +\left(\delta^{i}_{1}\delta^{j}_{1}+\delta^{i}_{2}\delta^{j}_{2}\right)\sigma_{\perp}(\Omega) +\varepsilon^{0ij3} \sigma_{H}(\Omega) ,
\label{sigma-ij}
\end{equation} 
where the longitudinal and transverse Hall conductivities are given by
\begin{eqnarray}
\sigma_{\parallel}(\Omega) &=& \frac{\mbox{Im}\left[ \Pi_{1}(\Omega;\bm{0}) \right]}{\Omega} \nonumber\\
&=& -  \frac{\alpha N_{f}}{\Omega \ell^2} \sum_{n,n^\prime=0}^{\infty} 
\sum_{\lambda,\eta=\pm 1} \Theta_{\lambda, \eta}^{n,n^{\prime}}(\Omega,0)
\frac{n_F(E_{n,p_z^{(+)}})-n_F(\lambda E_{n^{\prime},p_z^{(+)}}) }{\eta\lambda  \sqrt{ \Omega^2  \left( \Omega^2-k_{+}^2\right)} } k_{+}^2  \left( \delta_{n,n^{\prime}}+ \delta_{n-1,n^{\prime}-1}\right) ,
\label{sigma-parallel-0}
\\
\sigma_{\perp}(\Omega) &=&\frac{\mbox{Im}\left[\Pi_{4}(\Omega;\bm{0}) \right]}{\Omega} \nonumber\\
&=&-  \frac{\alpha N_{f}}{\Omega \ell^4} \sum_{n,n^\prime=0}^{\infty} 
\sum_{\lambda,\eta=\pm 1} \Theta_{\lambda, \eta}^{n,n^{\prime}}(\Omega,0)
\frac{n_F(E_{n,p_z^{(+)}})-n_F(\lambda E_{n^{\prime},p_z^{(+)}}) }{\eta\lambda  \sqrt{ \left( \Omega^2-k_{-}^2 \right) \left( \Omega^2-k_{+}^2\right)} }\left[ \Omega^2\ell^2-2(n+n^{\prime}) \right]
\left( \delta_{n,n^{\prime}-1}+\delta_{n-1,n^{\prime}} \right).
\label{sigma-perp-0}
\end{eqnarray}
Here we used explicit expressions for the component functions $\mbox{Im}\left[ \Pi_{i}(\Omega;\bm{0})\right] $, which are obtained from the definitions in Appendix~\ref{app:Tensor-structure} by taking the limit $\mathbf{k}\to \bm{0}$. In the derivation, we also took into account that the solutions of the energy conservation equation at $k_z=0$ are given by
\begin{eqnarray}
p_{z}^{(\pm)}&=& \pm \frac{1}{2\Omega}  \sqrt{ \left( \Omega^2- k_{-}^2 \right)
\left( \Omega^2- k_{+}^2 \right)}  ,
\label{pz-solution-k0}
\end{eqnarray}
and the corresponding quasiparticle energies are 
\begin{eqnarray}
  E_{n,p_z^{(\pm)}}  &=& - \frac{\eta }{2\Omega} \left[
\Omega^2+ 2(n-n^{\prime})|qB| \right], \label{E1-solution-k0}\\
  E_{n^{\prime},p_z^{(\pm)}}  &=& \frac{\lambda \eta }{2\Omega} \left[
\Omega^2- 2(n-n^{\prime})|qB| 
\right],
\label{E2-solution-k0}
\end{eqnarray}
which are the same for $p_{z}^{(-)}$ and $p_{z}^{(+)}$. 

The expressions for the longitudinal and transverse conductivities, defined by Eqs.~(\ref{sigma-parallel-0}) and (\ref{sigma-perp-0}), can be further simplified, i.e., 
\begin{eqnarray}
\sigma_{\parallel}(\Omega) &=& \frac{\alpha N_{f}}{ \ell^2} \sum_{n=0}^{\infty} \left(2-\delta_{n,0}\right)
\frac{4 M_{n}^2 \theta\left( \Omega^2  - 4 M_{n}^2\right)  }{  \Omega^2 \sqrt{\Omega^2-4 M_{n}^2} } 
\tanh\left(\frac{|\Omega|}{4T}\right) ,
\label{sigma-parallel}\\
\sigma_{\perp}(\Omega)  &=& \frac{2\alpha N_{f}\sinh\left(\frac{\Omega}{2T}\right)  }{\Omega \ell^4\left[\cosh\left(\frac{\Omega}{2T}\right) +\cosh\left(\frac{|qB|}{T\Omega}\right) \right]} \sum_{n=1}^{\infty} 
\frac{ \left[  2 (2n-1) -\Omega^2\ell^2\right] 
\theta\left[ (M_{n}-M_{n-1})^2 - \Omega^2 \right] }{
\sqrt{ \left[ (M_{n}-M_{n-1})^2 - \Omega^2 \right] \left[ (M_{n}+M_{n-1})^2 - \Omega^2 \right]} } \nonumber\\
&+& \frac{2\alpha N_{f}\sinh\left(\frac{|qB|}{T\Omega}\right)}{\Omega \ell^4 \left[\cosh\left(\frac{\Omega}{2T}\right) +\cosh\left(\frac{|qB|}{T\Omega}\right) \right]} \sum_{n=1}^{\infty} 
\frac{\left[ \Omega^2\ell^2 -  2(2n-1) \right]\theta\left[  \Omega^2-(M_{n}+M_{n-1})^2  \right] }{ \sqrt{\left[  \Omega^2-(M_{n}-M_{n-1})^2  \right] \left[  \Omega^2-(M_{n}+M_{n-1})^2  \right] } }  ,
\label{sigma-perp}
\end{eqnarray}
where we used the shorthand notation $M_{n}=E_{n,0}=\sqrt{2n|qB|+m^2}$. 

\begin{figure}[th]
\centering
\subfigure[]{\includegraphics[width=0.45\textwidth]{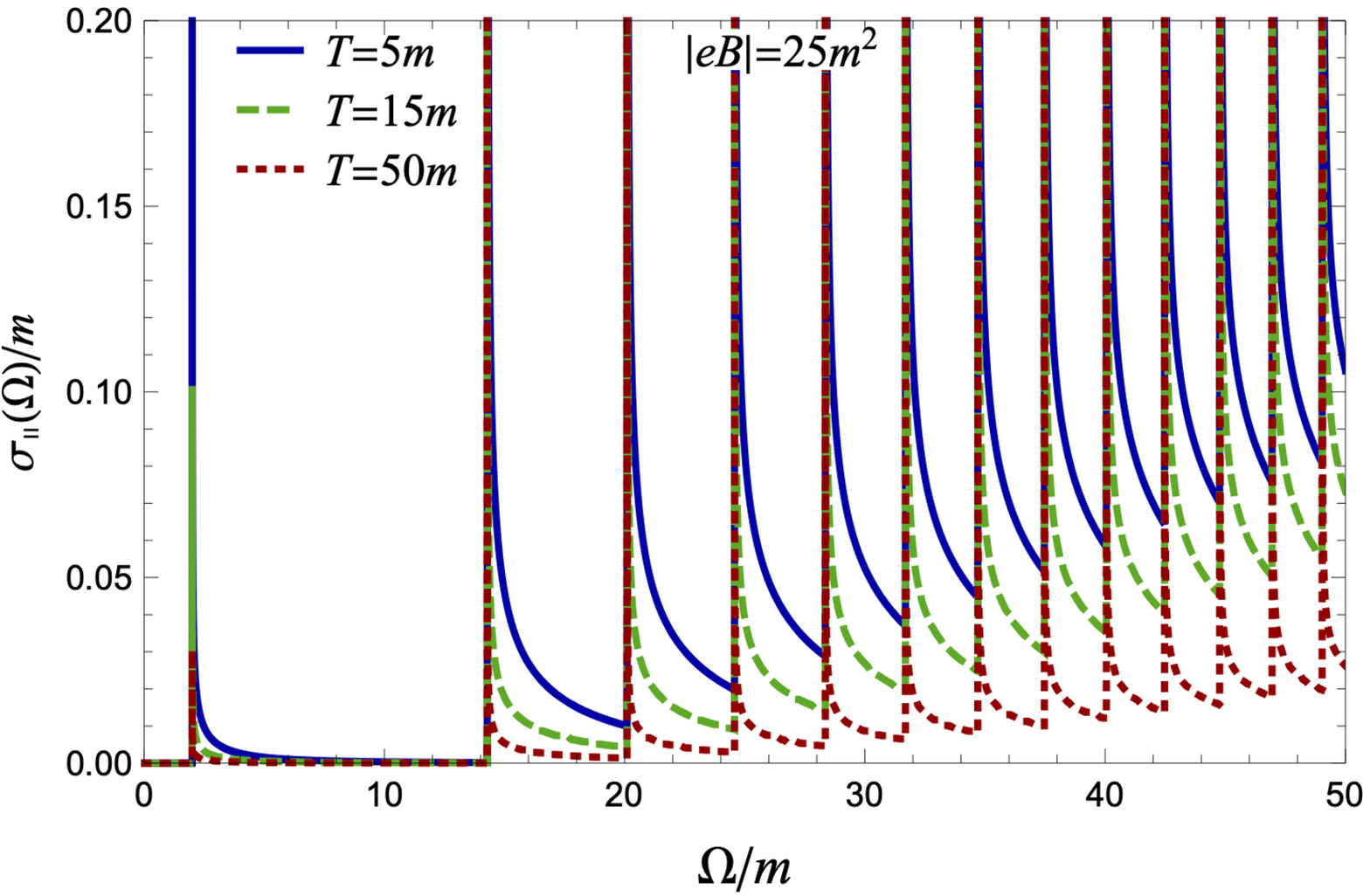}}
\hspace{0.05\textwidth}
\subfigure[]{\includegraphics[width=0.45\textwidth]{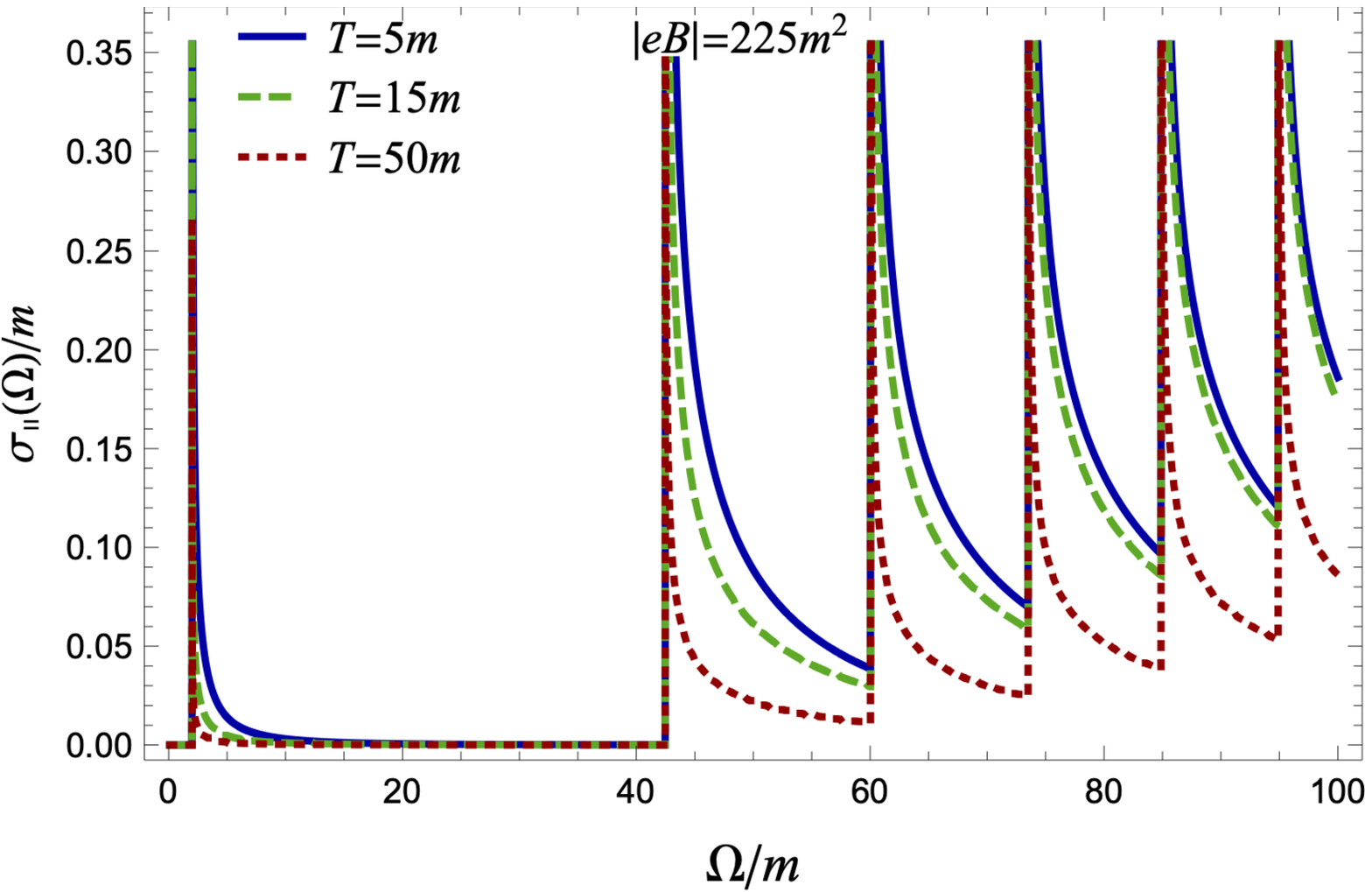}} 
\caption{Longitudinal magneto-optical conductivity as a function of frequency $\Omega$
for $|qB|=25m^2$ (panel a) and $|qB|=225m^2$ (panel b).}
\label{fig:CondParl}
\end{figure}

Note that the longitudinal conductivity  $\sigma_{\parallel}(\Omega)$ in Eq.~(\ref{sigma-parallel}) comes exclusively from the annihilation (or ``interband") transitions. Thus, while the conductivity is nonzero for  $\Omega\geq 2m$, it vanishes below the lowest frequency annihilation threshold, i.e., $\Omega<2m$. (Needless to say that the threshold goes away in the limit of massless fermions.) Numerically, the longitudinal conductivity as a function of frequency is relatively easy to calculate even in the limit of a weak magnetic field. This is because the sum in Eq.~(\ref{sigma-parallel}) terminates at a finite Landau-level index $n_{\rm max} =\left[(\Omega^2-4m^2)/(4|qB|)\right]$. The corresponding numerical results for the two choices of the magnetic field and three different values of temperature are shown in Fig.~\ref{fig:CondParl}. By comparing the energy dependence in the two panels, we see that the longitudinal conductivities for the weaker and stronger magnetic fields differ primarily by the separation scale between the neighboring peaks. 

It is interesting to note that the longitudinal part of conductivity, shown in Fig.~\ref{fig:CondParl}, tends to decrease with increasing the temperature of plasma. Such decrease should not be too surprising however. Mathematically, it comes from $\tanh[\Omega/(4T)]$ in Eq.~(\ref{sigma-parallel}). From a physics viewpoint, it is explained by the gradual depletion of the occupied low-energy states when the temperature grows. 

Unlike the longitudinal conductivity, the transverse conductivity $\sigma_{\perp}(\Omega)$ contains two different types of contributions, represented by the two sums in Eq.~(\ref{sigma-perp}). The first sum represents the contributions due to particle-particle and antiparticle-antiparticle (or ``intraband") transitions. As one can verify, they are nonzero only in the region of sufficiently small frequencies, i.e., $\Omega < \sqrt{2|qB|+m^2}-m$. Note that the quoted upper limit is determined by the largest vertical separation between neighboring Landau levels, i.e., the energy difference between the $0$th and $1$th levels at $p_z=0$. The second sum in Eq.~(\ref{sigma-perp}) captures the contributions due to annihilation (``interband") transitions. This one is nonvanishing only at sufficiently large frequencies, i.e., $\Omega > \sqrt{2|qB|+m^2}+m$. As one can see, the frequency windows for the two types of processes do not overlap and are separated by a gap of width $2m$. Interestingly, for both types of contributions, the sums over Landau levels terminate at $n_{\rm max} =\left[((2|qB|+\Omega)^2-4m^2\Omega^2)/(8|qB|\Omega^2)\right]$.  

\begin{figure}[th]
\centering
\subfigure[]{\includegraphics[width=0.45\textwidth]{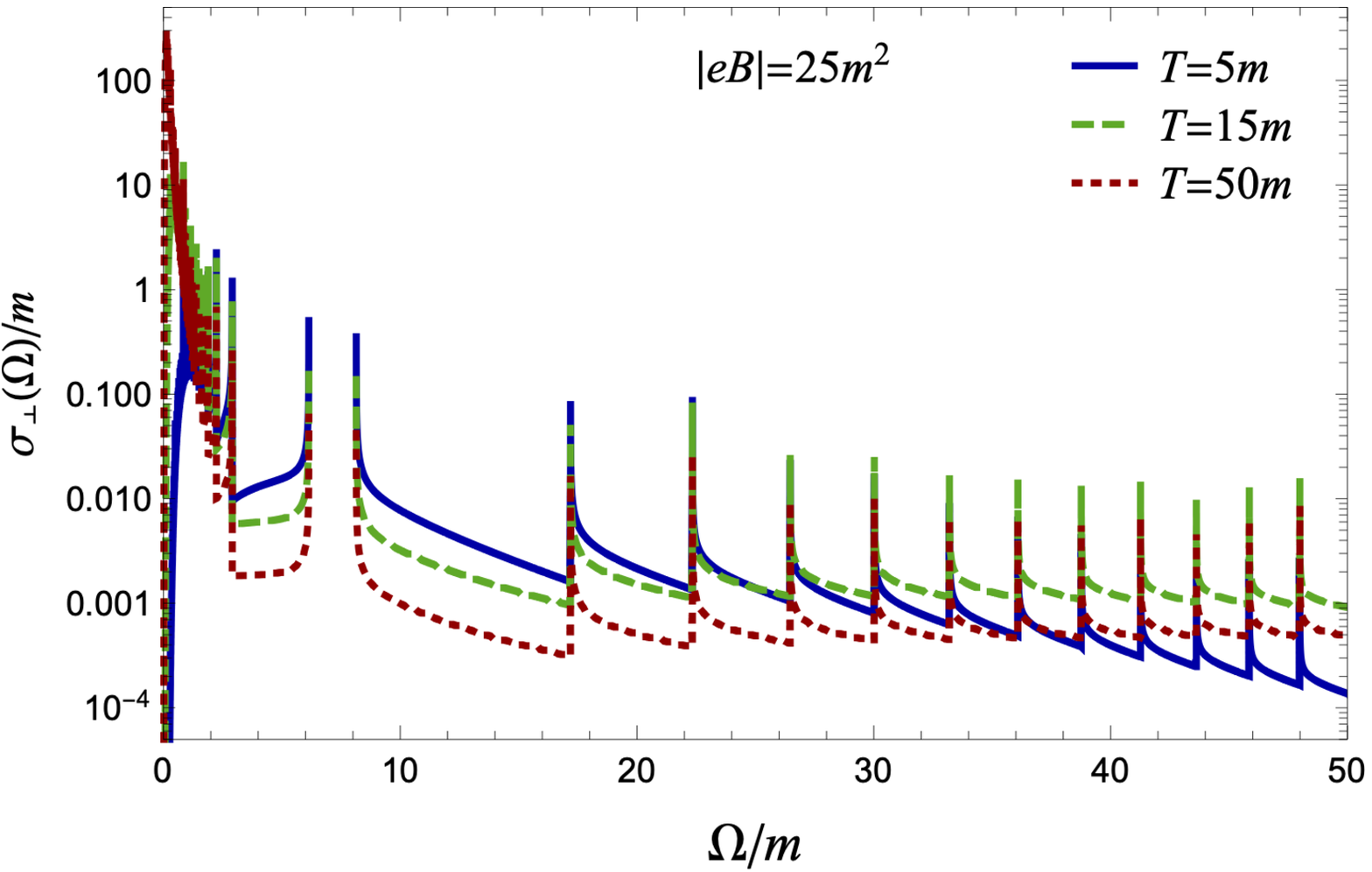}}
\hspace{0.05\textwidth}
\subfigure[]{\includegraphics[width=0.45\textwidth]{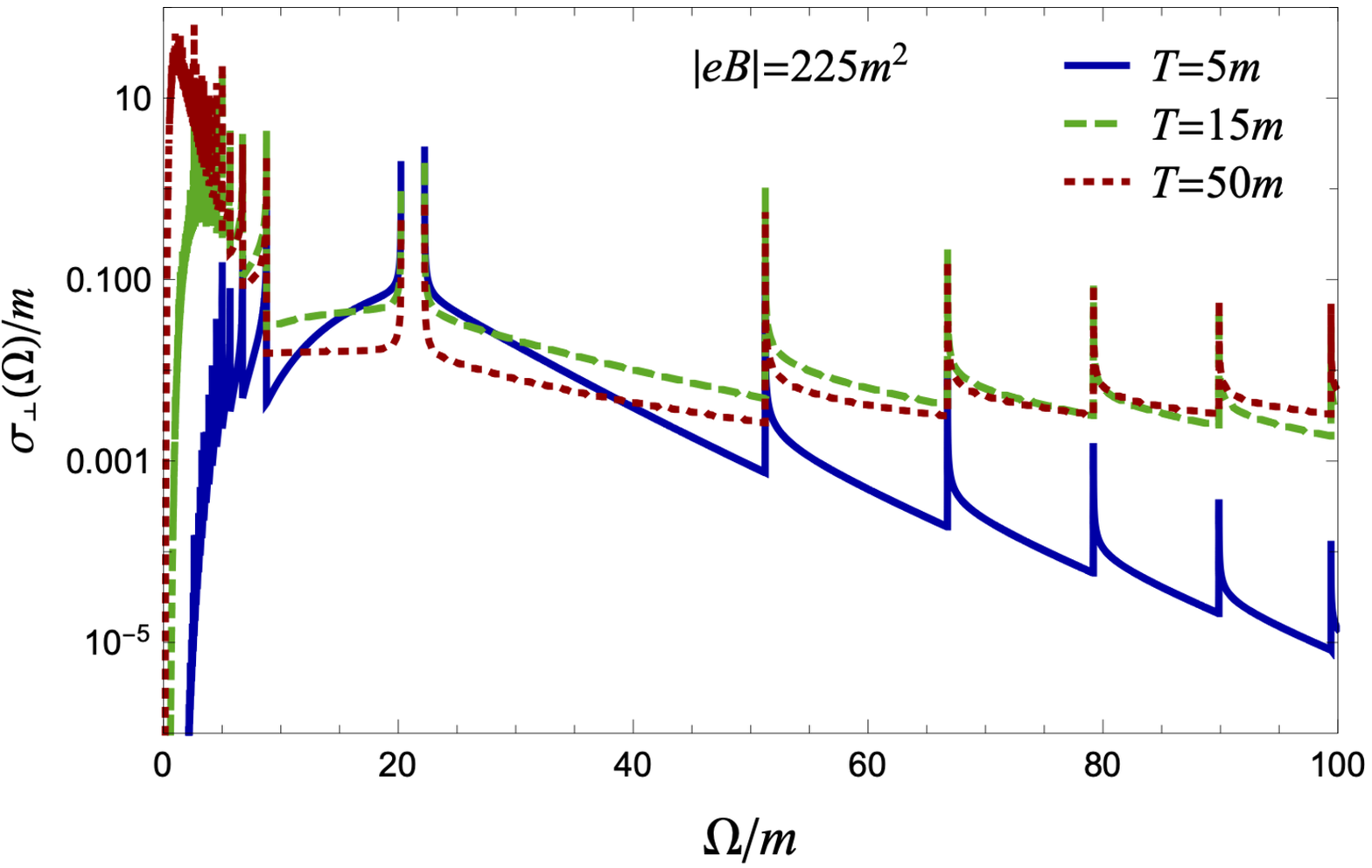}} \\
\subfigure[]{\includegraphics[width=0.45\textwidth]{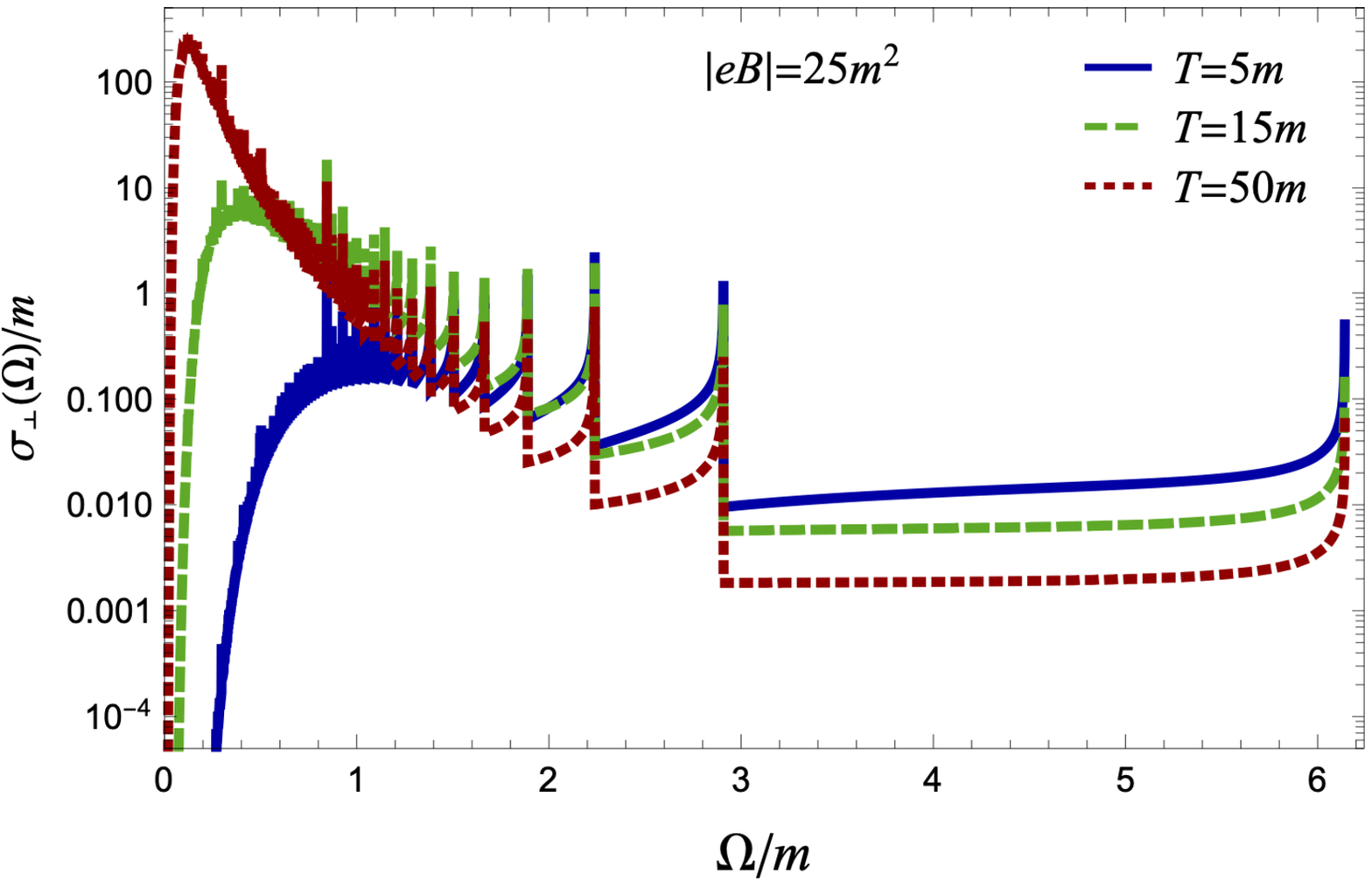}} 
\hspace{0.05\textwidth}
\subfigure[]{\includegraphics[width=0.45\textwidth]{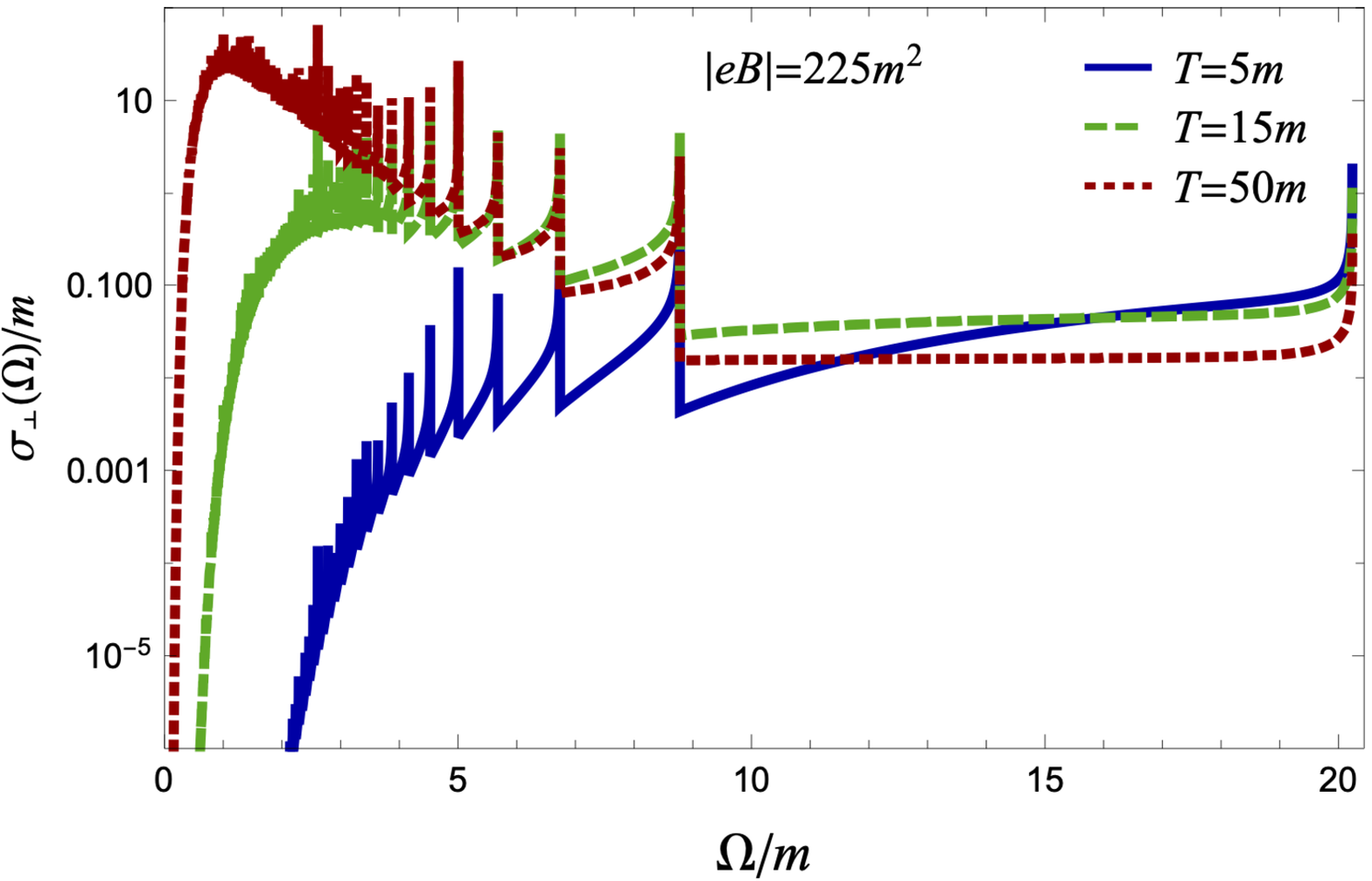}}
\caption{Transverse magneto-optical conductivity as a function of frequency $\Omega$ for $|qB|=25m^2$ (panel a) and $|qB|=225m^2$ (panel b). Panels (c) and (d) show the same results at small frequencies, where only the ``intraband" transitions contribute.}
\label{fig:CondPerp}
\end{figure}

The numerical results for the transverse conductivity $\sigma_{\perp}(\Omega)$ are presented in Fig.~\ref{fig:CondPerp}. Panels (a) and (b) display the overall dependence of the conductivity on frequency for the two fixed values of the magnetic field, $|qB|=(5m)^2$ and $|qB|=(15m)^2$, respectively. As predicted, only the intraband transitions contribute to $\sigma_{\perp}(\Omega)$ in the low-frequency region ($\Omega < \sqrt{2|qB|+m^2}-m$) and only the interband transitions contribute in the high-frequency region ($\Omega > \sqrt{2|qB|+m^2}+m$). In the intermediate window of frequencies, i.e., $\sqrt{2|qB|+m^2}-m<\Omega<\sqrt{2|qB|+m^2}+m$, the transverse conductivity vanishes identically. 

While it is easy to resolve and identify the annihilation peaks in the high-frequency regime in panels (a) and (b) of Fig.~\ref{fig:CondPerp}, the low-frequency region appears much more convoluted due to many nearly overlapping thresholds for particle-particle (antiparticle-antiparticle) transitions. Thus, to resolve the details of the transverse conductivity better, we show the low-frequency results separately in panels (c) and (d) for $|qB|=(5m)^2$ and $|qB|=(15m)^2$, respectively.  

The temperature dependence of the transverse conductivity in the high-frequency region can be understood qualitatively from the results in panels (a) and (b) of Fig.~\ref{fig:CondPerp}. Just above the annihilation threshold, the conductivities have an inverse ordering with the largest (smallest) value of conductivity obtained for the lowest (highest) temperature. At sufficiently high frequencies, however, the ordering gradually becomes normal, i.e., the largest (smallest) value of conductivity is obtained for the (lowest) temperature. The transition from the inverse-ordering ($\sigma_{\perp}(\Omega)$ increasing with $T$) to the normal-ordering regime ($\sigma_{\perp}(\Omega)$ decreasing with $T$) depends both on the magnetic field and the ratio $T/\sqrt{|qB|}$. 

As one can see from panels (c) and (d) of Fig.~\ref{fig:CondPerp}, the transverse conductivity mostly grows with temperature in the region of low frequencies. This is expected, of course, since the main part of the result comes from many transitions between neighboring Landau levels with sufficiently close energies and, thus, large indices $n$ (running up to about $n\sim |qB|/\Omega^2$). When the temperature increases, the occupation numbers of the relevant high-energy states grow and, as a result, $\sigma_{\perp}(\Omega)$ becomes larger. As we see from Figs.~\ref{fig:CondPerp}(c) and \ref{fig:CondPerp}(d), this trend gradually changes as the frequency increases and approaches the largest particle-particle threshold $\sqrt{2|qB|+m^2}-m$ from below. Near the highest particle-particle threshold, the ordering of results for the three values of temperature ($T=5m$, $T=15m$, and $T=50m$) is opposite. The width of the corresponding near-threshold region is not universal however. It changes depending on the ratio of energy scales set by the magnetic field and temperature. As one can see from the results shown in panels (c) and (d) of Fig.~\ref{fig:CondPerp}, the near-threshold region with inverse ordering of conductivities for the three fixed temperatures widens when $T/\sqrt{|qB|}$ grows. 

Similarly to the photon emission discussed in Sec.~\ref{sec:emission-QED}, both longitudinal and transverse magneto-optical conductivities have numerous singular peaks, associated with the threshold effects. From the analytical expressions in Eqs.~(\ref{sigma-parallel}) and (\ref{sigma-perp}), we see that the singularities are also of the same inverse-square-root type. As  argued in Sec.~\ref{interaction-effects}, the peaks are expected to smooth out when one takes quasiparticle interaction effects into account. 

Our results for the magneto-optical conductivity agree with those obtained in the quantum limit for Dirac semimetals in Refs.~\cite{Ashby:2013aa,Long:2018aa,PhysRevB.102}. Indirectly, such an agreement provides a valuable cross-check for the derivation of the photon polarization tensor in Sec.~\ref{sec:Polarization}. Unlike the conductivity, though, the polarization tensor extends to the case of nonzero photon momenta and has a considerably more complicated structure.

\section{Summary and Conclusions}
\label{sec:summary}

In this study, we utilized a coordinate representation for the fermion Green’s function in a constant magnetic field to derive a closed-form analytical expression for the absorptive part of the photon polarization tensor in a relativistic plasma. We found that the polarization tensor contains 4 different symmetric structures and 2 antisymmetric ones. Such an abundance of tensor structures is the result of the Lorentz symmetry breaking by the magnetic field and the thermal bath. As required by the gauge symmetry, all 6 tensor structures are transverse. We found that the antisymmetric parts of the tensor vanish at zero chemical potential ($\mu=0$), which was the main focus of this study. This is the consequence of the charge conjugation symmetry. It is expected, however, that both antisymmetric structures will be nonzero when $\mu\neq 0$ \cite{Wang:2021eud}.  

The derivation in this study was focused on the absorptive part of the photon polarization. The latter includes the imaginary parts of symmetric tensor structures and, in principle, the real parts of antisymmetric ones.  As we argued before \cite{Wang:2020dsr}, the absorptive part of the tensor is determined by the following three types of processes: (i) $e^{-}\rightarrow e^{-}+\gamma$, (ii) $e^{+}\rightarrow e^{+}+\gamma$, and (iii) $e^{-}+e^{+}\rightarrow \gamma$. In a quantum description, they correspond to particle-particle, antiparticle-antiparticle, and particle-antiparticle transitions between Landau levels, respectively. At the vanishing chemical potential, both particle-particle and antiparticle-antiparticle transitions contribute equally to the polarization tensor.

By utilizing the photon polarization tensor, we studied in detail the differential photon emission rate for a strongly magnetized thermal relativistic plasma. In this context, the particle-particle and antiparticle-antiparticle transitions can be viewed as splitting ($1\to2$) processes, while the particle-antiparticle ones as annihilation ($2\to1$) processes. All of them are affected by the Landau-level quantization. In particular, the photon rate as a function of the photon energy (for a fixed direction of emission) or as a function of the angular coordinate (for fixed photon energy) contains numerous inverse-square-root singularities. They are associated with the threshold effects in photon production. Since the inverse-square-root singularities are integrable, they cause no complications in studying rates integrated over the angular coordinates or the energy. Conceptually, the actual singularities are an artifact of the clean-limit approximation that ignores all particle interactions except for the interaction with the background magnetic field. As we showed, taking interactions into account will smooth out the threshold singularities. Quantitative effects of interactions can be captured by the quasiparticle width $\Gamma$. It the value of $\Gamma$ that also determines the width of an energy window near the threshold, where the singular behavior will be smeared out. 

In this study, we analyzed the angular dependence of the differential photon emission rate in a wide range of model parameters. In all regimes, ranging from a moderately relativistic to an ultrarelativistic plasma, the emission is highly anisotropic. In addition to its spiky dependence on the angular coordinate coming from the threshold effects, the emission has an overall ellipsoidal profile that changes gradually with the photon energy. In general, the profile is prolate at small energies ($\Omega \lesssim \sqrt{|qB|}$) and oblate at large energies ($\Omega \gtrsim \sqrt{|qB|}$). In other words, the dominant emission tends to be along the line of the magnetic field at small energies and perpendicular to the magnetic field at large energies. The prolate emission profile at small energies can be explained largely by the Landau-level quantization that strongly restricts the kinematics of photon emission. On the other hand, we speculate that the oblate profile at large energies is a natural consequence of the synchrotron radiation in the semiclassical limit \cite{Sokolov:1986nk}, which goes predominantly in the directions perpendicular to the magnetic field. Of course, the synchrotron radiation comes only from the splitting ($1\to 2$) processes. However, there are also contributions from the annihilation ($2\to 1$) processes. While the latter are subdominant at sufficiently high temperatures, they become important at low temperatures and high photon energies.

The combination of our results for several representative choices of the magnetic field and temperature provides a rather detailed picture of the photon emission. As one might expect, the photon rate tends to grow with increasing the temperature of the plasma. Concerning the energy dependence, however, we found that the rate has a nonmonotonic behavior. For small photon energies, $\Omega \ll \sqrt{|qB|}$, it is strongly suppressed by the Landau-level quantization. With increasing the energy, the quantization effects gradually relax and the rate grows. After reaching a maximum value at an intermediate value, however, the rate starts to decrease. The decreasing trend also persists at very large energies, where the suppression comes from the Fermi-Dirac distribution of charged particles with high energies.  

The other application of the absorptive part of the photon polarization tensor, presented in this paper, was the calculation of magneto-optical conductivity in a quantum limit of relativistic plasma. By making use of the explicit tensor structure, we calculated both longitudinal and transverse parts of the conductivity. As in the case of photon emission, the Landau-level quantization affects the main features of the frequency dependence of  optical conductivity. One of such characteristic features is the appearance of singular spikes associated with the Landau-level threshold effects. Of course, the singularities come from using the clean-limit approximation and should go away when particle   interactions are taking into account. 

In the case of the longitudinal component of conductivity $\sigma_{\parallel}(\Omega)$, we showed that only the annihilation (``interband") processes contribute. Accordingly, the result for $\sigma_{\parallel}(\Omega)$ is nonvanishing only when $\Omega>2m$. The transverse component of conductivity $\sigma_{\perp}(\Omega)$, on the other hand, comes from the splitting (``intraband") processes when $\Omega<  \sqrt{2|qB|+m^2}-m$ and the annihilation (``interband") ones when $\Omega>\sqrt{2|qB|+m^2}+m$. The two windows of frequencies are separated by a gap of width $2m$, where the transverse conductivity vanishes.

While we carefully studied the imaginary part of the photon polarization at $\mu=0$ in this paper, the case of nonzero $\mu$ was not addressed. It appears that the corresponding problem poses only a technical complication that can be overcome with a moderate effort. Such study is underway now \cite{Wang:2021eud}. A more difficult extension of the present work will be obtaining the real part of the photon polarization. We are currently exploring several promising approaches to address the problem.

\acknowledgements
The work of X.W. was supported by the start-up funding No.~4111190010 of Jiangsu University. The work of I.A.S. was supported by the U.S. National Science Foundation under Grant No.~PHY-1713950. 
 
\appendix

\section{Matsubara sums}
\label{MatsubaraSums}

In this Appendix, we present several general results for the fermionic Matsubara sums needed in the calculation of the photon polarization function in the main text. 

Let us start by quoting the following standard sums
\begin{eqnarray}
T\sum_{k=-\infty}^{\infty} 
\frac{1}{\left[\omega_k^2+a^2\right]\left[(\omega_k-\Omega_m)^2+b^2\right]} 
&=&\frac{(a+b)\left[1-n_F(a)-n_F(b)\right]}{2ab\left[(a+b)^2+\Omega_m^2\right]}
+\frac{(a-b)\left[n_F(a)-n_F(b)\right]}{2ab\left[(a-b)^2+\Omega_m^2\right]},
\label{Matsubara-sum-1} \\
T\sum_{k=-\infty}^{\infty} 
\frac{\omega_k(\omega_k-\Omega_m)}{\left[\omega_k^2+a^2\right]\left[(\omega_k-\Omega_m)^2+b^2\right]} 
&=&\frac{(a+b)\left[1-n_F(a)-n_F(b)\right]}{2\left[(a+b)^2+\Omega_m^2\right]}
-\frac{(a-b)\left[n_F(a)-n_F(b)\right]}{2\left[(a-b)^2+\Omega_m^2\right]},
\label{Matsubara-sum-2} \\
T\sum_{k=-\infty}^{\infty} \frac{i\omega_k}{\left[\omega_k^2+a^2\right]\left[(\omega_k-\Omega_m)^2+b^2\right]} 
&=&\frac{i\Omega_m\left[1-n_F(a)-n_F(b)\right]}{2b\left[(a+b)^2+\Omega_m^2\right]}
+\frac{i\Omega_m\left[n_F(a)-n_F(b)\right]}{2b\left[(a-b)^2+\Omega_m^2\right]},
\label{Matsubara-sum-3} \\
T\sum_{k=-\infty}^{\infty} \frac{i(\omega_k-\Omega_m)}{\left[\omega_k^2+a^2\right]\left[(\omega_k-\Omega_m)^2+b^2\right]} 
&=&-\frac{i\Omega_m\left[1-n_F(a)-n_F(b)\right]}{2a\left[(a+b)^2+\Omega_m^2\right]}
+\frac{i\Omega_m\left[n_F(a)-n_F(b)\right]}{2a\left[(a-b)^2+\Omega_m^2\right]}.
\label{Matsubara-sum-4}
\end{eqnarray}
where $n_F(\epsilon) = 1/\left[\exp(\epsilon/T)+1\right]$ is the Fermi-Dirac distribution function, and  $\omega_k =  (2k+1)\pi T$ and $\Omega_m = 2m\pi T$ are the fermionic and bosonic Matsubara frequencies, respectively. Note that distribution function satisfies the following property: $n_F(-x)=1-n_F(x)$.

By making use of Eqs.~(\ref{Matsubara-sum-1}) through (\ref{Matsubara-sum-4}), it is straightforward to derive the following Matsubara sum of a more general type:  
\begin{eqnarray}
&& T\sum_{k=-\infty}^{\infty} 
\frac{i\omega_k (i\omega_k-i\Omega_m) X + i\omega_k Y_1 +(i\omega_k-i\Omega_m) Y_2 +Z}
{\left[(i\omega_k)^2-a^2\right]\left[(i\omega_k-i\Omega_m)^2-b^2\right]} \nonumber\\
&&= \sum_{\lambda=\pm 1} 
\frac{ (a-\lambda b)\left[ n_F(a)-n_F(\lambda b)  \right]}{2 \lambda  ab\left[(a-\lambda b)^2+\Omega_m^2\right]}
\left[\lambda ab X+i\Omega_m \frac{aY_1+\lambda  bY_2}{a-\lambda b} +Z\right]\nonumber\\
&&= \sum_{\eta,\lambda=\pm 1} 
\frac{ n_F(a)-n_F(\lambda b)}
{4 \lambda  ab \left(a-\lambda b+\eta i\Omega_m\right)}
\left[\lambda ab X-\eta \left(a  Y_1+ \lambda  b Y_2 \right)+Z\right],
\label{Matsubara-sum}
\end{eqnarray}
where $X$, $Y_1$, $Y_2$, and $Z$ are arbitrary coefficient functions
independent of the fermion Matsubara frequency $\omega_k$. 

In the calculation of the photon polarization function in the main text, parameters 
$a$ and $b$ should be replaced with the Landau-level energies 
$E_{n,p_z}=\sqrt{p_z^2+m^2+2n|qB|}$ and 
$E_{n^{\prime},p_z-k_z}=\sqrt{(p_z-k_z)^2+m^2+2n^{\prime}|qB|}$, respectively.

\section{Integrals over transverse spatial coordinates}
\label{prop-use}

In this Appendix, we present the explicit results for all types of integrals over the transverse spatial coordinates, $\mathbf{r}_\perp$, that appear in the calculation of the photon polarization function. 

Instead of the Cartesian coordinates $x$ and $y$, it is convenient to use the polar coordinates $r_\perp\equiv |\mathbf{r}_\perp|$ 
and $\varphi_\perp$ in the $\mathbf{r}_\perp$ plane. Then, the results of the angular integrations are given in terms of the Bessel functions. In particular, one derives the following results for the integrals that appear in the calculation:
\begin{eqnarray}
\label{bes1}
\int d^2 \mathbf{r}_\perp e^{-i \mathbf{r}_\perp\cdot \mathbf{k}_\perp} f (r_\perp) 
&=&  2\pi \int_{0}^{\infty} r_\perp d r_\perp J_{0}\left(r_{\perp}k_\perp\right) f (r_\perp) ,  \\
\label{bes2}
\int d^2 \mathbf{r}_\perp e^{-i \mathbf{r}_\perp\cdot \mathbf{k}_\perp} r_\perp^{\mu} f (r_\perp) 
&=&  - 2\pi i \hat{k}_\perp^{\mu} \int_{0}^{\infty} r_\perp^2 d r_\perp J_{1}\left(r_{\perp}k_\perp\right) f (r_\perp) \nonumber\\
&=& - \pi i k_\perp^{\mu} \int_{0}^{\infty} r_\perp^3 d r_\perp \left[J_{0}\left(r_{\perp}k_\perp\right) +J_{2}\left(r_{\perp}k_\perp\right)\right] f (r_\perp)
,  \\
\label{bes3}
\int d^2 \mathbf{r}_\perp e^{-i \mathbf{r}_\perp\cdot \mathbf{k}_\perp} \mathbf{r}_\perp^{2} f (r_\perp) 
&=&  2\pi \int_{0}^{\infty} r_\perp^3 d r_\perp  J_{0}\left(r_{\perp}k_\perp\right) f (r_\perp),  \\
\label{bes4}
\int d^2 \mathbf{r}_\perp e^{-i \mathbf{r}_\perp\cdot \mathbf{k}_\perp} r_\perp^{\mu} r_\perp^{\nu} f (r_\perp) 
&=&  2\pi \int_{0}^{\infty} r_\perp^3 d r_\perp \left[ \left(
- \hat{k}_\perp^{\mu} \hat{k}_\perp^{\nu} 
-\frac{1}{2} g_{\perp}^{\mu\nu}\right)J_{2}\left(r_{\perp}k_\perp\right)-\frac{1}{2} g_{\perp}^{\mu\nu}J_{0}\left(r_{\perp}k_\perp\right)  \right]f (r_\perp),
\end{eqnarray}
where we used the following relation for the Bessel functions:
\begin{equation}
J_{1}\left(r_{\perp}k_\perp\right) = \frac{1}{2}r_{\perp}k_\perp\left[J_{0}\left(r_{\perp}k_\perp\right) +J_{2}\left(r_{\perp}k_\perp\right) \right].
\label{Bessel-identity}
\end{equation}
The remaining integrals over $r_\perp$ can be expressed in terms of the following dimensionless functions:
\begin{eqnarray}
\mathcal{I}_{0}^{n,n^{\prime}}(\xi)&=& \frac{1}{\ell^2} \int_{0}^{\infty} r_\perp d r_\perp e^{-\mathbf{r}_\perp^2/(2\ell^2)} J_{0}\left(r_{\perp}k_\perp\right)L_{n}\left(\frac{\mathbf{r}_{\perp}^2}{2\ell^{2}}\right)L_{n^\prime}\left(\frac{\mathbf{r}_{\perp}^2}{2\ell^{2}}\right) ,
\label{I0f}  \\
\mathcal{I}_{1}^{n,n^{\prime}}(\xi)&=&  \frac{1}{\ell^3} \int_{0}^{\infty} r_\perp^2  d r_\perp e^{-\mathbf{r}_\perp^2/(2\ell^2)} J_{1}\left(r_{\perp}k_\perp\right) L_{n} \left(\frac{\mathbf{r}_{\perp}^2}{2\ell^{2}}\right)L_{n^\prime}^{1}\left(\frac{\mathbf{r}_{\perp}^2}{2\ell^{2}}\right) ,
\label{I1f} \\
\mathcal{I}_{2}^{n,n^{\prime}}(\xi)&=&  \frac{1}{\ell^4} \int_{0}^{\infty} r_\perp^3  d r_\perp e^{-\mathbf{r}_\perp^2/(2\ell^2)} J_{0}\left(r_{\perp}k_\perp\right) L_{n}^{1}\left(\frac{\mathbf{r}_{\perp}^2}{2\ell^{2}}\right)L_{n^\prime}^{1}\left(\frac{\mathbf{r}_{\perp}^2}{2\ell^{2}}\right), 
\label{I2f} \\
\mathcal{I}_{3}^{n,n^{\prime}}(\xi)&=&  \frac{1}{\ell^4} \int_{0}^{\infty} r_\perp^3  d r_\perp e^{-\mathbf{r}_\perp^2/(2\ell^2)} J_{2}\left(r_{\perp}k_\perp\right) L_{n}^{1}\left(\frac{\mathbf{r}_{\perp}^2}{2\ell^{2}}\right)L_{n^\prime}^{1}\left(\frac{\mathbf{r}_{\perp}^2}{2\ell^{2}}\right),
\label{I3f} 
\end{eqnarray}
where $\xi = k_{\perp}^2\ell^{2}/2$.

It should be noted that $\mathcal{I}_{0}^{n,n^{\prime}}(\xi)$, $\mathcal{I}_{2}^{n,n^{\prime}}(\xi)$, 
and $\mathcal{I}_{3}^{n,n^{\prime}}(\xi)$ are symmetric under the interchange of indices $n$ and $n^{\prime}$.
In contrast, $\mathcal{I}_{1}^{n,n^{\prime}}(\xi)$ is not symmetric, but it can be rewritten in terms of the other functions. 
Indeed, by making use of Eq.~(\ref{Bessel-identity}) and the well-known identity for the Laguerre polynomials
$L_{n}^{\alpha}(x) = L_{n}^{\alpha+1}(x) -L_{n-1}^{\alpha+1}(x)$, one derives
\begin{equation}
\mathcal{I}_{1}^{n,n^{\prime}}(\xi) = \sqrt{\frac{\xi}{2}}\left[
\mathcal{I}_{2}^{n,n^{\prime}}(\xi) -\mathcal{I}_{2}^{n-1,n^{\prime}}(\xi)
+\mathcal{I}_{3}^{n,n^{\prime}}(\xi)-\mathcal{I}_{3}^{n-1,n^{\prime}}(\xi)
\right].
\label{I1-vs-I2-I3}
\end{equation}

One can also derive  the following relations:
\begin{eqnarray}
\mathcal{I}_{1}^{n,n^{\prime}}(\xi) &=&
\frac{n+n^{\prime}+2}{2\sqrt{2\xi}}\left[\mathcal{I}_{0}^{n,n^{\prime}}(\xi) -\mathcal{I}_{0}^{n+1,n^{\prime}+1}(\xi) \right]
+\frac{\sqrt{\xi}}{2\sqrt{2}}\left[\mathcal{I}_{0}^{n+1,n^{\prime}}(\xi) +\mathcal{I}_{0}^{n,n^{\prime}+1}(\xi) \right] 
-\frac{n-n^{\prime}}{\sqrt{2\xi}}\mathcal{I}_{0}^{n,n^{\prime}}(\xi) ,
\label{I1-via-I0}  \\
\mathcal{I}_{2}^{n,n^{\prime}}(\xi) &=&
\frac{n+n^{\prime}+2}{2}\left[\mathcal{I}_{0}^{n,n^{\prime}}(\xi) +\mathcal{I}_{0}^{n+1,n^{\prime}+1}(\xi) \right]
-\frac{\xi}{2}\left[\mathcal{I}_{0}^{n+1,n^{\prime}}(\xi) +\mathcal{I}_{0}^{n,n^{\prime}+1}(\xi) \right] 
\label{I2-via-I0}
,\\
\mathcal{I}_{3}^{n,n^{\prime}}(\xi) 
&=&\frac{n+n^{\prime}+2}{2}\left[\mathcal{I}_{0}^{n+1,n^{\prime}}(\xi) +\mathcal{I}_{0}^{n,n^{\prime}+1}(\xi) \right]
-\frac{(n-n^{\prime})^2}{2\xi}\left[\mathcal{I}_{0}^{n,n^{\prime}}(\xi) +\mathcal{I}_{0}^{n+1,n^{\prime}+1}(\xi) \right] .
\label{I3-via-I0}
\end{eqnarray}
Other random relations:
\begin{eqnarray}
\mathcal{I}_{1}^{n,n^{\prime}}(\xi) &=& \sqrt{\frac{2}{\xi}} (n^{\prime}+1) \mathcal{I}_{0}^{n,n^{\prime}}(\xi) 
-\frac{1}{\sqrt{2\xi}} \mathcal{I}_{2}^{n,n^{\prime}}(\xi) \nonumber\\
&=&\frac{n+n^{\prime}+2}{\sqrt{2\xi}}\mathcal{I}_{0}^{n,n^{\prime}}(\xi) 
-\frac{1}{\sqrt{2\xi}}\mathcal{I}_{2}^{n,n^{\prime}}(\xi) 
-\frac{n-n^{\prime}}{\sqrt{2\xi}}\mathcal{I}_{0}^{n,n^{\prime}}(\xi) 
\label{I1-I0-I2},\\
\mathcal{I}_{3}^{n,n^{\prime}}(\xi) &=& \sqrt{\frac{2}{\xi}} \left[
(n+1)\mathcal{I}_{1}^{n,n^{\prime}}(\xi) 
+(n^{\prime}+1) \mathcal{I}_{1}^{n^{\prime}+1,n}(\xi) \right] ,\\
\mathcal{I}_{3}^{n,n^{\prime}}(\xi) &=& \mathcal{I}_{2}^{n,n^{\prime}-1}(\xi) - 
\sqrt{\frac{2}{\xi}}(n-n^\prime -\xi)\mathcal{I}_{1}^{n^{\prime},n}(\xi) , \\
2(n+n^{\prime}+1-\xi) \mathcal{I}_{0}^{n,n^{\prime}}(\xi) &=& 
\mathcal{I}_{2}^{n,n^{\prime}}(\xi) +\mathcal{I}_{2}^{n-1,n^{\prime}-1}(\xi) , \\
\sqrt{ 2\xi } \mathcal{I}_{0}^{n,n^{\prime}}(\xi) &=& \mathcal{I}_{1}^{n,n^{\prime}-1}(\xi)
+ \mathcal{I}_{1}^{n^{\prime},n}(\xi)  
\label{I3-I2-I1}, \\
2(n-n^{\prime})\, \mathcal{I}_{0}^{n,n^{\prime}}(\xi) &=&\sqrt{ 2\xi }\left(
\mathcal{I}_{1}^{n^{\prime},n}(\xi) -  \mathcal{I}_{1}^{n,n^{\prime}}(\xi)  \right) ,
\\
\mathcal{I}_{1}^{n,n^{\prime}}(\xi) &=& \mathcal{I}_{1}^{n^{\prime},n}(\xi)  
+\sqrt{ \frac{2}{\xi} }(n^{\prime}-n)\, \mathcal{I}_{0}^{n,n^{\prime}}(\xi) , 
\label{I1-asymmerty} \\
\mathcal{I}_{1}^{n^{\prime},n}(\xi) &=& \sqrt{2\xi} \mathcal{I}_{0}^{n,n^{\prime}}(\xi) 
- \mathcal{I}_{1}^{n,n^{\prime}-1}(\xi)  , 
\label{I1-I1-I0-a} \\
\mathcal{I}_{1}^{n,n^{\prime}}(\xi) &=& \sqrt{2\xi} \mathcal{I}_{0}^{n,n^{\prime}}(\xi) 
- \mathcal{I}_{1}^{n^{\prime},n-1}(\xi)  , 
\label{I1-I1-I0-b} \\
\frac{2(n^{\prime}-n)}{\xi}\mathcal{I}_{0}^{n,n^{\prime}}(\xi) &=& 
\mathcal{I}_{2}^{n,n^{\prime}-1}(\xi)
-\mathcal{I}_{2}^{n-1,n^{\prime}}(\xi)
+\mathcal{I}_{3}^{n,n^{\prime}-1}(\xi) 
-\mathcal{I}_{3}^{n-1,n^{\prime}}(\xi) .
\label{I-2-3-asymmerty}
\end{eqnarray}

All of the above integrals can be calculated exactly in terms of the generalized Laguerre polynomials by using table integral \cite{Gradshtein}. The results read
\begin{eqnarray}
\mathcal{I}_{0}^{n,n^{\prime}}(\xi)&=& (-1)^{n+n^\prime} e^{-\xi} L_{n}^{n^\prime-n}\left(\xi\right) 
L_{n^\prime}^{n-n^\prime}\left(\xi\right) ,
\label{I0f-LL}  \\
\mathcal{I}_{1}^{n,n^{\prime}}(\xi)&=& \sqrt{2\xi} (-1)^{n+n^\prime} e^{-\xi} L_{n}^{n^\prime-n+1}\left(\xi\right) 
L_{n^\prime}^{n-n^\prime}\left(\xi\right) 
\label{I1f-LL} , \\
\mathcal{I}_{2}^{n,n^{\prime}}(\xi)&=&2 (-1)^{n+n^\prime}(n^\prime+1)
e^{-\xi} L_{n}^{n^\prime-n}\left(\xi\right) 
L_{n^\prime+1}^{n-n^\prime}\left(\xi\right) , 
\label{I2f-LL} \\
\mathcal{I}_{3}^{n,n^{\prime}}(\xi)&=&  2(-1)^{n+n^\prime} \xi
e^{-\xi} L_{n}^{n^\prime-n+1}\left(\xi\right) 
L_{n^\prime}^{n-n^\prime+1}\left(\xi\right).
\label{I3f-LL} 
\end{eqnarray}
Note that the above functions are expressed in terms of the exponential function $e^{-\xi}$ and a product of the Laguerre polynomials of degrees $n$ and $n^\prime$ (or $n$ and $n^\prime+1$). In numerical calculations,  it is the product of two polynomials that produces the largest evaluation errors. This is particularly detrimental when the values of $n$, $n^\prime$, and $|n-n^\prime|$ are all large. As we argue below, the problem can be relaxed partially by rewriting the corresponding functions equivalently as a product of the power function $\xi^{|n-n^\prime|}$ and the Laguerre polynomials of smaller degree(s) $\sim \mbox{min}(n,n^\prime)$. This can be achieved by utilizing the well-known identity for the Laguerre polynomials
\begin{equation}
L_{n}^{\alpha}(\xi) = (-\xi)^{-\alpha}\frac{(n+\alpha)!}{n!}L_{n+\alpha}^{-\alpha}(\xi) .
\label{Ln-alpha}
\end{equation}
By applying this to the first Laguerre polynomial in each product, the functions in Eqs.~(\ref{I0f-LL}) -- (\ref{I3f-LL}) can be rewritten in the following equivalent form: 
\begin{eqnarray}
\mathcal{I}_{0}^{n,n^{\prime}}(\xi)&=& \frac{(n^\prime)!}{n!} e^{-\xi}  \xi^{n-n^\prime}
\left(L_{n^\prime}^{n-n^\prime}\left(\xi\right)\right)^2 ,
\label{I0f-LL-form1}  \\
\mathcal{I}_{1}^{n,n^{\prime}}(\xi)&=& - \sqrt{2\xi} \frac{(n^\prime+1)!}{n!}e^{-\xi} \xi^{n-n^\prime-1} 
L_{n^\prime+1}^{n-n^\prime-1}\left(\xi\right) 
L_{n^\prime}^{n-n^\prime}\left(\xi\right) 
\label{I1f-LL-form1} , \\
\mathcal{I}_{2}^{n,n^{\prime}}(\xi)&=& 2 \frac{(n^\prime+1)!}{n!}e^{-\xi}  \xi^{n-n^\prime} 
L_{n^\prime}^{n-n^\prime}\left(\xi\right)
L_{n^\prime+1}^{n-n^\prime}\left(\xi\right) , 
\label{I2f-LL-form1} \\
\mathcal{I}_{3}^{n,n^{\prime}}(\xi)&=&  -2 \frac{(n^\prime+1)!}{n!} e^{-\xi} \xi^{n-n^\prime} 
L_{n^\prime+1}^{n-n^\prime-1}\left(\xi\right) 
L_{n^\prime}^{n-n^\prime+1}\left(\xi\right).
\label{I3f-LL-form1} 
\end{eqnarray}
These representations are particularly useful in numerical calculations when $n> n^\prime$. Similarly, by applying Eq.~(\ref{Ln-alpha}) to the second Laguerre polynomial in each product, we can rewrite functions $\mathcal{I}_{i}^{n,n^{\prime}}(\xi)$ as follows:
\begin{eqnarray}
\mathcal{I}_{0}^{n,n^{\prime}}(\xi)&=&  \frac{n!}{(n^\prime)!}e^{-\xi} \xi^{n^\prime-n}
\left(L_{n}^{n^\prime-n}\left(\xi\right)\right)^2 ,
\label{I0f-LL-form1}  \\
\mathcal{I}_{1}^{n,n^{\prime}}(\xi)&=&  \sqrt{2\xi}  \frac{n!}{(n^\prime)!} e^{-\xi} \xi^{n^\prime-n} 
L_{n}^{n^\prime-n+1}\left(\xi\right) 
L_{n}^{n^\prime-n}\left(\xi\right) 
\label{I1f-LL-form1} , \\
\mathcal{I}_{2}^{n,n^{\prime}}(\xi)&=& 2 \frac{(n+1)!}{(n^\prime)!} e^{-\xi}  \xi^{n^\prime-n}  L_{n}^{n^\prime-n}\left(\xi\right)L_{n+1}^{n^\prime-n}\left(\xi\right)  , 
\label{I2f-LL-form1} \\
\mathcal{I}_{3}^{n,n^{\prime}}(\xi)&=&  - 2 \frac{(n+1)!}{(n^\prime)!} e^{-\xi} \xi^{n^\prime-n} 
L_{n}^{n^\prime-n+1}\left(\xi\right) 
L_{n+1}^{n^\prime-n-1}\left(\xi\right).
\label{I3f-LL-form1} 
\end{eqnarray}
These are suitable when $n< n^\prime$.

For completeness, let us also discuss the properties of $\mathcal{I}_{0}^{n,n^{\prime}}$ and $\mathcal{I}_{2}^{n,n^{\prime}}$
in the limit of small $|\mathbf{k}_{\perp}| \ell$. This limit is of relevance for the photon emission in the regime of small $\Omega$, 
as well as for the photon emission in the direction of the magnetic field ($\theta \approx 0$) at an arbitrary $\Omega$. By making use a closed form representation for the generalized Laguerre polynomials \cite{Abramowitz}
\begin{eqnarray}
L_{n}^{\alpha}(\xi) &=&\sum_{i=0}^{n} \frac{(n+\alpha)! (-\xi)^i }{ i! (n-i)! (\alpha+i)!},
\end{eqnarray}
which is valid for $\alpha > -1$, and keeping only the terms up to linear order in $\xi = |\mathbf{k}_{\perp}|^2 \ell^2/2$, 
one obtains the following results:
\begin{eqnarray}
\mathcal{I}_{0}^{n,n^{\prime}}(\xi) &\simeq&  \delta_{n,n^{\prime}}
- \left[(2n+1)\delta_{n,n^{\prime}} -(n+1)\delta_{n,n^{\prime}-1}
- (n^{\prime}+1)\delta_{n-1,n^{\prime}}\right]\xi 
+O\left[\xi^2\right],
\\
\mathcal{I}_{2}^{n,n^{\prime}}(\xi)  &  \simeq & 2 (n+1)  \delta_{n,n^{\prime}}
 - 2 (n+1) (n^{\prime}+1)\left(2\delta_{n,n^{\prime}}- \delta_{n,n^{\prime}-1}-  \delta_{n-1,n^{\prime}} \right)
\xi +O\left[\xi^2\right],
\\
\mathcal{I}_{3}^{n,n^{\prime}}(\xi)  &  \simeq & (n+1)(n^{\prime}+1)\left(2\delta_{n,n^{\prime}}- \delta_{n,n^{\prime}-1}-  \delta_{n-1,n^{\prime}} \right)
\xi+O\left[\xi^2\right].
\end{eqnarray}
Note that a similar expansion for $\mathcal{I}_{1}^{n,n^{\prime}}(\xi)$ can be obtained by using Eq.~(\ref{I1-vs-I2-I3}). 
As we can see, functions $\mathcal{I}_{0}^{n,n^{\prime}}(\xi)$ and $\mathcal{I}_{2}^{n,n^{\prime}}(\xi)$
are diagonal in the Landau-level indices (i.e., $\sim \delta_{n,n^{\prime}}$) at the leading (zeroth) order in $|\mathbf{k}_{\perp}| \ell$. On the other hand, $\mathcal{I}_{3}^{n,n^{\prime}}(\xi)$ vanishes when $|\mathbf{k}_{\perp}| \ell\to 0$.

\section{Explicit expressions for $T_{i}^{\mu\nu} $ and  $I_{i}^{\mu\nu}$}
\label{AP-tr}

In the expression for the photon polarization function, there are several types of Dirac traces. For completeness, the corresponding results are presented here. For brevity of notation, we will use the following shorthand notations: 
$L_{n}^{m} \equiv L_{n}^{m}\left(\xi\right)$, $T_{i}^{\mu\nu} \equiv T_{i}^{\mu\nu}(\mathbf{k}_{\perp})$, and  $I_{i}^{\mu\nu} \equiv I_{i}^{\mu\nu}(\mathbf{k}_{\perp})$, where $i =1, 2, 3, 4$.

The four types of traces needed in the calculation are 
\begin{eqnarray}
\label{trt1}
T_{1}^{\mu\nu} &=&
\tr \left[ \gamma^\mu \left(p_\parallel \gamma_\parallel + m \right)
 \left( \mathcal{P}_{+}L_n +\mathcal{P}_{-}L_{n-1} \right)
\gamma^\nu \left((p_\parallel -k_\parallel )\gamma_\parallel + m \right)
\left( \mathcal{P}_{+}L_{n^\prime} +\mathcal{P}_{-}L_{n^\prime-1} \right)\right]  \nonumber\\
&=&2\left(p_\parallel^\mu (p_\parallel -k_\parallel )^\nu+(p_\parallel -k_\parallel )^\mu p_\parallel^\nu
 -g_\parallel^{\mu\nu} 
\left[p_\parallel  (p_\parallel -k_\parallel )-m^2 \right]  \right)
 \left(L_{n}L_{n^\prime}+L_{n-1}L_{n^\prime-1}\right)
\nonumber\\
&-&2g_\perp^{\mu\nu} 
\left[p_\parallel  (p_\parallel -k_\parallel )-m^2 \right]
\left(L_{n}L_{n^\prime-1}+L_{n-1}L_{n^\prime}\right)
+2i \ell^2 qF^{\mu\nu}
\left[p_\parallel  (p_\parallel -k_\parallel )-m^2 \right]
\left(L_{n}L_{n^\prime-1}-L_{n-1}L_{n^\prime}\right)
,\\
\label{trt2}
T_{2}^{\mu\nu} &=& \frac{i}{\ell^2}
\tr \left[ \gamma^\mu \left(p_\parallel \gamma_\parallel + m \right)\left( \mathcal{P}_{+}L_n +\mathcal{P}_{-}L_{n-1} \right)
\gamma^\nu  (\mathbf{r}_{\perp}\cdot\bm{\gamma}_{\perp}) L_{n^\prime-1}^{1} \right] \nonumber\\
&=&-\frac{2i}{\ell^2}
\left(p_\parallel^\mu r_\perp^{\nu}+ r_\perp^{\mu}p_\parallel^\nu \right)
\left( L_n +L_{n-1} \right)L_{n^\prime-1}^{1}
+ 2q
\left(p_\parallel^\mu  F^{\nu\rho} r_{\perp,\rho} -   F^{\mu\rho}r_{\perp,\rho}p_\parallel^\nu \right)
\left( L_n -L_{n-1} \right)L_{n^\prime-1}^{1}
,\\
\label{trt3}
T_{3}^{\mu\nu} &=& -\frac{i}{\ell^2}
\tr \left[ \gamma^\mu (\mathbf{r}_{\perp}\cdot\bm{\gamma}_{\perp}) L_{n-1}^{1}     
\gamma^\nu  \left((p_\parallel -k_\parallel )\gamma_\parallel + m \right)
\left( \mathcal{P}_{+}L_{n^\prime} +\mathcal{P}_{-}L_{n^\prime-1} \right)\right] \nonumber\\
&=&\frac{2i}{\ell^2}
\left((p_\parallel -k_\parallel )^{\mu} r_\perp^{\nu}+ r_\perp^{\mu}(p_\parallel -k_\parallel )^\nu \right)
\left( L_{n^\prime}  +L_{n^\prime-1} \right)L_{n-1}^{1}\nonumber\\
&+& 2q\left( (p_\parallel -k_\parallel )^\mu  F^{\nu\rho} r_{\perp,\rho}  - 
F^{\mu\rho}r_{\perp,\rho}(p_\parallel -k_\parallel )^\nu \right)
\left( L_{n^\prime} -L_{n^\prime-1} \right)L_{n-1}^{1}
,\\
\label{trt4}
T_{4}^{\mu\nu} &=&\frac{1}{\ell^4}
\tr \left[ \gamma^\mu (\mathbf{r}_{\perp}\cdot\bm{\gamma}_{\perp}) L_{n-1}^{1}    
\gamma^\nu  (\mathbf{r}_{\perp}\cdot\bm{\gamma}_{\perp}) L_{n^\prime-1}^{1} \right]
= \frac{4}{\ell^4} \left(g^{\mu\nu} \mathbf{r}_\perp^2+2 r_\perp^{\mu} r_\perp^{\nu}\right)  L_{n-1}^{1}  L_{n^\prime-1}^{1} ,
\end{eqnarray}
where $p_\parallel \gamma_\parallel \equiv p_0 \gamma^0 -p_z\gamma^3 = i\omega_k\gamma^0 -p_z\gamma^3$
and $(p_\parallel -k_\parallel )\gamma_\parallel \equiv (p_0-k_0) \gamma^0 -(p_z-k_z)\gamma^3 
= (i\omega_k-i\Omega_m)\gamma^0 -(p_z-k_z)\gamma^3$. The only nonzero components of the field strength tensor are 
$F^{12} = -F^{21} = -B$ or, equivalently, $F^{\mu\nu} = - \varepsilon^{0\mu\nu3} B$. Thus,  
$F^{\mu\rho} r_{\perp,\rho}  = (\mathbf{r}_\perp\times \mathbf{B})^{\mu}$.

By making use of the results in Appendix~\ref{prop-use}, one can also calculate the corresponding integrals over the transverse spatial coordinates, $\mathbf{r}_\perp$. The results are given by 
\begin{eqnarray}
I_{1}^{\mu\nu} &=& \int   d^2 \mathbf{r}_\perp e^{-i \mathbf{r}_\perp\cdot \mathbf{k}_\perp} e^{-\mathbf{r}_\perp^2/(2\ell^2)} T_{1}^{\mu\nu} =-4 \pi \ell^2 g_\perp^{\mu\nu} \left[p_\parallel  (p_\parallel -k_\parallel )-m^2 \right]
\left[ \mathcal{I}_{0}^{n,n^{\prime}-1}(\xi)+\mathcal{I}_{0}^{n-1,n^{\prime}}(\xi) \right]
 \nonumber\\
&+&4 \pi \ell^2 \left[p_\parallel^\mu (p_\parallel -k_\parallel )^\nu+(p_\parallel -k_\parallel )^\mu p_\parallel^\nu 
-g_\parallel^{\mu\nu} \left[p_\parallel  (p_\parallel -k_\parallel )-m^2 \right] \right]
\left[\mathcal{I}_{0}^{n,n^{\prime}}(\xi)+\mathcal{I}_{0}^{n-1,n^{\prime}-1}(\xi) \right] \nonumber\\
&+& 4 \pi i \ell^4 qF^{\mu\nu}
\left[p_\parallel  (p_\parallel -k_\parallel )-m^2 \right]
\left[\mathcal{I}_{0}^{n,n^{\prime}-1}(\xi)-\mathcal{I}_{0}^{n-1,n^{\prime}}(\xi)\right] , 
\label{I_if-1-appD}
\\
I_{2}^{\mu\nu}&=& \int   d^2 \mathbf{r}_\perp e^{-i \mathbf{r}_\perp\cdot \mathbf{k}_\perp} e^{-\mathbf{r}_\perp^2/(2\ell^2)} T_{2}^{\mu\nu}  
= - 4\pi \ell \left(p_\parallel^\mu \hat{k}_\perp^{\nu}+ \hat{k}_\perp^{\mu}p_\parallel^\nu \right)
\left[\mathcal{I}_{1}^{n,n^{\prime}-1}(\xi)+\mathcal{I}_{1}^{n-1,n^{\prime}-1}(\xi)\right] 
\nonumber\\
&-& 4 \pi i \ell^3  q \left(p_\parallel^\mu F^{\nu\rho} \hat{k}_{\perp,\rho}  -   F^{\mu\rho}\hat{k}_{\perp,\rho}p_\parallel^\nu \right)
\left[\mathcal{I}_{1}^{n,n^{\prime}-1}(\xi)-\mathcal{I}_{1}^{n-1,n^{\prime}-1}(\xi) \right],
\label{I_if-2-appD}
\\
I_{3}^{\mu\nu}&=& \int d^2 \mathbf{r}_\perp e^{-i \mathbf{r}_\perp\cdot \mathbf{k}_\perp} e^{-\mathbf{r}_\perp^2/(2\ell^2)} T_{3}^{\mu\nu} = 4\pi \ell 
\left((p_\parallel -k_\parallel )^{\mu} \hat{k}_\perp^{\nu}+ \hat{k}_\perp^{\mu}(p_\parallel -k_\parallel )^\nu \right)
\left[ \mathcal{I}_{1}^{n^{\prime},n-1}(\xi)+\mathcal{I}_{1}^{n^{\prime}-1,n-1}(\xi)\right]
\nonumber\\
&-& 4\pi i \ell^3 q
\left( (p_\parallel -k_\parallel )^\mu F^{\nu\rho} \hat{k}_{\perp,\rho} - F^{\mu\rho}\hat{k}_{\perp,\rho}(p_\parallel -k_\parallel )^\nu \right)
\left[ \mathcal{I}_{1}^{n^{\prime},n-1}(\xi)-\mathcal{I}_{1}^{n^{\prime}-1,n-1}(\xi) \right], 
\label{I_if-3-appD}
\\
I_{4}^{\mu\nu} &=& \int   d^2 \mathbf{r}_\perp e^{-i \mathbf{r}_\perp\cdot \mathbf{k}_\perp} e^{-\mathbf{r}_\perp^2/(2\ell^2)} T_{4}^{\mu\nu}
= 8\pi \left[ g_{\parallel}^{\mu\nu}  \mathcal{I}_{2}^{n-1,n^{\prime}-1}(\xi)
-\left(g_{\perp}^{\mu\nu}+ 2\hat{k}_\perp^{\mu} \hat{k}_\perp^{\nu}  \right)
\mathcal{I}_{3}^{n-1,n^{\prime}-1}(\xi)
 \right].
\label{I_if-4-appD}
\end{eqnarray}
The explicit expressions for functions $\mathcal{I}_{i}^{n,n^{\prime}}$ ($i = 1, 2, 3, 4$) are given in Eqs.~(\ref{I0f-LL}) -- (\ref{I3f-LL}) in Appendix~\ref{prop-use}.

After performing the Matsubara sums, substituting $i\Omega_m\to \Omega+i \epsilon$, and using the energy conservation condition (\ref{energy-conservation}), one finds that the $I_{i}^{\mu\nu}$ functions in the final result will be replaced by similar expressions where only the following replacements should be made:
\begin{eqnarray}
 p_\parallel ^\mu & \to &  \bar{p}_\parallel^\mu  \Big|_{p_z=p_{z}^{(\pm)}}
 = -\eta E_{n,p_z} \delta^{\mu}_{0} +p_z\delta^{\mu}_{3} \Big|_{p_z=p_{z}^{(\pm)}}
 \nonumber\\
 &=&\frac{1}{2}k_\parallel^\mu\left(\frac{2(n-n^{\prime})|qB|}{\Omega^2-k_z^2} +1\right)
 \pm\frac{1}{2}\tilde{k}_\parallel^\mu\sqrt{ \left(1-\frac{k_{-}^2}{\Omega^2-k_z^2} \right)
\left( 1-\frac{k_{+}^2}{\Omega^2-k_z^2}\right)}
 , \\
 (p_\parallel -k_\parallel )^\mu & \to &  (\bar{p}_\parallel -k_\parallel )^\mu   \Big|_{p_z=p_{z}^{(\pm)}}
 = -\eta \lambda E_{n^{\prime},p_z-k_z} \delta^{\mu}_{0} +(p_z-k_z)\delta^{\mu}_{3}   \Big|_{p_z=p_{z}^{(\pm)}}
  \nonumber\\
 &=&\frac{1}{2}k_\parallel^\mu\left(\frac{2(n-n^{\prime})|qB|}{\Omega^2-k_z^2} -1\right)
  \pm\frac{1}{2}\tilde{k}_\parallel^\mu\sqrt{ \left(1-\frac{k_{-}^2}{\Omega^2-k_z^2} \right)
\left( 1-\frac{k_{+}^2}{\Omega^2-k_z^2}\right)}
 ,\\
 p_\parallel  (p_\parallel -k_\parallel) & \to &   \bar{p}_\parallel (\bar{p}_\parallel -k_\parallel ) \Big|_{p_z=p_{z}^{(\pm)}}
 =  \lambda E_{n,p_z}E_{n^{\prime},p_z-k_z} -p_z(p_z-k_z) \Big|_{p_z=p_{z}^{(\pm)}} 
  \nonumber\\
 &=&  m^2+(n+n^\prime)|qB| -\frac{1}{2} k_\parallel^2 .
\end{eqnarray}
where $k_\parallel^\mu  = \Omega \delta^{\mu}_{0} +k_z \delta^{\mu}_{3} $ and 
$\tilde{k}_\parallel^\mu  = k_z \delta^{\mu}_{0} +\Omega \delta^{\mu}_{3} $.
Note that $k_{\parallel,\mu} \tilde{k}_\parallel^\mu =0$ and $\tilde{k}_{\parallel,\mu} \tilde{k}_\parallel^\mu = - k_\parallel^2$.
The definition of the transverse threshold momenta $k_{-}$ and $k_{+}$ are given in Eqs.~(\ref{k-minus}) and (\ref{k-plus}), respectively.
 
By using these results, we can derive the alternative expressions for tensors $I_{i}^{\mu\nu}$ tensors. In particular, one obtains
 \begin{eqnarray}
I_{1}^{\mu\nu}\Big|_{p_z=p_{z}^{(\pm)}}  &=& -4 \pi \ell^2 g_\perp^{\mu\nu} \left[ (n+n^\prime)|qB| -\frac{1}{2} k_\parallel^2 \right]
\left[ \mathcal{I}_{0}^{n,n^{\prime}-1}(\xi)+\mathcal{I}_{0}^{n-1,n^{\prime}}(\xi) \right]
\nonumber\\
&- &4 \pi \ell^2 g_\parallel^{\mu\nu} \left[ (n+n^\prime)|qB| -\frac{1}{2} k_\parallel^2 \right] 
\left[\mathcal{I}_{0}^{n,n^{\prime}}(\xi)+\mathcal{I}_{0}^{n-1,n^{\prime}-1}(\xi) \right]
\nonumber\\
&+& 4 \pi i \ell^4 qF^{\mu\nu}
\left[ (n+n^\prime)|qB| -\frac{1}{2} k_\parallel^2 \right]
\left[\mathcal{I}_{0}^{n,n^{\prime}-1}(\xi)-\mathcal{I}_{0}^{n-1,n^{\prime}}(\xi)\right] 
\nonumber\\
&+& 4 \pi \ell^2 A_{\pm}^{\mu \nu}
\left[\mathcal{I}_{0}^{n,n^{\prime}}(\xi)+\mathcal{I}_{0}^{n-1,n^{\prime}-1}(\xi) \right]  ,
\end{eqnarray}
where we used the following shorthand notation:
\begin{eqnarray}
A_{\pm}^{\mu \nu}
&=& -\frac{1}{2}k_\parallel^\mu k_\parallel^\nu
\left( 1- \frac{4(n-n^\prime)^2 (qB)^2}{(\Omega^2-k_z^2)^2} \right)
+\frac{1}{2} \tilde{k}_{\parallel}^{\mu}\tilde{k}_{\parallel}^{\nu}
\left( 1-\frac{4\left[m^2+(n+n^\prime)|qB|\right]}{\Omega^2-k_z^2} + \frac{4(n-n^\prime)^2 (qB)^2}{(\Omega^2-k_z^2)^2} \right)
\nonumber\\
&\pm & \left( k_{\parallel}^{\mu}\tilde{k}_{\parallel}^{\nu}+k_{\parallel}^{\nu}\tilde{k}_{\parallel}^{\mu}\right)
\frac{(n-n^{\prime})|qB|}{\Omega^2-k_z^2} 
\sqrt{1-\frac{4\left[m^2+(n+n^\prime)|qB|\right]}{\Omega^2-k_z^2} + \frac{4(n-n^\prime)^2 (qB)^2}{(\Omega^2-k_z^2)^2}} .
\end{eqnarray}
Note that the upper and lower signs correspond to $p_{z}^{(+)}$ and $p_{z}^{(-)}$, respectively.

Similarly, by combining the next two tensor functions together, one finds
\begin{eqnarray}
I_{2}^{\mu\nu}+ I_{3}^{\mu\nu}\Big|_{p_z=p_{z}^{(\pm)}}  &=& - 4\pi \ell B_{\pm}^{\mu \nu}
\left[\mathcal{I}_{1}^{n,n^{\prime}-1}(\xi)+\mathcal{I}_{1}^{n-1,n^{\prime}-1}(\xi)\right] 
\nonumber\\
&-& 4\pi \ell 
\left(k_{\parallel}^{\mu}\hat{k}_{\perp}^{\nu}+k_{\parallel}^{\nu}\hat{k}_{\perp}^{\mu} -B_{\pm}^{\mu \nu} \right)
\left[ \mathcal{I}_{1}^{n^{\prime},n-1}(\xi)+\mathcal{I}_{1}^{n^{\prime}-1,n-1}(\xi)\right] 
\nonumber\\
&- & 4 \pi i \ell^3 \frac{qB}{k_\perp} C_{\pm}^{\mu \nu}
\left[\mathcal{I}_{1}^{n,n^{\prime}-1}(\xi)-\mathcal{I}_{1}^{n-1,n^{\prime}-1}(\xi) \right]
\nonumber\\
&- & 4\pi i \ell^3  \frac{qB}{k_\perp} 
\left[ C_{\pm}^{\mu \nu} - \left( k_{\parallel}^{\mu}\tilde{k}_{\perp}^{\nu}-k_{\parallel}^{\nu}\tilde{k}_{\perp}^{\mu}\right) \right]
\left[ \mathcal{I}_{1}^{n^{\prime},n-1}(\xi)-\mathcal{I}_{1}^{n^{\prime}-1,n-1}(\xi) \right],
\label{I_if-2+3-appD}
\end{eqnarray}
where, by definition,
\begin{equation}
B_{\pm}^{\mu \nu}
= \frac{1}{2}\left( k_{\parallel}^{\mu}\hat{k}_{\perp}^{\nu}+k_{\parallel}^{\nu}\hat{k}_{\perp}^{\mu}\right)
\left(\frac{2(n-n^{\prime})|qB|}{\Omega^2-k_z^2} + 1 \right)
\pm \frac{1}{2}\left( \tilde{k}_{\parallel}^{\mu}\hat{k}_{\perp}^{\nu}+\tilde{k}_{\parallel}^{\nu}\hat{k}_{\perp}^{\mu}\right)
\sqrt{1-\frac{4\left[m^2+(n+n^\prime)|qB|\right]}{\Omega^2-k_z^2} + \frac{4(n-n^\prime)^2 (qB)^2}{(\Omega^2-k_z^2)^2}} ,
\end{equation}
and
\begin{equation}
C_{\pm}^{\mu \nu}
= \frac{1}{2}\left( k_{\parallel}^{\mu}\tilde{k}_{\perp}^{\nu}-k_{\parallel}^{\nu}\tilde{k}_{\perp}^{\mu}\right)
\left(\frac{2(n-n^{\prime})|qB|}{\Omega^2-k_z^2} + 1 \right)
\pm \frac{1}{2}\left( \tilde{k}_{\parallel}^{\mu}\tilde{k}_{\perp}^{\nu}-\tilde{k}_{\parallel}^{\nu}\tilde{k}_{\perp}^{\mu}\right)
\sqrt{1-\frac{4\left[m^2+(n+n^\prime)|qB|\right]}{\Omega^2-k_z^2} + \frac{4(n-n^\prime)^2 (qB)^2}{(\Omega^2-k_z^2)^2}} .
\end{equation}

Finally, the last one is given by
\begin{eqnarray}
I_{4}^{\mu\nu}\Big|_{p_z=p_{z}^{(\pm)}}  &=&  8\pi \left[ g_{\parallel}^{\mu\nu}  \mathcal{I}_{2}^{n-1,n^{\prime}-1}(\xi)
-\left(g_{\perp}^{\mu\nu}+ 2\hat{k}_\perp^{\mu} \hat{k}_\perp^{\nu}  \right)
\mathcal{I}_{3}^{n-1,n^{\prime}-1}(\xi)
 \right].
\end{eqnarray}

When calculating the Lorentz contracted expression for the polarization tensor $\mbox{Im} \left[\Pi^{\mu}_{\mu}\right] \sim \sum_{i=1}^{4} g_{\mu\nu}I^{\mu\nu}_{i} $, it is convenient to introduce the following scalar functions: $\mathcal{F}_i  = g_{\mu\nu}I^{\mu\nu}_{i}$. 
By making use of the original definition for $I^{\mu\nu}_{i}$, one finds that 
\begin{eqnarray}
\mathcal{F}_1&=&  8\pi \left[\frac{k_\parallel^2\ell^2}{2} -(n+n^\prime)\right]
\left( \mathcal{I}_{0}^{n-1,n^{\prime}}(\xi) +\mathcal{I}_{0}^{n,n^{\prime}-1}(\xi)  \right)
+8\pi \ell^2 m^2\left( \mathcal{I}_{0}^{n,n^{\prime}}(\xi) +\mathcal{I}_{0}^{n-1,n^{\prime}-1}(\xi)  \right)
\label{F1-orig}
,\\
\mathcal{F}_2&=& g_{\mu \nu} I_{2}^{\mu \nu}=0, \\
\mathcal{F}_3&=& g_{\mu \nu} I_{3}^{\mu \nu}=0,  \\
\mathcal{F}_4&=& = 16 \pi \, \mathcal{I}_{2}^{n-1,n^{\prime}-1}(\xi).
\label{F4-orig}
\end{eqnarray}
By adding together the nonvanishing functions $\mathcal{F}_i$, we derive 
\begin{equation}
\mathcal{F}_1 + \mathcal{F}_4 
= 8\pi \left(n+n^{\prime}+m^2\ell^2\right)\left[\mathcal{I}_{0}^{n,n^{\prime}}(\xi)+\mathcal{I}_{0}^{n-1,n^{\prime}-1}(\xi) \right]
+8\pi  \left(\frac{k_{\parallel}^2 -k_{\perp}^2}{2} \ell^2  -(n+n^{\prime})\right)
\left[\mathcal{I}_{0}^{n,n^{\prime}-1}(\xi)+\mathcal{I}_{0}^{n-1,n^{\prime}}(\xi) \right]. 
\end{equation}
Note that the second term simplifies when photons satisfy the on-shell condition $k_{\parallel}^2 = k_{\perp}^2 $.

\section{Tensor structure of $\mbox{Im} \left[\Pi^{\mu\nu} \right] $}
\label{app:Tensor-structure}

By making use of the expression for the polarization tensor in Eq.~(\ref{Im-Pol-fun}) and the definition of tensors $I_{i}^{\mu\nu}$ in Appendix~\ref{AP-tr}, we find that the imaginary part of $\Pi_R^{\mu\nu}(\Omega;\mathbf{k})$ has the following structure:
\begin{eqnarray}
\mbox{Im} \left[\Pi_R^{\mu\nu}(\Omega;\mathbf{k}) \right] &=&
  \left( \frac{k_\parallel^\mu k_\parallel^\nu }{k_\parallel^2} -  g_{\parallel}^{\mu\nu} \right)  \mbox{Im} \left[\Pi_{1}\right]
+ \left(g_{\perp}^{\mu\nu}+ \frac{k_\perp^\mu k_\perp^\nu}{k_\perp^2} \right) \mbox{Im} \left[\Pi_{2}\right]
\nonumber\\
&+& \left( \frac{k_\parallel^\mu \tilde{k}_\parallel^\nu +\tilde{k}_\parallel^\mu k_\parallel^\nu}{k_\parallel^2}  
+ \frac{\tilde{k}_\parallel^\mu k_\perp^\nu +k_\perp^\mu \tilde{k}_\parallel^\nu}{k_\perp^2}\right) \mbox{Im} \left[\Pi_{3}\right]
+\left(\frac{k_\parallel^\mu k_\perp^\nu +k_\perp^\mu k_\parallel^\nu}{k_\parallel^2 } + \frac{k_\perp^2}{k_\parallel^2 }  g_{\parallel}^{\mu\nu}  - g_{\perp}^{\mu\nu} \right) \mbox{Im} \left[\Pi_{4}\right] \nonumber\\
&+& \left( \frac{F^{\mu\nu} }{B}
+ \frac{k_\parallel^\mu \tilde{k}_{\perp}^\nu -\tilde{k}_{\perp}^\mu k_\parallel^\nu}{k_\parallel^2} \right) \mbox{Im} \left[\tilde{\Pi}_{5}\right]
+ \frac{\tilde{k}_\parallel^\mu \tilde{k}_{\perp}^\nu -\tilde{k}_{\perp}^\mu \tilde{k}_\parallel^\nu}{k_\parallel^2} \mbox{Im} \left[\tilde{\Pi}_{6}\right] ,
\label{ImPi-tensor-app0}
\end{eqnarray}
where we utilized the shorthand notations introduced in Eq.~(\ref{ggkkkk}). Note that $k_\mu \tilde{k}_{\perp}^{\mu} =k_{\mu} \tilde{k}_\parallel^\mu =0$ and $F^{\mu\nu} k_{\perp,\nu} =B \tilde{k}_{\perp}^{\mu}$. One can also check that 
$\tilde{k}_{\perp,\mu} \tilde{k}_{\perp}^{\mu} = k_{\perp,\mu} k_{\perp}^{\mu} = -k_\perp^2$ and 
$\tilde{k}_{\parallel,\mu} \tilde{k}_\parallel^\mu = -k_{\parallel,\mu} k_\parallel^\mu= - k_\parallel^2$.

To simplify the representation of the six component functions in the imaginary part of polarization tensor (\ref{ImPi-tensor-app0}), it is convenient to introduce the following operator:
\begin{equation}
\hat{\cal X} (\ldots)=
  \frac{\alpha N_{f}}{4\pi \ell^4} \sum_{n,n^\prime=0}^{\infty} 
\sum_{\lambda,\eta=\pm 1}\sum_{s=\pm 1} \Theta_{\lambda, \eta}^{n,n^{\prime}}(\Omega,k_z)
\frac{n_F(E_{n,p_z})-n_F(\lambda E_{n^{\prime},p_z-k_z}) }{\eta\lambda  \sqrt{ \left( \Omega^2-k_z^2-k_{-}^2 \right) \left( \Omega^2-k_z^2-k_{+}^2\right)} } (\ldots).
\label{hat-X-app}
\end{equation} 
Note that, with the help of the operator $\hat{\cal X}$, the polarization tensor in Eq.~(\ref{Im-Pol-fun}) can be rewritten in a compact form as $\mbox{Im} \left[\Pi_R^{\mu\nu}\right]  = \hat{\cal X}  \sum_{i=1}^{4}I_{i}^{\mu\nu} $. 

By using the operator in Eq.~(\ref{hat-X-app}), we can also write down the explicit expressions for the individual tensor component functions. In particular, the first four components, defining the symmetric tensor structures, are
\begin{eqnarray}
\mbox{Im} \left[\Pi_{1}\right] &=& 
8 \pi \hat{\cal X} \left(\frac{2(n-n^{\prime})^2}{k_\parallel^2\ell^2} -m^2\ell^2-(n+n^\prime)\right)
\left[\mathcal{I}_{0}^{n,n^{\prime}}(\xi)+\mathcal{I}_{0}^{n-1,n^{\prime}-1}(\xi) \right], 
\\
\mbox{Im} \left[\Pi_{2}\right] &=& -16 \pi \hat{\cal X} \, \mathcal{I}_{3}^{n-1,n^{\prime}-1}(\xi)
\nonumber\\
&=& 
-8 \pi \hat{\cal X}  \Bigg\{
(n+n^{\prime})\left[\mathcal{I}_{0}^{n,n^{\prime}-1}(\xi) +\mathcal{I}_{0}^{n-1,n^{\prime}}(\xi) \right]
-\frac{(n-n^{\prime})^2}{\xi}\left[\mathcal{I}_{0}^{n,n^{\prime}}(\xi) +\mathcal{I}_{0}^{n-1,n^{\prime}-1}(\xi) \right] 
\Bigg\}, 
\\
\mbox{Im} \left[\Pi_{3}\right] &=& \pm 4 \pi \hat{\cal X}  (n-n^{\prime}) 
\sqrt{\left(1-\frac{k_{-}^2}{k_\parallel^2}\right)\left(1-\frac{k_{+}^2}{k_\parallel^2}\right)}
\left[\mathcal{I}_{0}^{n,n^{\prime}}(\xi)+\mathcal{I}_{0}^{n-1,n^{\prime}-1}(\xi) \right],
\\
\mbox{Im} \left[\Pi_{4}\right] &=& -2\pi \hat{\cal X} \Bigg\{ \frac{k_\parallel^2\ell^2+2(n-n^{\prime})}{k_\perp \ell}
\left[\mathcal{I}_{1}^{n,n^{\prime}-1}(\xi)+\mathcal{I}_{1}^{n-1,n^{\prime}-1}(\xi)\right] 
+  \frac{ k_\parallel^2\ell^2 - 2(n-n^{\prime})}{k_\perp \ell}
\left[ \mathcal{I}_{1}^{n^{\prime},n-1}(\xi)+\mathcal{I}_{1}^{n^{\prime}-1,n-1}(\xi)\right] \Bigg\}
\nonumber\\
&=& - 2\pi \hat{\cal X}  \left\{k_\parallel^2\ell^2
\left[\mathcal{I}_{0}^{n,n^{\prime}-1}(\xi)+\mathcal{I}_{0}^{n-1,n^{\prime}}(\xi) \right]
-\frac{2 (n-n^{\prime})^2 }{\xi}
\left[\mathcal{I}_{0}^{n,n^{\prime}}(\xi)+\mathcal{I}_{0}^{n-1,n^{\prime}-1}(\xi) \right] \right\}.
\end{eqnarray}
Similarly, the component functions of the antisymmetric contributions read 
\begin{eqnarray}
\mbox{Im} \left[\tilde{\Pi}_{5}\right]  &=& -  2 \pi i s_\perp \hat{\cal X} \left( k_\parallel^2 \ell^2 - 2(n+n^\prime) \right)
\left[\mathcal{I}_{0}^{n,n^{\prime}-1}(\xi)-\mathcal{I}_{0}^{n-1,n^{\prime}}(\xi)\right] ,
\end{eqnarray}
and 
\begin{eqnarray}
\mbox{Im} \left[\tilde{\Pi}_{6}\right]  &=&  \mp 2 \pi i \hat{\cal X}  \frac{s_\perp  k_\parallel^2 \ell^2}{k_\perp\ell}  \sqrt{ \left(1-\frac{k_{-}^2}{k_\parallel^2} \right)
\left( 1-\frac{k_{+}^2}{k_\parallel^2}\right)}
\left[\mathcal{I}_{1}^{n,n^{\prime}-1}(\xi)-\mathcal{I}_{1}^{n-1,n^{\prime}-1}(\xi) 
+ \mathcal{I}_{1}^{n^{\prime},n-1}(\xi)-\mathcal{I}_{1}^{n^{\prime}-1,n-1}(\xi) \right]
\nonumber\\
&=& \mp 4\pi i   \frac{s_{\perp}}{k_\perp^2}  \hat{\cal X}  (n+n^\prime) \sqrt{ \left(k_\parallel^2- k_{-}^2  \right)
\left(k_\parallel^2- k_{+}^2 \right)}
\left[\mathcal{I}_{0}^{n,n^{\prime}}(\xi)-\mathcal{I}_{0}^{n-1,n^{\prime}-1}(\xi)\right].
\end{eqnarray}

\section{Lowest Landau level approximation}
\label{LLLAL}

In the lowest Landau level  approximation (i.e., $n=n^{\prime}=0$), the explicit result for the Lorentz-contracted imaginary part  of the polarization tensor reads
\begin{eqnarray}
\mbox{Im} \left[\Pi^{\mu}_{R,\mu} (\Omega;\mathbf{k}) \right]  &=&
- \frac{4 \alpha N_f  m^2 \theta\left(\Omega^2-k_z^2-4m^2\right)}{(\Omega^2-k_z^2)\ell^2 R_m} 
\frac{\sinh\left(\frac{\Omega}{2T}\right) e^{-k_{\perp}^2\ell^{2}/2}  }{\cosh\left(\frac{\Omega}{2T}\right)+\cosh\left(\frac{k_z R_m}{2T}\right)} ,
\label{ImPi-LLL}
\end{eqnarray}
where $R_m = \sqrt{1-4m^2/(\Omega^2-k_z^2)}$.
Note that the solutions to the energy conservation equation are given by 
\begin{eqnarray}
p_{z,0}^{(\pm)}&=& 
\frac{k_z}{2}
\pm \frac{\Omega}{2}\sqrt{1-\frac{4m^2}{\Omega^2-k_z^2} } .
\label{pz-roots-ky-LLL}
\end{eqnarray}
On these solutions, the charged particle energies are  
\begin{eqnarray}
E_{0, p_z^{s+}} =E_{0, p_z^{s-}-k_z} = \frac{\Omega+k_z R_m}{2}, \\
E_{0, p_z^{s+}-k_z} = E_{0, p_z^{s-}} =\frac{\Omega-k_z R_m}{2}  . 
\end{eqnarray}
In deriving the result in Eq.~(\ref{ImPi-LLL}), we used $L_{0}^{0}(x)=L_{0}^{-1}(x)=1$
and took into account that, in the lowest Landau level approximation, 
\begin{equation}
\mathcal{F}_{1} = 8 \pi m^2\ell^{2}e^{-k_{\perp}^2\ell^{2}/2} ,
\end{equation}
and $\mathcal{F}_{4}=0$.
As is easy to check, the  result in Eq.~(\ref{ImPi-LLL}) is consistent with the spectral function obtained in
the lowest Landau level approximation in Ref.~\cite{Bandyopadhyay:2016fyd}. With that being said, it is important to remember 
that, as emphasized in the main text, this approximation is not very reliable for calculating the photon production
rate even in very strong magnetic fields.


\begin{thebibliography}{99}
\bibitem{foot1} The absorptive part of the photon polarization function includes the imaginary part of the symmetric tensor structure and the real part of the antisymmetric tensor structure. Note, however, that the latter vanishes at $\mu=0$ because of the charge conjugation symmetry.
\end{thebibliography}
\end{document}